\documentclass[prl,twocolumn,noshowpacs,amssymb,amsmath,aps,floatfix,makecell,longbibliography]{revtex4-2}
\usepackage[colorlinks=true,linkcolor=blue,citecolor=blue,urlcolor=blue]{hyperref}
\usepackage{graphicx,bm}
\usepackage{epstopdf}
\usepackage{bm}
 
\begin{document}

\title{Ballistic-hydrodynamic phase transition in flow of two-dimensional electrons }

\date{\today}

\author{A.~N.~Afanasiev}
\author{P.~S.~Alekseev}
\author{A.~A.~Greshnov}
\author{M.~A.~Semina}

\affiliation{Ioffe Institute, St.~Petersburg 194021, Russia}

\begin{abstract}
Phase transitions are characterized by a sharp change in the type of dynamics of microparticles, and their description usually requires quantum mechanics. Recently, a peculiar type of conductors was discovered in which two-dimensional (2D) electrons form a viscous fluid. In this work we reveal that such electron fluid in high-quality samples can be formed from ballistic  electrons via a phase transition. For this purpose, we theoretically study the evolution of a ballistic flow of 2D weakly interacting electrons with an increase of magnetic field and trace an emergence of a fluid fraction at a certain critical field. Such restructuring of the flow manifests itself in a kink in magnetic-field dependencies of the longitudinal and the Hall resistances. It is remarkable that the studied  phase transition has a classical-mechanical origin and is determined by both the ballistic size effects and the electron-electron scattering. Our analysis shows that this effect was apparently  observed in the recent transport experiments on 2D electrons in graphene and high-mobility GaAs quantum wells.
\end{abstract}

\maketitle

{\em 1. Introduction.} Frequent electron-electron collisions  in high-quality conductors can lead to realization  of the hydrodynamic regime  of charge transport~\cite{Gurzhi}. This regime  was  recently  reported  for high-quality graphene~\cite{graphene,Levitov_et_al,graphene_3,rrecentnest,rrecentnest2}, Weyl semimetals~\cite{Weyl_sem_1,Weyl_sem_2},
 and GaAs quantum wells~\cite{exps_neg_3,exps_neg_1,exps_neg_2,exps_neg_4,example,Gusev_1,Gusev_2,Gusev_2_Hall,je_visc,
 exp_GaAs_ac_1,exp_GaAs_ac_2,exp_GaAs_ac_3,Alekseev_Alekseeva,new1,new2}. Formation of the electron fluid  was detected by   a specific dependence of the resistance on the sample width~\cite{Weyl_sem_1};   by observation of   the negative nonlocal resistance~\cite{graphene,Levitov_et_al,Gusev_2,new1}, the giant negative magnetoresistance~\cite{exps_neg_3,exps_neg_1,exps_neg_2,exps_neg_4,example,Gusev_1,Gusev_2_Hall,Weyl_sem_2,je_visc,new2}, and the magnetic resonance at double cyclotron frequency~\cite{exp_GaAs_ac_1,exp_GaAs_ac_2,exp_GaAs_ac_3,Alekseev_Alekseeva}.

Much attention was paid to  the transition between the hydrodynamic  and non-hydrodynamic   regimes of electron transport.  In Refs.~\cite{rrecentnest,rrecentnest2} precise measurements of  profiles of the Hall electric field and the current density of 2D electron flow in graphene stripes were performed. This allowed  to detect the Ohmic, the hydrodynamic, and the ballistic flows at varying  temperature, electron density and magnetic field. In particular, a peculiar nonmonotonic  magnetoresistance  reflecting the ballistic and the hydrodynamic transport regimes was observed~\cite{rrecentnest,rrecentnest2}.   Similar magnetoresistance was detected in long samples of    high-quality GaAs quantum wells~\cite{Gusev_1,Gusev_2_Hall}, that, apparently, also evidences the ballistic-hydrodynamic transition. In Refs.~\cite{new1,new2}  the transitions from the ballistic to the hydrodynamic regimes in GaAs quantum wells with changing of the sample geometry, temperature and magnetic field  were vividly demonstrated.

A  theory of  2D electron flow  in samples with macroscopic obstacles was  constructed  in Ref.~\cite{Levitov_et_al_2}. In the absence of magnetic field, the ballistic-hydrodynamic  transition   occurs is such system with changing the interparticle scattering rate and  has the type of  a smooth
crossover. A numerical theory of the  ballistic-Ohmic transition and the hydrodynamic transport for 2D electrons in stripes in a perpendicular  magnetic field was developed in Ref.~\cite{Scaffidi}.  At weak interparticle and disorder scattering rates, the longitudinal and  the Hall resistances of a stripe as functions of magnetic field $B$ exhibit   kinks at the field $B=B_c$ above which the diameter of the electron cyclotron orbit,  $2R_c$, becomes smaller than the sample width, $W$. In Ref.~\cite{pohozaja_statja} the same system was theoretically studied  in more details.
 It was shown that
 an increase of the curvature of the Hall electric field characterizes the transition from the ballistic to the hydrodynamic  regimes.


In this Letter,  we demonstrate that the hydrodynamic regime of transport of 2D electrons in high-quality stripes is formed from the ballistic regime via a genuine phase transition with the increase of magnetic field $B$. For this purpose,  first, we reveal that in the lower vicinity of the critical field, $ 0 < B_c - B \ll B_c $, the momentum relaxation due to collisions of electrons with the stripe edges becomes  strongly  suppressed due to the ballistic size effects, thus even weak electron-electron scattering begins to be important for the flow formation. Second, we show that in the upper vicinity of $B_c$,  $ 0 < B - B_c  \ll B_c $, the  emerging  ``central'' electrons, which are not scattered at the edges, critically change the type of the electron distribution and become the nucleus of a collectivized  fluid phase. We develop a mean field model based on the classical kinetic equation to describe these critical transport regimes of the ballistic-hydrodynamic phase transition. Our analysis of the results of  experiments~\cite{rrecentnest,rrecentnest2} evidences  that formation of the hydrodynamic regime from the ballistic one was realized in them via such phase transition.

Using the developed approach, we also obtain new results on  the ballistic transport in stripes at low magnetic fields, $B \ll B_c $. In particular, we  show that  the interplay of the ballistic effects and the interparticle scattering induces a strongly nonuniform  electron flow and a non-trivial character of  the Hall effect.

\begin{figure}[t!]
	\includegraphics[width=0.6\linewidth]{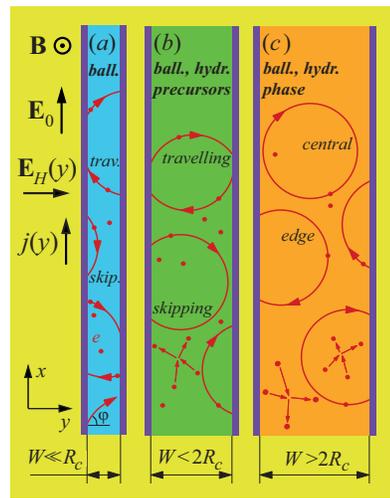}
	\caption{
	Two-dimensional electrons in a long   sample at low, $ W \ll R_c$  (a); intermediate, $ W \sim R_c$, $ W <2 R_c$ (b); and moderately high, $ W \sim R_c$, $ W >2 R_c$ (c)  magnetic fields. The ballistic-hydrodynamic phase transition occurs at the critical field $B_c$ corresponding to $2R_c=W$. In its lower vicinity,  $0<B_c-B \ll B_c$, electrons moving along the skipping orbits close to complete cyclotron circles undergo slow momentum relaxation. They are precursors of hydrodynamic flow. Above the critical point, $B>B_c$, a group of the central electrons appears those do not collide with the edges, representing a nucleus of the hydrodynamic fluid phase.
	}
	\label{Fig1}
\end{figure}

{\em 2.  Ballistic regime.} We consider a flow of 2D degenerate electrons in a long sample with rough edges in a perpendicular magnetic field~$\mathbf{B}$ (see Fig.~\ref{Fig1}).  Electrons are diffusively  scattered on the rough sample edges leading to momentum relaxation. In the bulk of the sample, electrons collide with each other and their total momentum is conserved. We assume that the  rate $\gamma $ of the electron-electron scattering is weak:  $W  \ll l $, where $l=v_F /\gamma$ is the  mean free path and $v_F$ is the Fermi velocity. We describe the transport in this system by the non-equilibrium part $\delta f(y,\varphi, \varepsilon) $ of the distribution function $f=f_F+\delta f$ determined by the linearized kinetic equation:
 \begin{equation}
 \label{1}
 v_F \cos \varphi \,  \frac{\partial \:  \delta f }{\partial y } + \frac{e}{m}  \, \mathbf{E} \cdot \frac{\partial f_F }{\partial \mathbf{v} }
  -\omega_c \,  \frac{\partial \: \delta  f }{\partial  \varphi  } = \mathrm{St} [ \, \delta f\, ]
\end{equation}
with the diffusive boundary conditions at $y=\pm W/2$. Here $\varepsilon$ is the electron energy, $\varphi$ is the angle of the electron  velocity $ \mathbf{v}/v_F = (\cos \varphi \, , \, \sin \varphi )  $,   $f_F$  is the Fermi distribution, $ \omega_c  = v_F /R_c $  is the cyclotron frequency, $e$ and $m$ are the electron charge and mass, $\mathbf{E} = \mathbf{E}_0 +\mathbf{E}_H$ is the total electric field, ${\bf E}_0$ is the applied field, ${\bf E}_{H}$ is the Hall field, $\mathrm{St}[\delta f ] = -\gamma \, (\delta f - \hat{P}[\delta f] ) $ is the simplified interparticle collision operator,  in which $\hat{P}$ is the projector onto the zeroth and first harmonics of $\delta f$ by $\varphi$.

At the magnetic fields $B$ below the critical field,  $B<B_c$, when $ 2R_c >W$, each electron is predominantly scattered at the edges.  The transport is ballistic in  almost all such $B$, and the interparticle collisions can constrain the time which electrons spend on the ballistic trajectories. Our analysis~\cite{SI} based on Eq.~(\ref{1}) shows that the ballistic regime has a fine structure, namely, it is divided into the   three  subregimes.

In the first ballistic subregime of low fields $B \ll B_c (W/l)^2$, the electron trajectories are almost straight. Their maximum length is limited by the interparticle scattering length $l$. This subregime was studied in Refs.~\cite{we_6,we_6_2} and in this work~\cite{SI}. Depending on the arrangement  of trajectories, electrons are divided on the two groups: the ``traveling electrons'' which, after scattering on an edge, reach the other edge or scatter in the bulk, and the ``skipping electrons'' which return to the same edge after scattering on it [see Fig.~\ref{Fig1}(a,b)].  Most of electrons belong to the first type. Electrons of the second type are located in  the edge vicinities,  $W/2-|y| \lesssim l^2/R_c$, and their velocity angles are  $\varphi \approx \pm \pi/2$.

In the central part of the sample,  $W/2-|y| \gg l^2/R_c$,  the Hall electric field $E_H (y)$ is related to the dynamics of the ``traveling electrons''~\cite{SI}. The resulting local Hall resistance, $\varrho _{xy} (y)  =E_H (y) /j (y)$,  turns out  to be one half of the Hall resistance  of macroscopic Ohmic samples~\cite{we_6_2}:
\begin{equation}
\label{2}
    \varrho_{xy} = R_H^{(0)}   B / 2
  \: , \quad R_H^{(0)} = 1/(n_0ec) \:,
\end{equation}
where $j (y)$ is the current density, $n_0$ is the electron density, and $c$ is the velocity of light. In Ref.~\cite{we_6_2} this result was obtained from a straightforward solution of kinetic equation~(\ref{1}). In this work we reveal~\cite{SI}   the physical essence of result (\ref{2}). Namely, the value  $E_H$  yielding Eq.~(\ref{2}) corresponds to the balance of the Hall force $eE_H$ and the component of the Lorentz force $ \Delta F_{L,y} (t) = eB  a_x t/c $, averaged over all the traveling electrons in the region $W/2-|y| \gg l^2/R_c$. Here $a_x = eE_0/m$  is the acceleration  of electrons by the field  $E_0$ and  $t = t(y,\varphi)$ is the time passed since the scattering at the edge.

\begin{figure}[t!]
	\includegraphics[width=0.98\linewidth]{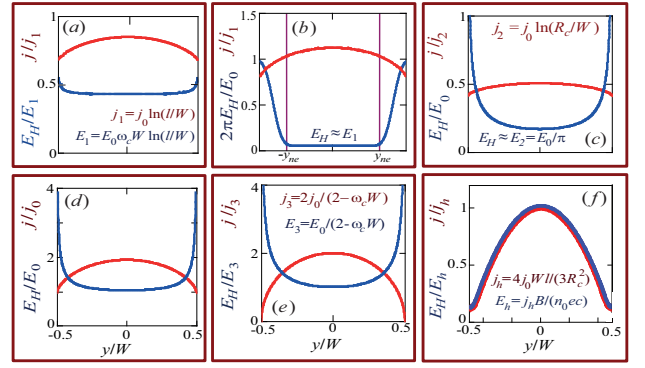}
	\caption{
	Current density $j(y)$ and  Hall electric field $E_H(y)$ at various magnetic fields $B$:
(a) the first  ballistic subregime, in the limit $B\to 0$ [only the flow in the central part of the sample,
	$W/2-|y| \gg  l^2/R_c $,  is shown];
(b) the middle part of the first ballistic subregime, $R_c \gg l^2 /W $  [schematically; violet lines depict   the boundaries of the near-edge regions where skipping electrons propagate, $y_{ne} \approx \pm ( W/2 - l^2/R_c  )$];
 (c) the second ballistic subregime, $ W \ll R_c   \ll l^2 /W $;
 (d)    the middle part of the third ballistic subregime, $ 0  <  2R_c /W-  1   \sim 1  $;
 (e)~the upper part of the third
	ballistic subregime near the critical field, $W/l \ll 2R_c/W-1 \ll 1   $;
and
(f)    the hydrodynamic regime with a Poiseuille flow,   $W \gg R_c $ (schematically).
 	}
 	\label{Fig2}
\end{figure}

In the vicinities of the edges, $W/2-|y| \lesssim  l^2/R_c$, the Hall field $E_H (y) $ and the current density $ j (y) $ are strongly affected by the skipping electrons [see Fig.~\ref{Fig2}(b)]. The resulting values of $E_H (y) $ and $ j (y) $ determine the resistances  $\varrho_{xx} $ and  $ \varrho_{xy} $ of the whole sample, provided it is sufficiently long and straight~\cite{SI}.

In the second ballistic subregime, $B_c(W/l)^2 \ll B \ll B_c$, electron trajectories become substantially bent. Their maximal  length is now limited by their geometry.  This subregime was studied in Refs.~\cite{Scaffidi} by numeric solution of Eq.~(\ref{1}) and in Refs.~\cite{pohozaja_statja} by its analytical solution accounting only for the departure term $-\gamma\, \delta f $  in the operator $\mathrm{St}$. The resulting longitudinal and Hall resistances exhibit the singular behavior, $ \varrho_{xx,xy}(B) \sim 1/\ln ( \sqrt{R_c /  W}\,)$,  originating from the shortening of the longest ballistic trajectories  with the increase of $B$~\cite{SI}.

In the third ballistic  subregime,  $ B_c  W/l \ll B_c -B \lesssim B_c  $, the number of the skipping electrons becomes relatively large: comparable with or even greater than the number of traveling ones [see Fig.~\ref{Fig1}(b)]. To our knowledge this subregime has not been noticed and studied yet. In order to satisfy the condition $j_y=0$ of the absence of the transverse current, the $\mathbf{E}_0\times \mathbf{B}$-drift contribution  related to all electrons, $j^{(0)}_y = n_0 e c E_0/B$, is compensated by the excess and the deficiency of non-equilibrium traveling electrons
 with $v_y(t) >0 $ and  with $v_y(t) <0 $.  Non-equilibrium   skipping electrons do not compensate $j^{(0)}_y$,
  as $v_y(t) >0 $ and  $v_y(t) <0 $ symmetrically for each skipping  trajectory.
   The diffusive reflection of electrons from the edges occurs with equal probabilities for all~$\varphi$.
    Thus, at $   2R_c/W  - 1  \ll 1  $,  when the skipping electrons dominate,
    the whole electron density strongly increases  (as compared with  the  case  $    2R_c/W  - 1  \sim  1  $)
    in order to compensate  $j^{(0)}_y $
     by the relatively small part of the traveling electrons.

This dynamics is described by the distribution~\cite{SI}:
\begin{equation}
 \label{3}
  \delta f (y,\varphi, \varepsilon )  =  [E_0 / \, ( \omega_c  u )]\:  \chi(y,\varphi)  \:  f_F ' (\varepsilon)
   \:,  \;\;
\end{equation}
where the behavior of the factor $\chi $, $\chi(y,\varphi) \approx  1$ at $-\pi + \varphi_-  < \varphi < \varphi_+ $ and $\chi(y,\varphi) \approx  -1$ at $\varphi_+  < \varphi < \pi +\varphi_ - $, reflects the domination of the skipping electrons
\{here $\varphi_{\pm }    ( y ) = \arcsin [1- (W /2\pm y)/R_c] $\}.
 The small parameter  $ u (B) = (2/\pi)( 2- W/R_c ) \ll  1 $ in Eq.~(\ref{3})
  shows how close is $B$ to the critical field $B_c$.
  The resulting current density  and the Hall field   in the main order by $u $ take the form:
\begin{equation}
 \label{4}
  j(y)= 2 r(y) j_0   / \, (\pi u)
  \:, \qquad
   E_H(y)=   2 E_0/ \, [  \pi \, r(y)   \, u] \:,
\end{equation}
where  $j_0 = n_0 e^2 E_0 W/ (v_F m)   $ and   $r(y) = \sqrt{1-(y/R_c)^2}$.  The magnitudes of $  j(y)$ and $   E_H(y)$ rapidly increase as $B$ approaches $B_c$. The averaged  resistances $\varrho _{xx}  = E_{0} / \langle j(y) \rangle  $ and $\varrho _{xy}  = \langle E_H(y) \rangle   / \langle j(y) \rangle  $ determined by $\delta f $ after Eq.~(\ref{3})  in the two main orders by $\sqrt{u}$  are:
\begin{equation}
 \label{5}
  \varrho _{xx}  (B)=  2   \varrho _0  u
  \:, \qquad
   \varrho _{xy} (B) =  R^{(0)}_H  B  \, F(u) \:,
\end{equation}
where $ \varrho _0 =  E_0 / j_0 $ and  $ F(u) = 1 - \sqrt{ u / \pi } $. The vanishing of  $\varrho _{xx}$ as $\sim u $  reflects
 the transitional character
 of the ballistic electron dynamics at $B\to B_c$ [see Fig.~\ref{Fig1}(b)].

The evolution of   $j(y)$ and $E_H (y)$ in the  ballistic subregimes with the increase of $B$ are shown in Figs.~\ref{Fig2}(a-e).

{\em 3. Phase transition.} At the fields $B$ in the upper and the lower vicinities  of
$B_c$, $|B -B_c| \ll B_c $,  most of  electrons are the ``edge electrons''
 those move along the skipping trajectories those hit one of  the edges.  In the upper vicinity, when $ W > 2R_c$,
  a small group of the ``central electrons'' arises those never touch the edges    [Fig.~\ref{Fig1}(c)].

In the nearest lower vicinity of $B_c$,
$  0 <  B_c -B \lesssim B_c W /l$,
 the imbalance i densities of the left-edge and the right-edge skipping  electrons increases  dramatically,
   as  ballistic distribution function~(\ref{3}) and values~(\ref{4}) diverge by $u \to 0 $ at  $W/l  \to 0$.
   Therefore the electron dynamics at such $B$ is to be controlled not only by their scattering on the edges, but also by  interparticle collisions.

To describe such semiballistic flow, first, we calculate~\cite{SI} the trial distribution function, similar to purely ballistic function~(\ref{3}), but additionally accounting  for the departure term $-\gamma \, \delta f $ in the operator $\mathrm{St}$ in Eq.~(\ref{1}):
\begin{equation}
  \label{6}
  \delta f_d (y,\varphi, \varepsilon ) =\frac{ E_0 /\omega_c  }{    u + W/l   }\:  \chi(y,\varphi)  \,  f_F ' (\varepsilon)
 \:.
\end{equation}
The averaged current density and the Hall field corresponding to $\delta f_d$ rapidly increase  at $B \to B_c$ up to the values limited by
 the slow scattering rate $\gamma  = v_F/l$:
\begin{equation}
  \label{7}
  j_d=  \frac{  j_0 /2 }{  u  +  W/l  }
  \:, \quad
 E_{H,d}= \, \frac{  E_0 \, F(u)   }{  u   +W/l }\:.
\end{equation}

Second, to describe the flow at $  0<  B_c -B \lesssim B_c W /l$, we need to account for the effect of the  arrival term $\gamma \hat{P} [\delta f ] $  in Eq.~(\ref{1}).   Indeed,   the departure term  $-\gamma\, \delta f  $  dominates in the first ballistic subregime \cite{we_6}, both the departure and the arrival terms are relatively small in the second and the third ballistic  subregimes \cite{pohozaja_statja,SI},  whereas in a well-formed hydrodynamic flow at $W\gg R_c$   they are close one to other~\cite{we_6}. An estimate shows that for function $ f_d $~(\ref{6}) at $  0<  2R_c -W \lesssim W^2/ l$ these two terms  have the values of the same order of magnitude. In this connection, we propose a mean field model based on the approximation of the arrival term $\gamma \hat{P}[\delta f]$~\cite{SI} by its averaged by $y$ value, whose main part is: $ \gamma \,  \sin \varphi  \, j \,/ (n_0/m) $. After this substitution, the external field $E_0$ in kinetic equation~(\ref{1}) is changed on the effective one: $ \tilde{E}_0 = E_0 + \gamma j/ (n_0/m) $. As a result, the self-consistent distribution~$f$ and averaged current density $j$ are given by semiballistic formulas~(\ref{6}) and
 (\ref{7}) with $E_0 \to \tilde{E}_0$. For $j$ we obtain:
\begin{equation}
 \label{8}
    j= \frac{1}{2} \,  \frac{  j_0 + j \, W/ \, l  }{  u  + W /l }
 \end{equation}
This mean-field-type equation accounts for the redistribution of momentum between the skipping electrons in their collisions with each other, while
 formulas ~(\ref{6}) and  (\ref{7})  imply  the relaxation of momentum
  in  scattering of electrons in the bulk. The solution of Eq.~(\ref{8}) is $ j=  j_0  / [ 2u  + W /l ] $. To find the Hall field near the critical point, we should substitute the renormalization  $E_0 \to  \tilde{E}_0 $  in the  semiballistic value $E_{H,d}$~(\ref{7}), that   yields: $  E_H =   E_0  F(u)  / [ u   +  W / (2l)]  $.

In the upper vicinity of  the transition point,   $  0<  B -B_c \ll B_c $, when the relative density of the central electrons
  is small,  $\alpha_c =(W-2R_c)/W \ll  1$, each edge electron is
   still scattered predominantly on  the edges and on  the other  edge electrons.
       Similarly as for the flow in the lower vicinity of $B_c$,
     $  0<  B _c-B \ll B_c $  the distribution function of these electrons,   $\delta f_e$,
   is  given  by  a formula based on the semiballistic distribution~$  \delta f _d  $~(\ref{6}).
   Therefore  the departure and the arrival terms of $\mathrm{St} $ in Eq.~(\ref{1})
   with this $ \delta f _{e} $
 are also of the same order of magnitude.  To account the arrival term $\gamma P[\delta f ]$,
 we  again substitute it by its averaged value,
  which is mainly proportional to  the averaged  current $ j=j_e+j_c $
   corresponding to $ \delta f = \delta f _ e + \delta f _ c $.
     As a result, the function  $\delta f_e$ is given by Eq.~(\ref{6}) at $u=0$ with
  the sample width  $W  $  changed on the width $\tilde{W}= 2R_c $ of the subregion with the edge electrons and the
   renormalized electric field  $ E_0 \to \tilde{E}_0 = E_0 + \gamma (j_e+j_c)/ (n_0/m) $.
 Correspondingly, for the current component  $j_e$
we should  use   Eq.~(\ref{8}) at $u=0$   with  $j=j_e+j_c$ and the density factor $ \alpha _e = \tilde{W}/W$.

All the central electrons have almost coinciding trajectories
and are scattered mainly by the edge ones [see Fig.~1(c)].
   Thus the   flow of the central electrons  is similar to an  Ohmic one, and
   their component $j_c$ is given by the Drude formula
   with the density factor $ \alpha_c   $ and the same $ \tilde{E}_0  $.
 In the distribution of the central electron  $\delta f_c $
 the first angular harmonic dominate, unlike the  semiballistic function
 $\delta f _b $~(\ref{6}),  which  is discontinuous in $\varphi$
 and, thus, contains many comparable harmonics by $\varphi$~\cite{SI}.

   We arrive the mean field equations for $j_e$ and $j_c$~\cite{SI}:
\begin{equation}
\label{9}
\begin{array}{c}
   j_e=  \alpha_e \, (   \,  j_{cr}    +    j_e+j_c    \,    ) / \,  2
     \:,
 \\
    j_c = \alpha_c  \, (   \,  j_{cr}     +   j_e+j_c     \,  ) \:,
 \end{array}
\end{equation}
where  $j_{cr}   = j_0 l/W $. These  equations are similar by their  meaning to the one-component equation~(\ref{8}), but accounts for the appearance
  at $B>B_c$ of the two electron  species. Solution of~(\ref{9}) yields: $j = (1+2 \alpha_c ) j_{tr}  $.

The Hall field is also related to the edge and the central electrons: $ E_H = E_{H,e}+E_{H,c}$. The first term is calculated by Eq.~(\ref{7}) at $u=0$ with the factor $\alpha _e $ and  the substitutions $E_0 \to \tilde{E}_0$ and  $W \to \tilde{W}$.  According to the Ohmic-like form of the distribution $\delta f_c$, the term $E_{H,c}$ is given by the Drude formula $ E_{H,c} = \omega_c j_c /(n_0/m)$. As a result, we obtain:
 $ E_H = (1 + 2 \alpha_c )(l/R_c)   E_0 $.

\begin{figure}[t!]
	\includegraphics[width=1.0\linewidth]{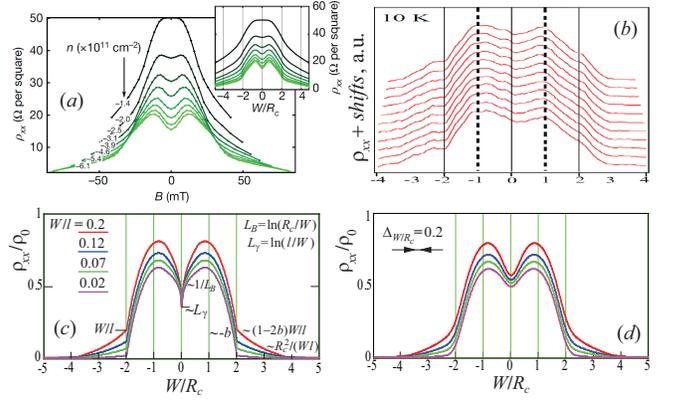}
	\caption{ Longitudinal resistance  $\varrho_{xx}$  of long samples as a function of magnetic field $B \propto W/R_c$.
 Panels~(a) and~(b) present experimental results for graphene stripes and are taken from
	Refs.~\cite{rrecentnest} and \cite{rrecentnest2}, respectively. Different curves correspond to varying 2D electron densities, controlled by the gate voltages.
 Distortions of the curves in panel~(b), asymmetric in $B$, can be due to
some contribution from the Hall resistance $\varrho_{xy}$ in the measured data.  (c): Results of our theory  for several the interparticle scattering rates $\gamma$. In panel (d) we
	plot the  curves from panel (c), smoothed by convolution with a Gaussian weight function
$ G_\Delta (B)$
with the width $\Delta_{ W/R_c} $, that simulates contribution from sample corners,  several sections of a long sample with
	varying widths, and other imperfections.
	}
 	\label{Fig3}
\end{figure}

The described change in electron dynamics above and below the critical field is reflected in kinks in the obtained  magnetic dependencies   $j(B)$ and $E_H(B)$. Next, for $\varrho_{xx}$ in the main orders by $|b| \ll 1$ and $W/l \ll 1 $ we obtain~\cite{SI}:
\begin{equation}
\label{10}
  \frac{   \varrho _{xx} (B)  }{ \varrho _ {cr} } =
    1
    -  b\,
 \left\{
  \begin{array}{l}
   8l  /  ( \pi W ) \:, \;\;\;\; b< 0  \,, \; |b| \ll 1
 \\
 2 \:, \quad  \; 0 <b \ll 1
 \end{array}
 \right.
 .
 \end{equation}
where $\varrho _ {cr} = E_0 / j_ {cr}$ and  $ b = (B -B _c )/  B_c$. For the Hall resistance $\varrho_{xy}(B)$  in the main order by $W / l  $ and the two first orders by $\sqrt{|b|} $ we obtain the same result as in Eqs.~(\ref{5}) at $ b<0$, while  at $b> 0$  the above formulas for  $j$ and $E_H$ yield:  $\varrho_{xy}(B)  \equiv R_H^{(0)}B $. The kinks in the obtained   longitudinal and the Hall resistances at $B=B_c$  evidence  that the formation of the hydrodynamic flow from the ballistic one is realized via a phase transition.

With the increase of $B$ and $\alpha_c $,
 the  collisions between the central electrons  become important, therefore $j_c$ becomes non-uniform by $y$.
   The hydrodynamic-ballistic flow at $\alpha_c \sim 1 $ was numerically
 studied in Refs.~\cite{Scaffidi,pohozaja_statja}.
At $\alpha_c \gg 1 $ the central electrons dominate everywhere except the edges vicinities, $ W/2-|y| \sim R_c$, and the Poiseuille flow $j_c(y) \sim (W/2)^2 - y^2$ is formed [see Fig.~\ref{Fig2}(f)].
  The resulting resistance $\varrho_{xx}$ is determined by the viscosity:  $\varrho_{xx} \sim \eta_{xx} /W^2 $,
  $ \eta_{xx} \sim \gamma /\omega _c^2 $,
   while the Hall resistance  $\varrho_{xy}$
   is  close  to $\varrho_{xy } ^{(0)} =R_H^{(0)} B$~\cite{je_visc}.

In Fig.~\ref{Fig3} we compare the results of experiments \cite{rrecentnest,rrecentnest2}  on 2D electron transport in high-quality graphene stripes with our theoretical results. Both the theoretical and the experimental resistances $\varrho_{xx}$ have similar profiles, including the minimum at $W /R_c \ll 1$, the maximum at $ W /R_c\sim 1 $, and the kink at  $ W  /R_c = 2 $. Convolution of the calculated dependencies  $\varrho_{xx} (B)$ with a weight function
 $ G_\Delta (B)$, simulating the imperfection of the sample, leads to a very good agreement of the shapes of the observed and calculated curves [compare (a) and (d)]. In Supplementary material~\cite{SI} we also compare our results with  preceding theories~\cite{Scaffidi,pohozaja_statja} and other related experiments~\cite{exps_neg_3,exps_neg_1,exps_neg_2,Gusev_2_Hall}. Numerical  solution~\cite{Scaffidi} of Eq.~(\ref{1}) for the stripes
 in which the scattering on disorder dominates leads to the dependencies $\varrho_{xx,xy}(B)$ almost identical to the ones calculated within our theory based on Eq.~(\ref{1}) with $\mathrm{St} =\mathrm{St}_{dis}$ (see details in~\cite{SI}).
Our analytical results for the second ballistic subregime coincide with the ones obtained in Ref.~\cite{pohozaja_statja}.
 The longitudinal and Hall resistances
   observed in Ref.~\cite{Gusev_2_Hall} in long high-quality samples of GaAs quantum wells are in a good agreement with the calculated  dependencies~$\varrho_{xx,xy}(B)$ ~\cite{SI}.

{\em 4. Conclusion.} The phase transition between the ballistic and the hydrodynamic regimes of transport  has been revealed and studied for 2D electrons in long samples in a magnetic field. Analysis of magnetotransport experiments~\cite{rrecentnest,rrecentnest2,Gusev_2_Hall} on high-quality stripes of graphene and GaAs quantum well shows that this transition was apparently observed in them.

Similar magnetic-field-induced  ballistic-hydrodynamic phase  transitions
   may be possibly realized  also  in materials with other geometries of 2D electron flows~\cite{SI}.

\begin{acknowledgments}
We  thank A.~I.~Chugunov, L.~E.~Golub, and A.~V.~Shumilin for fruitful discussions. The study was supported by  the Russian Foundation for Basic Research (Grant  No. 19-02-00999) and by the Foundation for the Advancement of Theoretical Physics and Mathematics ''BASIS''.
\end{acknowledgments}


 \bibliography{bib_3}


\clearpage

\setcounter{equation}{0}
\setcounter{figure}{0}
\setcounter{section}{0}
\makeatletter
\renewcommand{\thefigure}{S\arabic{figure}}
\renewcommand{\thesection}{S\Roman{section}}
\renewcommand{\theequation}{S\arabic{equation}}

\onecolumngrid
\begin{center}

{\large {\bf Supplementary material to ``Ballistic-hydrodynamic phase transition  in  flow  of two-dimensional electrons'' }
\linebreak
}

{
 A. N. Afanasiev, P. S. Alekseev, A. A. Greshnov, and M. A. Semina
\linebreak \linebreak}
{\small
Ioffe Institute, St.~Petersburg 194021, Russia
\linebreak
}
\end{center}

{\small
Here we present the details of our theoretical model~(Sec.~1), the solution of its equations
 for the ballistic regime and the derivation of  the properties of the resulting ballistic flow~(Sec.~2),
 and the   detailed report on the construction of the mean field theory for
  the ballistic-hydrodynamic phase transition~(Sec.~3). We also compare in detail our results with the results of
preceding theoretical and experimental studies~(Sec.~4).
\linebreak
\linebreak
\linebreak}
\twocolumngrid

\section{1. Model}
\label{Sec:S1}

In order to study the transitions from the ballistic regime of 2D electron transport to the hydrodynamic or the Ohmic ones, we consider  a flow in a long sample with the width $W$ and the length $L \gg W$ with the rough longitudinal edges (see Fig~\ref{Fig1} in the main text). Herewith we use the simplified forms of the  electron-electron and the disorder collision integrals,  allowing the analytical solution   of the kinetic equation.  Such model has been developed in Refs.~\cite{we_6, we_6_2} to study the ballistic transport of 2D interacting electrons  in the limit of very low magnetic field, $B \to 0$ (here and below~[1], [2], [3], and so are the references the main text).

We seek for the linear response of 2D electrons to a homogeneous electric field $ {\bm E}_0||x $ in the presence of external magnetic field ${\bm B}$ perpendicular to the sample  plane (see Fig.~\ref{Fig1} in the main text).   The corresponding distribution function of 2D electrons acquires a nonequilibrium part: \begin{equation} \delta f(y,\mathbf{p}) = -f_F'(\varepsilon) f(y, \varphi,\varepsilon) \:,\end{equation}
 where $f_F (\varepsilon )$  is the Fermi distribution function, $\varepsilon = p^2/(2m)$ is the electron energy,
 $\varphi$ is the angle between the electron velocity $\mathbf{v} = v(\varepsilon) [\, \sin \varphi , \cos \varphi \, ]$ and the normal to the left sample edge (see Fig.~\ref{Fig1} in the main text),  $\mathbf{p}=m \mathbf{v} $ is the electron momentum, $m$ is the electron mass, and the factor $f (y,\varphi,  \varepsilon )$ is linear in $E_0$: $f \sim E_0$. The dependence of $\delta f$  on the coordinate $x$ is absent since $L \gg W$. We also omit below the energy dependence of the electron velocity $ v(\varepsilon)= \sqrt{2\varepsilon/m} $ and of the factor $f(y,\varphi, \varepsilon)$ in the nonequilibrium part of distribution function $\delta f(y,\mathbf{p})$.
 This simplification of $f$ is valid for 2D degenerated
  electrons interacting by Coulomb's law  for the viscous~\cite{Alekseev_Dmitriev} and, apparently, the ballistic regimes of charge transport.

Hereinafter,  we use the units in which the absolute value of the electron velocity, $v(\varepsilon) \equiv v_F$, and of the electron  charge, $e$, are set to be unity. So coordinate, time, and reciprocal electric field, $1/E_0$, have the same units.

The kinetic equation for the nonequilibrium distribution function $f(y , \varphi)  $ takes the form:
\begin{equation}
\label{kin_eq}
\cos\varphi \, \frac{\partial f}{ \partial y }
 - \sin \varphi \, E_0 - \cos \varphi \, E_H -
  \omega_c \, \frac{\partial f}{ \partial \varphi } =
 \mathrm{St} [f]
 \:,
\end{equation}
where  $\omega_c =eB /mc $ is the cyclotron frequency, $E_H$ is the Hall electric field arising due to redistribution of electrons in the presence of magnetic field, and the collision integral $\mathrm{St}[f] $ describes both momentum-conserving electron-electron collisions and dissipative scattering by bulk disorder:
\begin{equation}
 \label{St_N__St_U}
 \begin{array}{c}
  \displaystyle
\mathrm{St}  [f]  = - \gamma \, f +
 \gamma_{ee} \hat{P} [f] + \gamma' \hat{P}_0 [f]
   \: ,
 \end{array}
\end{equation}
 where $\gamma_{ee}$ and $\gamma'$ are electron-electron  and disorder scattering rates,  $\gamma = \gamma_{ee} + \gamma' $ is the total scattering rate, $\hat{P}$ and $\hat{P}_0 $ are the projector operators of the functions $f(\varphi)$ onto the subspaces $\{1,e^{\pm i \varphi} \}$ and $\{ 1 \}$,  respectively. Such collision integral conserves perturbations of the distribution function corresponding to a nonequilibrium density. It also describes the conservation of momentum in the inter-particle scattering, when there is no disorder ($\gamma' = 0 $).

We consider that the longitudinal sample edges are rough. Thus the scattering  of electrons on them is diffusive  and  the boundary conditions for the distribution function take the form~\cite{Beenakker_Houten_obz,we_6_2}:
\begin{equation}
\label{bound_cond_1}
 \begin{array}{c}
   \displaystyle
   f(- W/2, \varphi)=c_l \, , \;\; - \pi/2 < \varphi < \pi / 2
   \:,
   \\
   \\
   \displaystyle
   f( W / 2, \varphi) = c_r \, , \;\;  \;\; \pi/2 < \varphi < 3 \pi /2
   \:,
   \end{array}
\end{equation}
(see also Fig.~\ref{Fig1} in the main text). Here the quantities $c_{l} = c_{l} [f] $ and  $c_{r} = c_{r} [f]$ are proportional to the $y$ components of the incident particle flows on the left $(y= - W/2)$ and the right $(y=  W/2)$ sample edges:
\begin{equation}
  \label{c_12}
  \begin{array}{c}
  \displaystyle
  c_l =  - \frac{1}{2} \int _{\pi/2} ^{3\pi/2}
   d \varphi' \: \cos\varphi' \, f(-W/2, \varphi')
     \: ,
   \\
   \\
   \displaystyle
   c_r = \frac{1}{2} \int _{-\pi/2} ^{\pi/2}
   d \varphi' \: \cos\varphi' \, f(W/2, \varphi')
   \:.
   \end{array}
\end{equation}
These boundary conditions indicate that (i) the probability of the electron reflection from the rough edges is independent on the reflection angle $\varphi$ and (ii) the transverse component of the electron flow,
\begin{equation}
\label{jy_def}
j_y(y) = \frac{n_0}{\pi m}
 \int _0 ^{2\pi} d \varphi' \: \cos \varphi' \, f(y, \varphi')
 \:,
\end{equation}
vanishes at the edges, $j_y|_{y=\pm W/2} =0 $  [thus, it is zero everywhere in the sample, $j_y \equiv 0$, due to the continuity equation $\mathrm{div}\,  \mathbf{j} = j' _y= 0$].

The longitudinal current density along the sample is:
 \begin{equation}
 \label{def_j}
  j(y) = \frac{n_0}{\pi m} \int _{0} ^{2 \pi }
  d \varphi' \: \sin\varphi'  \, f(y ,\varphi' )
 \:.
 \end{equation}
If an electric current flows through a sample in a magnetic field, a perturbation of the charged density and the Hall electric field arise due to the magnetic Lorentz force. Both these effects are described by the zeroth ($m=0$) angular harmonic of the distribution function:
\begin{equation}
\label{zero_harm_def}
 f^{m=0} (y) = \frac{1}{2\pi} \int _0 ^{2\pi}
d\varphi'
 \:
f(y,\varphi')
 \:.
\end{equation}

\begin{figure}[t!]
	\includegraphics[width=1.0\linewidth]{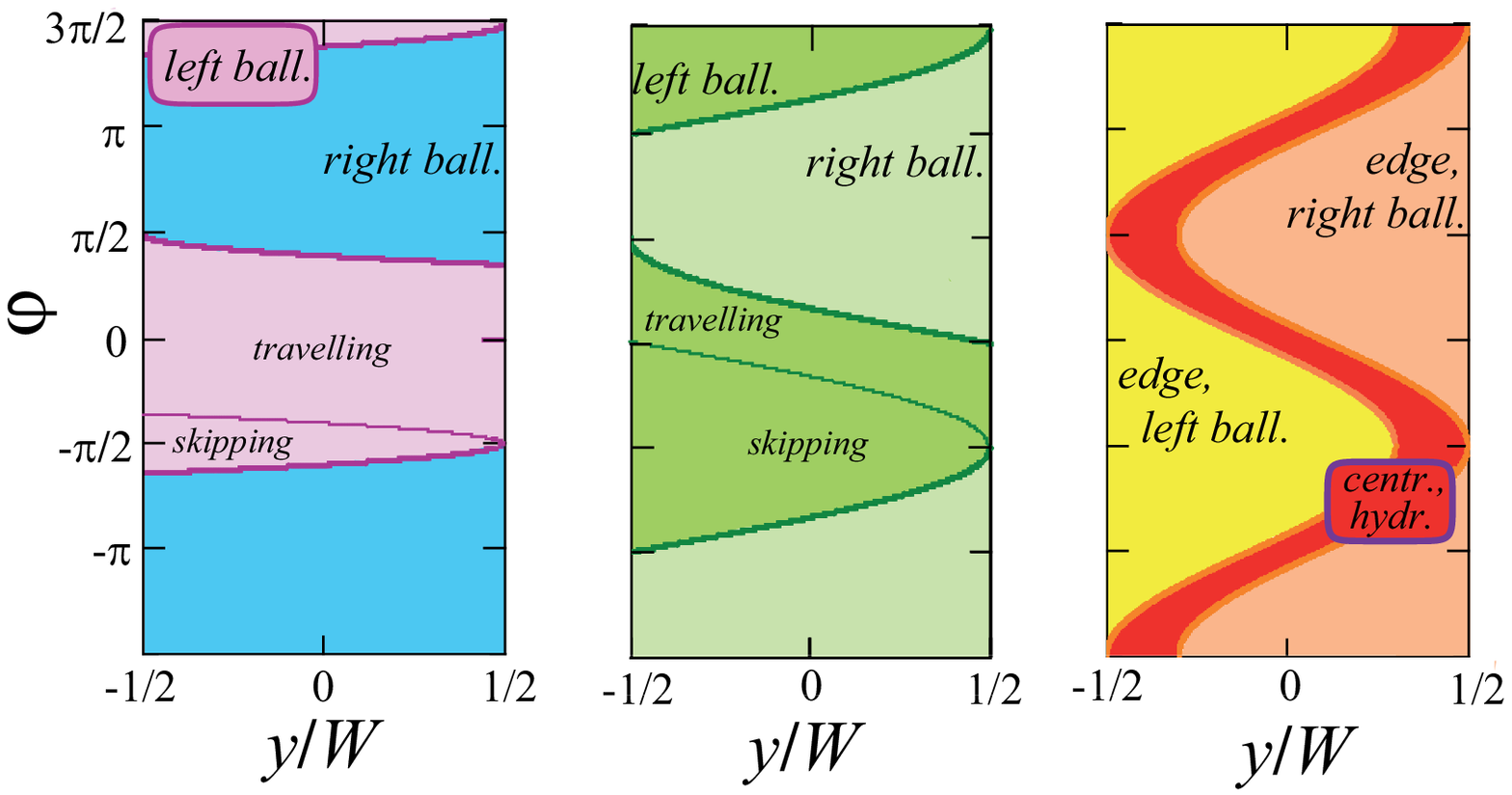}
	\caption{
	Regions in the plane $(y,\varphi)$ with the ballistic  electrons, reflected from the left and the right sample edges (in all panels),  as well as  with the central hydrodynamic
	electrons,  those do not scatter on the edges  [red region in panel (c)].
 The cases of narrow, $W\ll R_c$ (a); intermediate, $W\sim  R_c$ (b); and wide, $W>2R_c$ (c) samples are shown.
	}
	\label{FigS1}
\end{figure}

Figures~\ref{FigS1}(a,b) shows the regions in the $(y,\varphi)$-plane corresponding to the ballistic motion of electrons reflected from the right and from the left sample edges. Here and below in the current work we imply  that the electron mean free path $l=1/\gamma$ is much longer than the sample width.  Therefore, most of electrons are not scattering in the bulk and move inside the left  or the right regions
 by the ballistic trajectories between collisions with the edges.

For narrow samples, $W \ll R_c$, the left and the right ballistic regions are close to the rectangles $[-\pi/2,\, \pi/2]\times [-W/2, \,W/2] $ and $[\pi/2,\, 3\pi/2]\times [-W/2,\, W/2] $ [see Fig.~\ref{FigS1}(a)]. For wider samples, $W \sim R_c$,  $W < 2 R_c$,   the boundaries of the left and right region  $\varphi_{\pm} (y)$ begin to significantly depend on coordinate $y$ [see Fig.~\ref{FigS1}(b)]. The boundary curves
 $ \varphi_+(y) = \arcsin [  1 - \omega_c (W/2 + y ) ] $
  and
$ \varphi_-(y) + \pi $,  where $\varphi_-(y) = \arcsin [  1 - \omega_c (W/2 - y ) ] $
 coincide the electron trajectories  those are touching the edges tangentially. Therefore the distribution function  $f(y,\varphi)$ is not  well defined at $\varphi = \varphi_{+}(y), \, \varphi_{- }(y)  + \pi $ and can have a discontinuity at these curves. At magnetic fields above the critical field, $W>2R_c$, the central electrons those do not scatter on the edges arise in the region filled with red in Fig.~\ref{FigS1}(c).  Herewith the edge electrons in the left and right regions (yellow and pink) still scatter mainly on the edges.

It is useful to rewrite kinetic equation \eqref{kin_eq} in the form:
\begin{equation}
\label{kin_eq_with_gamma}
 \begin{array}{c}
 \displaystyle
\Big[\cos\varphi \, \frac{\partial }{ \partial y }
  + \gamma \Big] \widetilde{f} - \sin\varphi \,E_0
 =
 \\
 \displaystyle
 = \gamma_{ee} \hat{P} [\widetilde{f} \, ]
 +
 \gamma' \hat{P}_0 [\widetilde{f} \, ] +
 \omega_c \, \frac{\partial \widetilde{f}}{ \partial \varphi }
  \:,
  \end{array}
\end{equation}
where we have introduced the function:
\begin{equation}
 \label{f_tilda}
  \widetilde{f} (y,\varphi)= f(y,\varphi)  +  \phi(y) \:.
\end{equation}
Here $\phi$  is the electrostatic potential of the Hall electric field: $E_H = -\phi'$. Indeed, it follows from Eq.~(\ref{kin_eq}) that the Hall potential  $\phi(y)$ plays the same role in the transport equation as its progenitor, the zero harmonic of the distribution function $f^{m=0}(y)$~(\ref{zero_harm_def}) being proportional   to the  inhomogeneous density perturbations.  Therefore it is reasonable to introduce the function $ \widetilde{f} (y,\varphi)$~(\ref{f_tilda}) in order to take into account $\phi(y)$ and $f^{m=0}(y)$ within the same framework.

The zero harmonic $ \widetilde{f} ^{m=0}(y)$ of the generalized distribution function~(\ref{f_tilda}) takes the form:
\begin{equation}
\widetilde{f}^{m=0}(y)=\delta \mu (y) +  \phi(y)
\:,
\end{equation}
where $\delta \mu$ is the perturbation of the electron chemical potential.  In the case of sufficiently slow flows, the values  $\delta \mu (y) $ and $  \phi(y)$  are related by the electrostatic relations~\cite{Alekseev_Alekseeva}. For the considered case of 2D degenerated electrons, $\phi$ is usually much greater than the related perturbation of the chemical potential $\delta \mu$~\cite{Alekseev_Alekseeva}. Thus the Hall electric field is calculated just by the formula $ E_H (y) \approx   - [ \widetilde{f} ^ {m=0} ]'(y) $.

For brevity, we further omit the tilde in the function  $ \widetilde{f} $ and write $f \equiv  \widetilde{f}$.

\section{2. Ballistic regime}
\label{Sec:S2}

\subsection{2.1. Solution of kinetic equation in ballistic regime}
\label{Sec:S2.1}

In this section we consider  the 2D electron transport in a long sample in relatively  weak magnetic fields when the diameter of the cyclotron circle is larger than the sample width, $2R_c>W$.  Provided the bulk mean free path is much longer than the sample width, $l=1/\gamma \gg W$,  the dominant mechanism of scattering of most of electrons is collisions with the sample edges and the ballistic transport is realized.

In Ref.~\cite{we_6}  the kinetic equation~(\ref{kin_eq_with_gamma}) in the limits  $\gamma W \gg 1$ and $\gamma W \ll 1$ at zero magnetic  field  was analyzed. It was demonstrated  that
 in the first limit $\gamma W \gg 1$   the hydrodynamic regime is realized
 in the central region of the sample, $  W/ 2 - |y|    \gg 1/\gamma $.  Herewith the  departure  and the arrival terms,
  $ - \gamma  _{ee} f $ and
  $ \gamma  _{ee} \hat{P} [f]$,
  are of the same order of magnitude,   and Eq.~\eqref{kin_eq_with_gamma}
  is transformed into the Navier-Stokes equation for $ j(y)$.

It was also shown in Ref.~\cite{we_6} that in the ballistic regime, $\gamma W \ll 1$,
 the arrival terms $\gamma  _{ee} \hat{P} [f]$ and $\gamma ' \hat{P}_0 [f]$
 in the right-hand side    of Eq.~\eqref{kin_eq_with_gamma}
 are much smaller than
 the terms $\cos \varphi \, \partial  f / \partial y $ and  $\sin \varphi  \, E_0$
  in its left-hand side
 in the factor of $  \gamma W \ln[1/(\gamma W)] $,
  therefore the first   terms can be taken into account by the perturbation theory. Herewith the departure term $\gamma f$ in the left-hand side
  plays role of regularization of the kinetic equation near the angles $\varphi \approx \pm  \pi/2 $,   where the cosine factor in the term $\cos \varphi \, \partial  f / \partial y $ is close to zero.

In Ref.~\cite{we_6_2} the similar result for the arrival terms in Eq.~(\ref{kin_eq_with_gamma})   at a nonzero magnetic field was obtained. Namely, in the weak magnetic fields,  $\omega_c \ll \gamma^2 W $,  the terms $\gamma  _{ee} \hat{P} [f]$ and $\gamma  ' \hat{P}_0 [f]$ should be treated as perturbations in the equations for the first-, $f_ 1\sim \omega_c $,  and the second-order, $f_ 2 \sim \omega_c^2 $,  corrections by magnetic field to   the distribution function $f(y,\varphi)$ [the corrections $f_{1,2}$ are responsible for the Hall effect and the magnetoresistance].

Next, it is reasonable to expect that in narrow samples, $\gamma W  \ll 1$, in the intermediate magnetic fields below the critical field:
\begin{equation}
 \omega_c \sim 1/W \: , \quad \omega_c   <2 /W
 \:,
\end{equation}
the vanishing of the term $\cos \varphi \, \partial  f / \partial y $ at  $|\varphi |\approx \pi/2 $  is ``healed'' by the magnetic field term $ \omega_c \, \partial  f / \partial \varphi $ and, thus,     the electron flow at any $\varphi $  is mainly determined by  the scattering on the edges. The neglect of the collision terms in kinetic equation  (\ref{kin_eq_with_gamma}) leads to the estimates:
\begin{equation}
   \label{estim}
   f \sim E _0/ \omega_c
     \, , \qquad
   j \sim j_0
      \, ,\qquad
   E_H \sim E_0
   \:.
    \end{equation}
where  $j_0= n_0 E_0 W / m  $. According to  estimate (\ref{estim}) of $f$, both  the arrival  and the departure terms in
  $\mathrm{St}$ are  proportional to $(\gamma /\omega_c)\,E_0 $, thus both they
   are much smaller than the other ballistic terms of Eq.~(\ref{kin_eq_with_gamma}).
  In this way, the scattering of electrons in the bulk  leads only to small corrections to  the distribution function, proportional to $\gamma /\omega_c \ll 1 $.

Below we will see that estimates~(\ref{estim}) are valid  in the middle part of interval  $0<\omega_c W<2$, which
 excludes the very weak fields,  $0<\omega_c W \ll 1$,
 as well as
 the lower vicinity  of the critical point, $ 0 < 2 -\omega_c W  \ll 1$. In the first subinterval,
 the maximum length of  the ballistic trajectories, determining $j_0 $ and $E_H$,  is achieved
  for the group of electrons moving almost along the sample~\cite{pohozaja_statja,we_6,we_6_2}.
   In the second excluded subinterval, the
 ballistic size effects
 lead to the suppression of momentum  relaxation in the scattering of electrons on the edges. As a consequence, the
 collisions of electrons in the bulk becomes more important than at $2-\omega_c W \sim 1 $
 [we will show below that the role of the bulk scattering becomes  comparable with the one of
 the purely ballistic scattering at $2-\omega_c W \sim \gamma W $].

In this way, the ballistic regime is realized  in the interval  $0<\omega_c W \lessapprox 2 $ (until $  2 -\omega_c W  \gg \gamma W  $) and
 is  divided on the three following subregimes. The first one is:
\begin{equation}
\label{first}
\mathrm{(i)} \;\; \omega_c \ll \gamma^2 W \:.
 \end{equation}
Magnetotransport in this subregime  was partly studied in Refs.~\cite{we_6,we_6_2}.  The  length $l_b^{(2)} =\sqrt{R_c W}  $ of the maximal segment of a cyclotron circle inscribed in the stripe is longer that the bulk scattering length, $l_b^{(2)}  \gg  l = 1/\gamma$, therefore the longest ballistic trajectories are confined by $l$. The second subregime is:
 \begin{equation}
 \mathrm{(ii)} \;\; \gamma^2 W  \ll \omega_c  \ll  1/W
  \end{equation}
  This case was partly studied in Refs.~\cite{Scaffidi,pohozaja_statja}.
  The geometric parameter $ W/R_c$ is small as compared with unity  and
 $l_b^{(2)}  \ll 1/\gamma$, therefore the maximal trajectory length is limited by the  purely ballistic dynamic and is equal to $l_b^{(2)}$.
   The bulk scattering provides  small corrections to $j$ and $E_H$.
   The third subregime is:
   \begin{equation}
  \mathrm{(iii)} \;\;  \omega _c \sim   1/W
  \, , \quad
   2 -\omega_c W  \gg \gamma W
  \end{equation}
As far as we know, this subregime was not noticed and studied yet.
 In  it, $W/  R_c \sim 1 $, thus the trajectory lengths are limited by $W$. The bulk scattering still provides only small corrections, but
in the upper part of this interval, $\gamma W \ll 2 -\omega_c W \ll 1   $, a peculiar geometry of trajectories
 leads to a suppression of   the ballistic momentum  relaxation (for details see Sec.~3.1). Effects of the last type are usually referred as
 the  ballistic size effects.

According to the results of~\cite{Scaffidi,pohozaja_statja,we_6,we_6_2}    for the subregimes (i), (ii) and the above estimates characterizing  the subregime (iii),  in order to  develop a unified description of the ballistic transport one should use kinetic equation~(\ref{kin_eq_with_gamma}) with only the departure collision term retained:
\begin{equation}
\label{kin_eq_B_main}
   \left[\cos\varphi \,  \frac{\partial }{ \partial y }
  + \gamma \right] f
  -\sin \varphi \, eE_0
 =   \omega_c \frac{\partial f}{ \partial \varphi }
 \:.
\end{equation}
This equation is solved by the method of characteristics  for the first-order differential equations.
 Such solution was constructed in recent publication~\cite{pohozaja_statja}.

 In our work, using a slightly different approach, we obtain the ballistic distribution function $f(y,\varphi)$    similar  to the one obtained in Ref.~\cite{pohozaja_statja}. We use this solution  to study the regimes those were not considered   in Ref.~\cite{pohozaja_statja}: the first ballistic subregime (i),  $\omega_cW \ll \gamma^2 W^2$ (partly studied in Refs.~\cite{we_6},\cite{we_6_2}) and the third  ballistic subregime (iii),  especially, its right singular part: $ \gamma W \ll 2- \omega_cW \ll 1 $.

The solution of Eq.~(\ref{kin_eq_B_main}) with boundary conditions (\ref{bound_cond_1}) and (\ref{c_12})
 is a discontinuous function with  the domains of continuity  shown in Figs.~\ref{FigS1}(a,b). For the distribution function
 of the  electrons reflected
 from the left edge, whose trajectories lie in the interval:
\begin{equation}
   \label{left}
  \begin{array}{c}
   \displaystyle
    -\pi  +  \varphi_-(y) \,  <\,   \varphi   \,  <\,    \varphi_+(y)
    \:,
     \end{array}
\end{equation}
 we use  the notation:  $ f(y,\varphi) =  f_+(y,\varphi) $
\{in equation~(\ref{left}) we introduced the values: $\varphi_\pm(y) = \arcsin [  1 - \omega_c (W/2 \pm y ) ] $\}.
 For the electrons reflected from the right edge, whose trajectories are located in the interval:
 \begin{equation}
   \label{right}
   \begin{array}{c}
   \displaystyle
      \varphi_+(y)   \,  <\,   \varphi  \,  <\,   \pi  +   \varphi_- (y)
    \:,
      \end{array}
\end{equation}
 we use  the analogous notation:  $f(y,\varphi)= f_-(y,\varphi) $.   The functions $f_\pm$ are found by the
  standard, but lengthy calculations of the method of characteristics. We obtained:
\begin{equation}
\label{f_gen}
   \begin{array}{c}
   \displaystyle
f_{\pm}(y,\varphi) = \frac{E_0}{ \gamma^2 +\omega_c ^2 } \, \Big[ \,
 \omega_c \cos \varphi + \gamma \, \sin \varphi   \, +
 \\ \\
 \displaystyle
 +\:  e^ {  \gamma  \varphi  / \omega_c } \: Z_{\pm}(\sin \varphi  +\omega_c y)
\, \Big]
 \:,
   \end{array}
\end{equation}
where the  $y$-independent terms  $ \omega_c \cos \varphi $ and $\gamma  \sin \varphi$ are particular solutions of Eq.~(\ref{kin_eq_B_main}) those corresponds
 to the usual Drude
 formulas for a homogeneous Ohmic flow, whereas the term  $ e^{\gamma\varphi /\omega_c} Z_{\pm}(X)$ is a general solution of kinetic equation~(\ref{kin_eq_B_main})
 without the field term $\sin \varphi E_0$,  allowing to satisfy  the proper boundary conditions (\ref{bound_cond_1}).

By substituting Eq.~(\ref{f_gen}) into boundary conditions~(\ref{bound_cond_1}), we obtain the explicit form of $Z_{\pm}(X)$:
\begin{equation}
   \label{Z_pl}
   \begin{array}{c}
   \displaystyle
Z_{+}(X) = e^{- \frac{\gamma  }{\omega_c } \, \arcsin (X + \omega_c W /2 )} \, \Big[ \,\,c_l -
   \\\\
   \displaystyle
   - \gamma\,(X + \omega_c W /2) - \omega_c \sqrt{ 1 - (X + \omega_c W /2)^2}  \:\Big]
   \:,
   \end{array}
\end{equation}
\begin{equation}
   \label{Z_mi}
   \begin{array}{c}
   \displaystyle
Z_{-}(X) = e^{- \frac{\gamma  }{\omega_c } \, \big [\, \pi -\arcsin (X - \omega_c W /2 ) \big ] } \,
\Big[ \,\,c_r -
   \\\\
   \displaystyle
   - \gamma\,(X - \omega_c W /2) + \omega_c \sqrt{ 1 - (X - \omega_c W /2)^2 }  \:\Big]
   \:.
   \end{array}
\end{equation}
The coefficients  $c_l $ and $c_r$ in these formulas are determined from balance relations~(\ref{c_12})
 of the boundary conditions.  The resulting linear equations for $c_l $ and $c_r$  takes the form:
\begin{equation}
   \label{syst_c12_exact}
   \left(
   \begin{array}{cc}
      I_{ll} & I_{lr} \\
      I_{rl} & I _{rr}
    \end{array}
   \right)
    \left(
    \begin{array}{c}
        c_l \\
        c_r
      \end{array}
   \right) =- \left(\begin{array}{c}
           I_l \\
           I _r
         \end{array}
   \right)
   \:,
\end{equation}
where the coefficients in the first line of the matrix are expressed via the integrals:
\begin{equation}
 \label{I_ll}
I_{ll} = 2 + \int  _{ -\pi + \varphi _ 0 } ^{ - \pi /2 }
 d\varphi \: \cos \varphi \, e^{ \, \frac{\gamma  }{\omega_c }  \, (\pi + 2 \varphi ) }
 \:,
\end{equation}
\begin{equation}
    \label{I_lr}
  \begin{array}{c}
  \displaystyle
I_{lr} =
\int  \limits  _ {  \pi /2 } ^ { \pi + \varphi _ 0 }
d\varphi \,  \cos \varphi
  \: e^{ \,  \frac{\gamma  }{\omega_c }\,\big[   \varphi - \pi
+ \arcsin ( \sin \varphi - \omega_c W  )   \big] }
,
  \end{array}
\end{equation}
while the first components  of the right-hand vector is:
\begin{equation}
 \label{I_l}
\begin{array}{c}
\displaystyle
I_l
 = \frac{\pi \omega_c}{2 } +
\int  _{  \varphi _ 0 } ^{ \pi /2 }
 d\varphi \, \cos \varphi
 \times
\\\\
\displaystyle
 \times \,
  e^{ \, \frac{\gamma  }{\omega_c }  \,( 2 \varphi -\pi ) }
 ( \omega_c \cos \varphi - \gamma \sin  \varphi  )
 -
  \\\\
 \displaystyle
 -
 \int    _ {  - \pi /2 } ^ { \varphi _ 0 }
d\varphi \: \cos \varphi \: e^{ \, \frac{\gamma  }{\omega_c } \, \big[\,    \varphi
- \arcsin ( \sin \varphi + \omega_c W  )  \,  \big] }   \times
\\\\
 \displaystyle
\big[
 \omega_c \sqrt{1- (\sin \varphi +\omega_c W )^2 }
  +
  \gamma \, ( \sin  \varphi + \omega_c W)
  \big]
  \: .
 \end{array}
\end{equation}
 The other coefficients in Eq.~(\ref{syst_c12_exact}),  $I_{rr} $,  $I_{rl} $ and $I_r$, are related to $I_{ll}$, $ I_{lr}$, and $ I_l$ by
the formulas: $  I_{rr} =- I_{ll}$, $   I_{rl} =- I_{lr}$,  $ I_r = I_l $.
In Eqs.~(\ref{I_ll})-(\ref{I_l}) we introduced the notation:  $\varphi _0 = \arcsin (1- \omega_c W )$.

At general values of the parameter $\omega_cW$,   integrals (\ref{I_ll})-(\ref{I_l}) can be calculated only numerically. However, the explicit expressions for  these integrals and the resulting values
$c_{l,r}$,  $j(y) $, and $E_H(y)$ can be obtained in the limiting cases:  $\omega_c W \ll (\gamma W)^2$ [the first ballistic subregion (i)]; $(\gamma W)^2 \ll \omega_c W \ll 1 $ [the  second ballistic subregion (ii)]; and $ \gamma W \ll 2-\omega_cW \ll 1 $ [the right singular part of the third ballistic subregion (iii)].

\subsection{2.2. Ballistic transport in moderate magnetic fields}
\label{Sec:S2.2}

The estimates from Sec.~2.1 show that  in the interval of magnetic fields   $ (\gamma W)^2   \ll \omega_c W  \lesssim 1$ provided that  $ 2- \omega_c W \gg \gamma W  $ [the second  and the third ballistic subregimes]  the electron flow in the main order by $\gamma$ is determined by taking into account only the action of the external fields and the scattering on the rough sample edges.
The interparticle scattering provides only small corrections to all values, proportional to $\gamma W \ll 1$.

In this way,   the asymptote of Eq.~(\ref{f_gen}) by  $\gamma \to 0$  provides the distribution function describing
 the flow in the main order by   $\gamma W$:
\begin{multline}
\label{f_bez_gamma}
f_{\pm} (y, \varphi )= \widetilde{c} _{l,r}
 + \frac{E_0}{\omega_c }
 \Big\{ \,
 \cos \varphi
 \,
 \mp
 \\
 \mp
 \sqrt{
  1-\Big[\sin \varphi + \omega_c \, \Big(y \pm
  \frac{ W}{2}\Big ) \, \Big]^2
 }
 \; \Big\},
 \,
\end{multline}
where  $\widetilde{c} _{l,r}  = E _0  c _ {l,r} / \omega_c^2 $.
 Linear system ~(\ref{kin_eq_B_main})  is degenerate at $\gamma =0$, thus its
solution $\widetilde{c}_{l,r}$  can be determined up to a constant  $c_0$. Note that at $ \gamma > 0 $ function~(\ref{f_gen}),
  with $\widetilde{c}_{l,r}$   calculated from non-degenerate system~(\ref{kin_eq_B_main}),
 implies   a weak artefact relaxation of the electron density due to the neglect in Eq.~(\ref{kin_eq_B_main})  of the arrival terms $\gamma _{ee} P [f]$ and $\gamma ' P_0 [f]$.  Imposing the symmetric condition   $c _ {l}+ c _ {r} =0 $
 corresponding to $c_0 =0 $,  from Eq.~(\ref{syst_c12_exact}) we obtain:
\begin{equation}
\label{c_bez_gamma}
\begin{array}{c}
\displaystyle
  \widetilde{c} _{l,r}
  =\mp
  \frac{E_0}{\omega_c}
  \frac{U - V}{2\,(2-\omega_c W)}
  \:,
   \end{array}
\end{equation}
where $
  U=
  \arccos (1-\omega_cW )\:,
  $ and  $V=
  (1-\omega_cW )\sqrt{\omega_cW \, (2-\omega_c W)}$.
Solution~(\ref{f_bez_gamma})-(\ref{c_bez_gamma}) was recently obtained in Ref.~\cite{pohozaja_statja}.

A description of the profiles  of the current $j(y)$ and  the Hall field $E_H(y)$ corresponding to Eq.~(\ref{f_bez_gamma})
  as well as  a detailed justification of the applicability of Eq.~(\ref{f_bez_gamma}) in the moderately weak magnetic fields,  $ (\gamma W)^2   \ll \omega_c W  \lesssim 1$,  will be published in another publication.
 In this work we present only  the simplest  properties of the purely ballistic  flow.

   In the limit $  \omega_c W \ll 1 $ distribution~(\ref{f_bez_gamma})
   at the angles  $\sqrt{\omega_c (W/2 \pm y)} \ll ||\varphi | - \pi/2| \ll 1 $ takes  the form:
\begin{equation}
\label{f_pm_near_1}
f_{\pm} (y, \varphi )
    =
   \widetilde{c} _{l,r}
  +E_0
  \Big[
  \,
  y_\pm \,
   \frac{ \sin \varphi }{ \cos \varphi  }
   +
     \frac{  \omega_c \, y_\pm^2  }{2  \cos ^3\varphi  }
   \,
   \Big]
   \,.
\end{equation}
where and $y_\pm = y \pm W/2 $ and  $ \widetilde{c} _{l,r}  =\pm E_0 \sqrt{3 \omega_c W^3 }/2$.
From Eqs.~(\ref{f_bez_gamma}) and (\ref{f_pm_near_1}) we obtain for
 the current density $j(y)$ in the whole interval  of the moderate magnetic fields
  $ (\gamma W)^2   \ll \omega_c W  \lesssim 1$ [provided  $2-\omega_c W \gtrsim 1 $,
  that is until the vicinity of $B_c$]:
\begin{equation}
  \label{j_full_ball}
  j\sim j_0 \, \ln ( \,  1 \, /   \sqrt{\omega_c W} \, \Big)
  \:,
\end{equation}
 where $\sqrt{\omega_c W} = W/l_b^{(2)} $  is the maximum angle   between the sample direction $x$ and
 the longest ballistic trajectories at $y=\pm W/2$ (see Fig.~1 in the main text).
 Equation~(\ref{j_full_ball})  expresses the fact that the main contribution to $j$ comes
 from such  trajectories with  the lengths $\sim l^{(2)}_b$.   At $\omega_c W \sim 1 $ estimate~(\ref{j_full_ball}) coincides with Eq.~(\ref{estim}).

In the whole interval    $ (\gamma W)^2   \ll \omega_c W  \lesssim 1$ [provided  $2-\omega_c W \gtrsim 1 $]
 distributions~(\ref{f_bez_gamma}) and
 (\ref{f_pm_near_1}) leads to the above estimate~(\ref{estim})
 for the Hall field:  $   E_H\sim E_0$.

In the lower vicinity the transition point, $ \gamma W  \ll  2 - \omega_cW  \ll 1 $ [the right part of the third ballistic subregime],  the coefficients $\widetilde{c} _{l,r}$ diverge as $1/(2 - \omega_c W ) $, thus the main part of the distribution function is :
\begin{equation}
\label{c_near_2}
 f_\pm (y,\varphi )  =
  \pm \,  \frac{\pi E_0}{2 \, \omega_c \, (2-\omega_cW)}
 \:.
\end{equation}
The other terms of Eq.~(\ref{f_bez_gamma}) have the smaller order of magnitude: $\sim E_0 / \omega_c $.
 Function~$f_\pm$~(\ref{c_near_2}) describes the imbalance between the densities of the   electrons reflected from the left and the right edges,
  a part of which,
   the ``travelling electrons'' those  reach the opposite edges,
   compensates the $\mathbf{E}_0 \times \mathbf{B} $-drift contribution
   in  $ j_y = 0  $
  [the term $ E_0 \cos \varphi / \omega_c $ in Eq.~(\ref{f_bez_gamma})].

The current  $j$ and  the Hall field $E_H$  corresponding to $f_{\pm}$ with~(\ref{c_near_2}) also  diverge as $1/(2-\omega_cW)$
 at the near-transition region  $ \gamma W  \ll  2 - \omega_cW  \ll 1 $. The exact formulas for  them will be presented below in Sec.~3 with
taking into account also a weak scattering in the bulk.

\subsection{ 2.3. Ballistic transport in very weak  magnetic fields }
\label{Sec:S2.3}

In the first ballistic subregime,   $\omega_c \ll \gamma^2 W$,   the continuity domains of $f_{\pm}(y,\varphi)$ given by Eqs.~(\ref{left}) and (\ref{right}) become  close    to $-\pi/2 < \varphi < \pi/2$ and  $\pi/2 < \varphi < 3 \pi/2$ at any $y$ [see Fig.~\ref{FigS1}(a)]. Most of electrons are the ``travelling'' ones whose trajectories of  are slightly bent lines starting  on one edge and ending on another.

In Refs.~\cite{we_6, we_6_2} a solution of Eq.~(\ref{kin_eq_B_main}) in the limit $\omega_c \to 0$  based on the perturbation theory by the magnetic field term $\partial f/\partial \omega _c $ was constructed.   Up to the second order in $\omega_c$, such solution has the form:
\begin{equation}
 \label{power_dec}
 f=f_0+f_1+f_2
 \:,
 \end{equation}
where  $f_0$ is the distribution function in zero magnetic field,  while $ f_1\sim \omega_c  $ and  $ f_2 \sim  \omega_c^2  $.

In this subsection, first, we refine the applicability of the perturbation  approach of Refs.~\cite{we_6, we_6_2}.   Comparison of the magnetic field term, $\omega_c \partial f /   \partial  \varphi $, in the kinetic equation
 with the other terms, $\cos \varphi \,  \partial f /   \partial  y  $ and $-\gamma \, f $, at the angles $|\varphi| \to \pi/2$ shows that  the effect from a magnetic field can be treated as a perturbation for the travelling  electrons in the central bulk part of the sample:
 \begin{equation}
\label{bulk}
W/2 - |y| \gg \omega_c /\gamma ^2
\:.
 \end{equation}
 At such $y$, power decomposition~(\ref{power_dec}) for the distribution function is valid,
 which leads to the results obtained in~\cite{we_6, we_6_2}
 for the contributions of region~(\ref{bulk}) to the transport characteristics of the  sample.

 However, in the very vicinities of the sample edges,
 \begin{equation}
 \label{near_edg}
 W/2 - |y| \lesssim \omega_c /\gamma ^2
 \:,
  \end{equation}
the magnetic field term $\omega_c \partial f /    \partial  \varphi $   cannot be treated as a perturbation for few ``skipping'' electrons with  $| \varphi |\approx  \pi /2 $, which are returning to the same edge after the scattering on it (see Fig.~1 in the main text). Their  flow is almost collisionless, being similar to the one studied  in Sec.~2.2.
Criteria (\ref{bulk}) and (\ref{near_edg}) follow from
 the comparison of
the two possible limitations
 of the ballistic  trajectories lengths
 at $|\varphi| \to \pi/2$:  the size $\sqrt{R_c ( W/2 - |y| )}$ of the cyclotron circle segment with the height   $ ( W/2 - |y| )$  or the bulk scattering length, $l=1/\gamma$.

  In Refs.~\cite{we_6, we_6_2} the possibility of the formation of the near-edge regions where the perturbation theory by $\omega_c$ becomes not applicable was missed.

In realistic samples, the near-edges regions can be  formed only in not too small magnetic field  in  the samples with
  long and straight edges.
  Namely, the near-edge layer width    $ \omega_c /\gamma ^2 $ must be
  larger than the size of edges roughnesses.  At the very vicinity of $B=0$,  $\omega_c < \omega_c^{\star} \ll \gamma^2 W$,   the flow in the near-edge layers becomes controlled by a particular shape of  roughnesses. At such $\omega_c$,   the bulk  contributions to the current and the Hall field, described in~\cite{we_6, we_6_2},
     may   dominate and determine the sample resistances $\varrho_{xx} $ and $ \varrho_{xy} $.

Second, in this subsection we derive the  asymptote of the function $f_\pm$~(\ref{f_gen}) at $\omega_c \ll \gamma^2 W$
by the small parameter $\omega_c/ (\gamma^2 W)$
 and the resulting contributions to $j(y)$ and $E_H(y)$. Such asymptote corresponds to the travelling electrons and
   provides the main part  of  $f_\pm$  in the bulk region~(\ref{bulk}) at any $\varphi$ [thus the main contributions in $j$ and $E_H$ there]
   as well as
   the values of  $f_\pm$
   in the near-edge regions~(\ref{near_edg}) at the velocity angles $| \pi/2-|\varphi|| \gg \sqrt{\omega_c(W/2 - |y|)}$.
  The same  asymptotic form of $f_\pm$, $j$, and $E_H$  in the bulk region  were obtained in Refs.~\cite{we_6,we_6_2}
within the perturbation  approach:
 a direct solution of the kinetic equation in the limits $\gamma W \ll1 $
 and $\omega_c  \ll \gamma^2 W$.

\begin{figure}[t!]
	\includegraphics[width=1.0\linewidth]{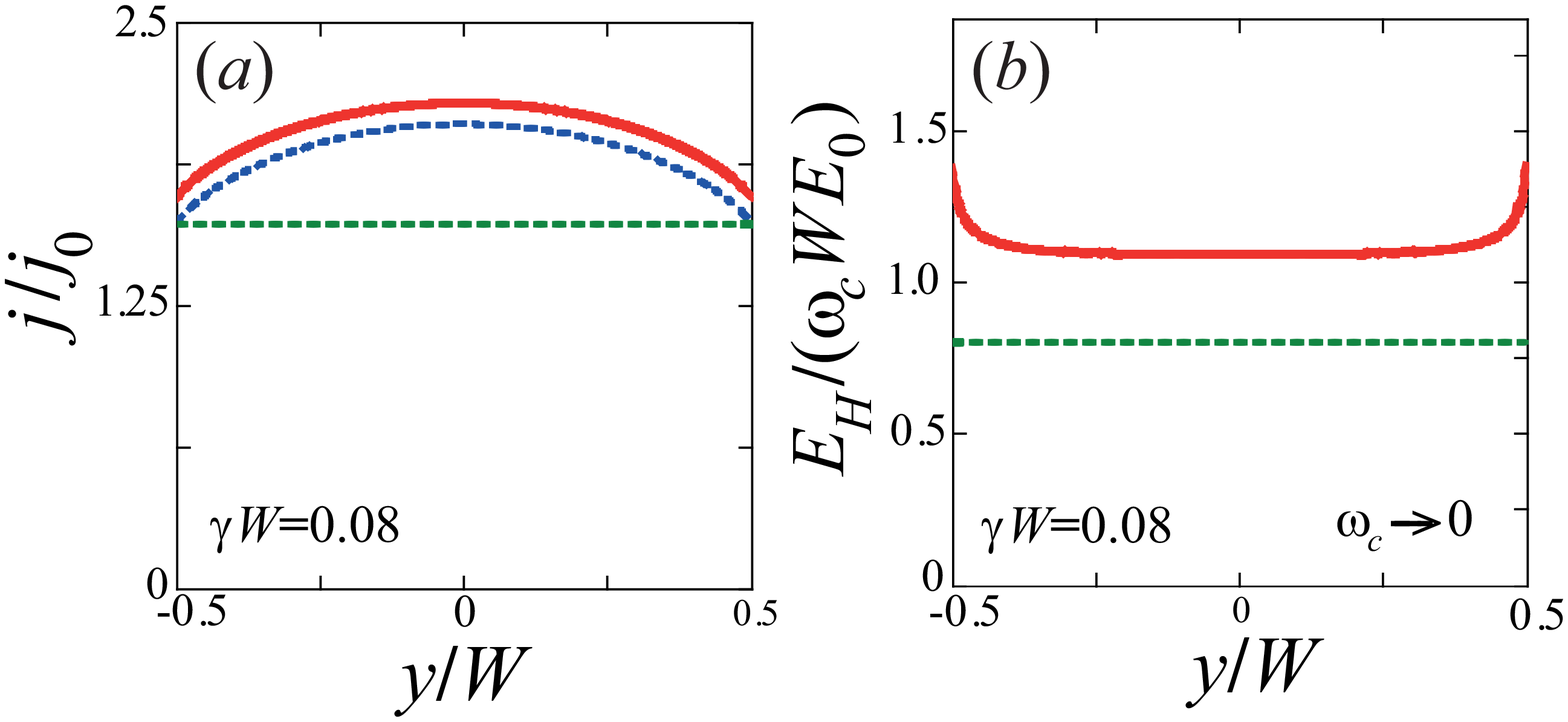}
	\caption{
	Current density (a) and Hall field (b) in the bulk part of a  sample, $W/2 -|y| \gg \omega_c/\gamma^2$, at $\gamma W = 0.08$  and in the limit $\omega_c \to 0 $ [the left part of  the first ballistic subregime]. Red
	curves present  numerical results obtained from Eqs.~(\ref{f_pm}), (\ref{f_1}), and  (\ref{delta_f_1}). Blue curve in panel (a) correspond to
	analytical solution (\ref{j_0_gamma}), while green curve shows the main contribution to the current density $j=j_{\gamma}$.  Green curve in panel (b) is plotted according to Eq.~(\ref{E_H_main}).
	}
	\label{FigS2}
\end{figure}

In the zeroth order by $\omega_c$, functions $f_{\pm}$~(\ref{f_gen}) take the well-known form~\cite{we_6},\cite{Beenakker_Houten_obz}:
\begin{equation}
\label{f_pm}
	f_{0,\pm} (y,\varphi) = E_0\,\frac{\sin\varphi  }{\gamma}\Big[1-\exp \Big(\displaystyle -\gamma  \, \frac{y \pm W/2  }{ \cos \varphi } \Big) \Big] \:,
\end{equation}
with the zero coefficients $c_{l,r}$. The flow density corresponding to Eq.~(\ref{f_pm}) in the leading order by the parameter $\gamma W$ is homogeneous, while an $y$-dependence emerges in the next order by~$\gamma W$~\cite{we_6}:
\begin{equation}
 \label{j_0_gamma}
 \begin{array}{c}
 \displaystyle
 j(y)  =j _{\gamma}  + \Delta j(y)\:,
 \\
 \\
 \displaystyle
  \frac{j _{\gamma}   } {j_0} =\frac{2}{\pi}\,
 \ln\Big( \frac{1}{\gamma W }\Big)
 \:,\;\; \;\;\;
 j_0 =\frac{n_0 E_0 W }{  m} \:,
 \\
 \\
 \displaystyle
 \frac{ \Delta j(y)  } {j_0}= -\frac{2}{\pi}
  \Big[
 \Big(\frac{1}{2}+\frac{y}{W}\Big) \, \ln \Big(
 \frac{1}{2}+\frac{y}{W}
 \Big)
 +
 \\
 \\
  \displaystyle
 +
 \Big(\frac{1}{2}-\frac{y}{W}\Big) \, \ln \Big(
 \frac{1}{2} - \frac{y}{W}
 \Big)
 \Big]
 \:.
 \end{array}
 \end{equation}
The contribution  $\Delta j(y)$ has infinite derivatives at the sample edges $y=\pm W/2 $ [see Fig.~\ref{FigS2}(a)].

The logarithmic divergence  of  $j_{\gamma}$ in Eq.~(\ref{f_pm})  by $\gamma W $
 originates from the travelling  electrons with the velocity angles   in the diapason:  $||\varphi | - \pi/2 | \lesssim \delta_m^{\pm}(y) $,
  where $\delta_{m}^{\pm} (y) = \gamma \, (W/2 \pm y) \ll  1$.  Such electrons move almost parallel to the sample
  and spend much longer time between scattering at the edges,
   acquiring  much  larger contribution to their velocities $v_x  $ due to acceleration by~$ E _0 $,
   as compared with other electrons with the angles  $\varphi \sim 1 $.

Direct calculations  yield that  the first order term in the expansion of $f_{\pm} $~(\ref{f_gen}) coincides with the function $f_1$ obtained in Ref.~\cite{we_6_2} in a perturbation solution of Eq.~(\ref{kin_eq_B_main}). Such $f_1$ satisfies the nontrivial boundary conditions (\ref{bound_cond_1}) with $c_{l,r}\neq 0$ and can be written as:
  \begin{equation}
  \label{f10}
  f_1=f_1^z+ f^b_1
  \:,
  \end{equation}
  where the first term
\begin{equation}
 \label{f_1}
 \begin{array}{c}
 \displaystyle
f_{1,\pm} ^z  (y,\varphi) =  \omega_c  E_0
 \Big\{
 \frac{\cos \varphi }{\gamma^2 }
 - \exp \Big[- \gamma \,   \frac{y  \pm  W/2  }
 { \cos \varphi }\Big]
\\
\\
\displaystyle  \times
 \Big[ \frac{\cos \varphi }{\gamma^2 }
  + \frac{ y \pm W/2 }{\gamma} -
  \frac{ \sin ^2 \varphi }{ 2 \cos ^3 \varphi
} \, \Big(y  \pm \frac{W}{2} \, \Big)   ^2
 \,  \Big] \, \Big\}
\end{array}
 \end{equation}
is the solution of the inhomogeneous kinetic equation ($E_0 \neq 0$) with zero boundary conditions, while the term
\begin{equation}
    \label{delta_f_1}
    f^b_{1,\pm} ( y , \varphi) = \mp \omega_c  E_0 \frac{W}{4\gamma}
    \exp \Big(  -\gamma \frac{ y \pm W/2 }{\cos \varphi} \Big) ,
\end{equation}
 being the solution of the kinetic equation with  $E_0 = 0$, ensures that boundary conditions~(\ref{bound_cond_1}) are met. The function $ f^b_{1,\pm}$  is much smaller than $f_{1,\pm}^z $ at $||\varphi | -  \pi/2 | \lesssim \delta_m ^{\pm}(y)$, however $ f^b  _{1,\pm} $ and $f_{1,\pm}^z $ give comparable contributions to $j_y$   in boundary condition~(\ref{bound_cond_1}) require that $j_y(y=\pm W/2)=0$.

Combining of Eqs.~(\ref{f_1}), (\ref{delta_f_1}),  and (\ref{zero_harm_def}),  we obtain for the zero harmonic of $f_1$:
\begin{equation}
   \label{bulk_result}
   f^{m=0} (y)  = -E_0
   \,
   \frac{ \omega_c y W }{\pi}
   \,
   \ln[1/(\gamma W)]
 \:.
\end{equation}
for the Hall  field in the central bulk region (\ref{bulk}) in the leading order by $\gamma W$  \cite{we_6_2}:
\begin{equation}
 \label{E_H_main}
 E_H (y)
 =
  E_0 \, \frac{ \omega_c W  }{\pi}
 \, \ln\Big(\frac{1}{\gamma W}\Big)
 \:.
\end{equation}
The numerically calculated exact profile $E_H(y)$ corresponding to Eqs.~(\ref{f_1}) and (\ref{delta_f_1}) differs from this analytical formula (\ref{E_H_main}) on the values of the order of $E_0 \omega_c W  $  [see Fig.~\ref{FigS2}(b)]. It is noteworthy that the perturbation theory result~(\ref{E_H_main}) is linear by magnetic field, like it takes place for the Hall field in bulk conductors.

Formulas~(\ref{j_0_gamma}) and (\ref{E_H_main}) lead to the following expression for the local Hall resistance
 $\varrho_{xy} (y)= E_H (y) / j(y) $:
\begin{equation}
 \label{R_H}
 \varrho_{xy} \approx  \frac{1}{2} \varrho_{xy} ^{(0)}
  \:, \qquad
   \varrho_{xy} ^{(0)} = \frac{m \,  \omega_c}{n_0  } = \frac{B}{n_0ec}
 \:,
\end{equation}
where $ \varrho_{xy} ^{(0)} $ is the conventional Hall resistance for the Ohmic and the hydrodynamic flows of charged particles at  low temperatures~\cite{je_visc}.

The second-order correction $f_2 \sim \omega_c^2$ to the electron distribution function was calculated in Ref.~\cite{we_6}.
 At the velocity  directions being close to the sample direction, $||\varphi| - \pi/2| \ll 1$, the function  $f_2  $ in the central region~(\ref{bulk}) in the main order by $1/(\gamma W) $ has the form:
\begin{equation}
 \label{f_2}
\begin{array}{c}
\displaystyle
f_2 (y,\varphi) =
 \frac{\omega_c ^2  E_0  \, (y\pm W/2)^3 }{ 2 \, \cos ^5\varphi  }
 \times  \\
\\
\displaystyle
\times
\Big[ \, 1 -
 \frac{\gamma \, (y\pm W/2)  }{4\, \cos \varphi} \, \Big]
\exp\Big[ - \frac{\gamma \,  (y\pm W/2) }{\cos \varphi}
  \Big]
 \:.
 \end{array}
\end{equation}
 This correction leads to the following  magnetic-field dependence of the current density in region~(\ref{bulk}):
\begin{equation}
  \label{final}
 j (y) \approx j_\gamma + j_2
 \,,\;\;\;
  j _2= \frac{3  n _0 E_0  }{2 \pi m}
  \frac{ \omega_c^2  }{ W \gamma^4 }
  \:.
\end{equation}
 The origin of correction~(\ref{MR_B_straight_gamma}) consists in a small increase of the mean length of the trajectories of the travelling electrons due to the action of the weak magnetic field (see discussion and Fig.~1 in Ref.~\cite{we_6}).

The positive correction $j _2 $~(\ref{final}) to the current $j_\gamma$  leads to a small negative magnetoresistance
 of the bulk region:
\begin{equation}
  \label{MR_B_straight_gamma}
 \frac{ \varrho_{xx} (B) - \varrho_{xx}(0) }{  \varrho_{xx}(0)}  =
   - \frac{ 3 \omega_c^2   }
 { \displaystyle 4 \gamma^4 W^2 \ln[1/(\gamma W )]}
  \:,
\end{equation}
where  $\varrho_{xx} = E_0 /j(y)$. It was discussed in Refs.~\cite{we_6,we_6_2} that for not too long samples, $ W \ll L \ll 1/\gamma $, the bulk scattering rate $\gamma$ in this formula is replaced on the reciprocal sample length, $1/L$, and magnetoresistance (\ref{MR_B_straight_gamma}) becomes temperature-independent. Moreover, it was discussed that result~(\ref{MR_B_straight_gamma}) is applicable even for the narrow short samples with the lengths $L\sim W \ll 1/\gamma$. In the last case, the rate  $\gamma$ should be replaced just on $1/W$:
\begin{equation}
  \label{MR_B_straight_gamma_shrt}
 \frac{ \varrho_{xx} (B) - \varrho_{xx}(0) }{\varrho_{xx}(0)}
 \sim
   -   \omega_c^2 W^2
  \:.
\end{equation}
In such short samples, only the ballistic subregimes $\omega_c W \ll 1  $ and  $ 1 \lesssim \omega_c W \lessapprox 2  $,
 analogous the first and the third subregimes for the long samples, are realized,
  so result~(\ref{MR_B_straight_gamma_shrt}) is valid until $ \omega_c \ll  1/W$.

Third, we study the flow  in the near-edge regions~(\ref{near_edg}).

  It can be seen from  Eq.~(\ref{f_gen}) that for $y$ in~(\ref{near_edg})
  one of the components of the distribution function $f_{\pm}$~(\ref{f_gen})
  [namely, $f_+$ at  $y\approx -W/2 $ and $f_-$ at $ y\approx W/2$]
  takes the form of the purely ballistic function $f_{\pm}$~(\ref{f_bez_gamma})
   of the second subregime with the coefficients  $\widetilde{c}_{l,r}$ corresponding
     the low-B-limit function $f^b_{1,\pm}$~(\ref{delta_f_1}):
\begin{equation}
\widetilde{c}_{l,r} = \mp \omega_cE_0W/(4 \gamma)
\:.
\end{equation}
Such components~$f_{\pm}$
 at the angles $\varphi$ near to the sample direction:
 \begin{equation}
\label{delta_pm_near_edg}
\begin{array}{c}
\displaystyle
| \pi/2 - |\varphi| | \lesssim
 \delta _\pm (y)
 \,,
\;\;\;
  \delta _\pm (y)  =  \sqrt{\omega_c (W/2 \pm y )}
 \,,
 \end{array}
\end{equation}
has finite values $\sim E_0\sqrt{W/\omega_c} $, while at the  angles
 $  \delta_\pm (y)  \ll |\, \pi/2 - |\varphi| \,| \ll 1   $
 are close to asymptotic formula~(\ref{f_pm_near_1}). The last one
 is equal to Eqs.~(\ref{f_pm}) and (\ref{f10}) in the main order by  $\gamma W \ll 1 $ [note that  $\delta_\pm (y) \gg \delta_m^ \pm (y) = \gamma \, (W/2 \pm y) $ in the near-edge layers].
  The other components of the distribution function, $f_+$ at $y\approx W/2 $ and  $f_-$ at $y\approx -W/2 $,
  are still described by Eqs.~(\ref{f_pm}) and (\ref{f10})
   with the characteristic angles $\delta _{m}^{\pm} (y) $, separating the regions of large and small magnitudes of $f_\pm$.

Analysis of electron trajectories shows that such distribution  functions~$f_{\pm}$~(\ref{f_gen}) in the layers, $y\approx \mp W/2 $, at the angles $| \pi/2-|\varphi|| \lesssim \delta _\pm (y)$  describe  the purely ballistic skipping  electrons.

\begin{figure}[t!]
	\includegraphics[width=1.0\linewidth]{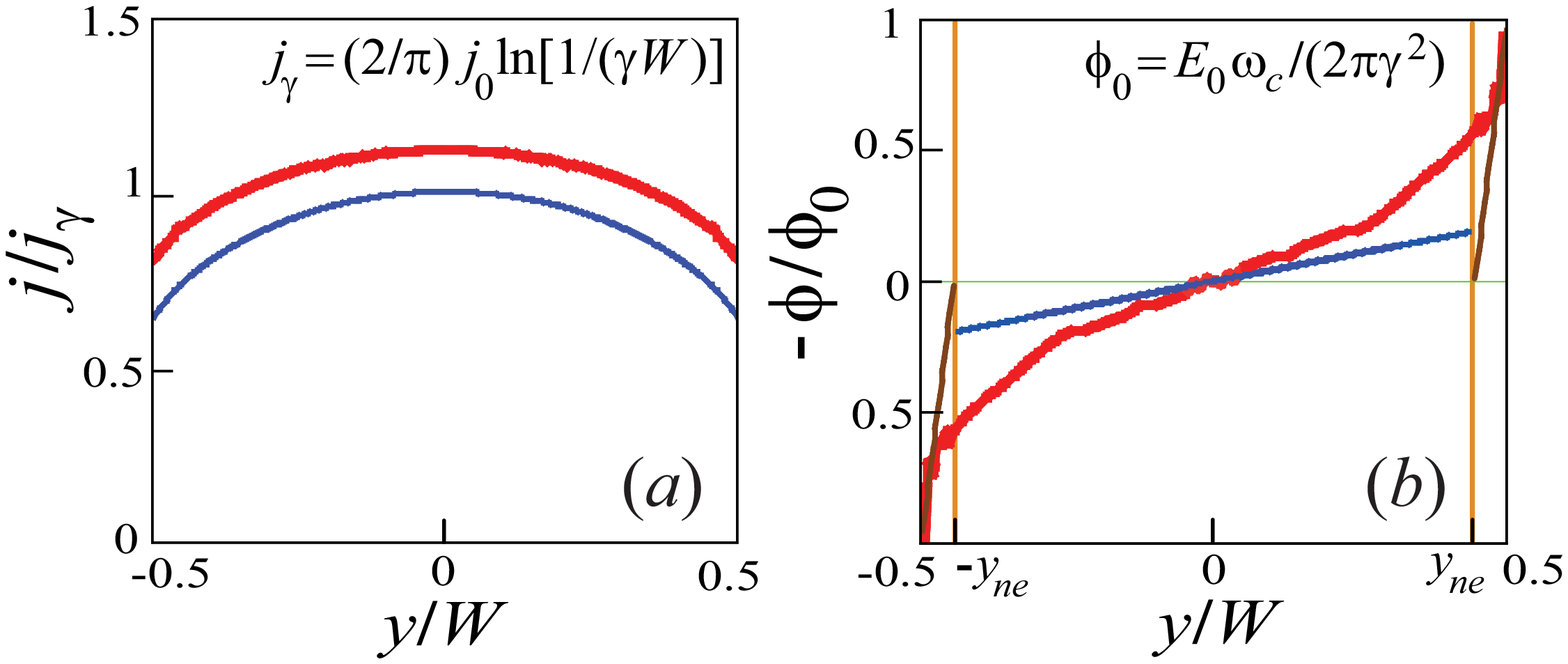}
	\caption{
	Current density $j(y)$ [a] and the potential $\phi_H(y)$ of the Hall field $ E_H(y) = - \phi_H'(y) $ [b] at a finite magnetic  field in the first ballistic subregime, $\omega_c \ll \gamma^2 W$, in the
	whole  sample, $ |y| < W/2$. The vertical  dashed lines depict the boundaries $\pm y_{ne}$, $y_{ne} = W/2 - \omega_c /\gamma^2  $ of the near-edge regions~(\ref{near_edg}). The
	parameters of the flow are: $\omega_c W =0.01$, $\gamma W =0.3$. Red curves on both panels present  numerical results
 obtained by general formula~(\ref{f_gen}). Blue curves on both
	panels  present the results of the perturbation theory for the bulk region, Eqs.~(\ref{j_0_gamma}) and (\ref{E_H_main}). Brown curves on panel  (b)  show the  non-perturbation result~(\ref{f_m_0_new}) for the near edge-regions.
 	}
	\label{FigS3}
\end{figure}

 The zero harmonic $f^{m=0}$ of the  described distribution $f_{\pm}$ in the near-edge regions $y \approx \pm W/2$~(\ref{near_edg}) takes the form:
\begin{equation}
\label{f_m_0_new}
f^{m=0} (y) \approx \mp \frac{  E_0 \, \omega_c }{2\pi} \, \Big[ \, \frac{1}{\gamma^2} - C \,  \frac{ W/2 \mp  y   }{\omega_c} \, \Big]
\:,
\end{equation}
 where $C \sim 1$ is a numeric coefficient.
Such $f^{m=0}$ is much larger than its value in  the bulk region, Eq.~(\ref{bulk_result}).  Both the terms in expression
$f^{m=0}(y)$~(\ref{f_m_0_new})
 originate the biggest singular terms $ \sim 1/ \cos ^3 \varphi$ in the $f_+ $ and $f_-$ components given by Eqs.~(\ref{f_1}) and (\ref{f_pm_near_1}).
 For $y$ in bulk region~(\ref{bulk}), such part  in $f^{m=0} (y)$  corresponding to the  $ \sim 1/ \cos ^3 \varphi$-terms
 in $f_\pm$ vanishes,
 as for the bulk function $f_{1,\pm}^z$~(\ref{f_1}) the contributions in $f^{m=0} (y)$ of this order from $f_{1,+}$ and $f_{1,-}$
  compensate each other.

So the anomalous behavior of $f^{m=0}  (y)$ in the near-edges layers [compare~(\ref{bulk_result}) and (\ref{f_m_0_new})] is related with the   decompensation
of the divergent contributions ($\sim 1/\gamma^2$)  in $f^{m=0}  (y) $  from the travelling electrons reflected from one edge   in the vicinities of the opposite edge  due to the appearance of the skipping electrons.

 Equation~(\ref{f_m_0_new}) leads to the Hall field in near-edge regions~(\ref{near_edg}), similar to one in the subregime~(ii) in the whole sample:
 \begin{equation}
   \label{E_H_ed}
    E_H(y)\sim E_0\,,\qquad W/2-|y| \lesssim \omega_c/\gamma^2
    \:.
 \end{equation}
In this way, the field $E_H(y)$   is  strongly enhanced in the near-edge regions and becomes non-analytical by $\omega_c$ [see Eqs.~(\ref{E_H_main}), (\ref{E_H_ed}) and
 Fig.~\ref{FigS3}(b)]. The resulting Hall voltage $U_H =\phi(W/2) - \phi(-W/2) $ takes the form:
    \begin{equation}
   \label{U_H}
   U_H = - E_0 \, \omega_c \, / \,  (  \pi \gamma^2)
   \:,
 \end{equation}
 that correspond to an anomalously large Hall resistance $   \varrho_{xy}  = U_H/(W j)$:
  \begin{equation}
   \label{ro_H_near_edg}
   \varrho_{xy}  \, = \,  \varrho_{xy}^{(0)} \, / \,  \{ \,  2\gamma^2 W^2 \ln[1/(\gamma W)] \, \}
      \:.
 \end{equation}

 From a similar analysis of the $m=1$ angular harmonic of general distribution $f_\pm$~(\ref{f_gen})
in near-edge regions $y \approx \mp W/2 $~(\ref{near_edg}),
  we obtain for the  current density in the main order by~$\delta _ m^\pm (y)\sim \gamma W$ and~$ \delta _\pm (y) \sim \sqrt{ \omega_c W}  $:
  \begin{equation}
  \label{j_new}
 \begin{array}{c}
  \displaystyle
   j(y) \approx
 \frac{2 j_0}{\pi}  \, \Big\{ \,
\Big( \, \frac{ 1 }{2}  \pm \frac{ y }{W} \Big)\ln\Big[  \, \frac{1}{ \sqrt{\omega_c (W/2  \pm y)  }  } \, \Big]
+
\\\\
 \displaystyle
+
\Big( \, \frac{ 1 }{2}  \mp \frac{ y }{W} \Big)\ln\Big[  \, \frac{1}{ \gamma\, (W/2  \mp y)   }  \, \Big]
    \, \Big\}
  \:.
  \end{array}
 \end{equation}
 The logarithms in both the first and the second term are estimated as $\ln(\gamma/\omega_c)$  at typical $y$ in layers~(\ref{near_edg}).  For the deviation of the averaged  current density, $j=\int _{-W/2} ^{-W/2}  j(y)\, dy /W$, from its value in zero magnetic field:
  \begin{equation}
  \label{delta_j_def}
  \begin{array}{c}
  \displaystyle
   \delta j_e \,=\,   j -j_\gamma -  \frac{1}{W}  \, \int   _{-W/2} ^{W/2} dy \, \Delta j(y) \sim
  \\\\
  \displaystyle
    \sim j_0  \int  _0^{ \Delta \xi } d \xi \,\xi
    \,\Big[ \, \ln \Big( \, \frac{ 1 } { \sqrt{ \omega_c  W \xi   }  }  \, \Big)
  -
  \ln \Big(\, \frac{ 1 } { \gamma W \xi }  \, \Big) \,
   \Big]
     \:,
     \end{array}
    \end{equation}
we obtain from   Eqs.~(\ref{j_0_gamma})  and (\ref{j_new}): 
  \begin{equation}
  \label{delta_j_res}
  \begin{array}{c}
    \displaystyle
\delta j_e
     \sim \,
       - j_0 \, \frac{\omega_c^2}{\gamma^4  W^2}
     \ln \Big( \,\frac{ \gamma  }{  \omega_c } \, \Big)
     \:.
     \end{array}
    \end{equation}
  In Eq.~(\ref{delta_j_def}) we introduced the values
    $\xi = 1/2 - |y|/W$ and  $\Delta \xi  = \omega_c  /   (\gamma^2   W )$.
Derived correction $\delta j_e$~(\ref{delta_j_res}) leads to a positive nonanalytical by $\omega_c$ magnetoresistance:
\begin{equation}
  \label{MR_positive}
 \frac{ \varrho_{xx} (B) - \varrho_{xx}(0) }{  \varrho_{xx}(0)}  \sim
    \frac{  \displaystyle  \omega_c ^2 \, \ln( \, \gamma /\omega_c \,  )     }
 { \displaystyle  \gamma^4 W ^ 2 \,  \ln[1/(\gamma W )] }
  \:,
\end{equation}
Such magnetoresistance  accounts the non-perturbative in $B$ effect of the  skipping electrons on the flow in the near-edge regions.
 Namely,
   the marginal angles   $\delta _m ^{\pm }(y) $ of the travelling electrons, which determines
    the main contribution to $j(y)$ and    $E_H(y)$  in the bulk region,
    are changed in the near-edge regions on  the purely ballistic angles    $\delta _{\pm} (y)   $~(\ref{delta_pm_near_edg}),
     $ \delta _{\pm } (y) \gg \delta _m ^ {\pm}(y) $,  separating the trajectories  of the skipping and the travelling electrons.

In realistic samples, depending on their geometry both the bulk or  the edge contributions to magnetoresistance and the Hall effect
  may appear.
  For long samples of very high quality,
   where the described above skipping trajectories are realized  near the longitudinal edges,
  the
 resistances  $\varrho_{xx}$ and $\varrho_{xy}$ will be determined by near-edge regions~(\ref{near_edg}),
   as contributions~(\ref{ro_H_near_edg}), (\ref{MR_positive})  from those regions
 to the magnetic field-dependent parts of $j$ and $E_H$ are  greater than
  from  bulk contributions~(\ref{E_H_main}), (\ref{MR_B_straight_gamma}).
  For samples with a more irregular geometry (not very long and straight), the contribution from the near-edge regions is to be  suppressed, thus bulk  magnetoresistance (\ref{MR_B_straight_gamma}) and bulk   Hall effect (\ref{R_H}) may be observed.

   We also note that  the particular experimental setup, in particular,
 the exact positions $y_{i}$ of the electrical contacts will determine the manifestation
 of the bulk or of the near-edge contributions
  to $\varrho_{xx}$ and $\varrho_{xy}$  even in long high-quality stripes.

Fourth, we   present in this subsection   an elementary ``kinematic'' derivation of equations~ (\ref{E_H_main}) and (\ref{R_H}),
being one of the main result of Ref.~\cite{we_6_2}.
 Such derivation elucidates why the local ballistic Hall resistance  at $\omega_c \to 0$
  in the bulk of a sample  is equal to one half of the conventional Hall resistance  of Ohmic samples,
   $\varrho_{xy}^0 = B/(n_0 e c)$.

Formulas~(\ref{j_0_gamma}), (\ref{E_H_main}), and  (\ref{final}) for the integral characteristics of the flow in the bulk region (\ref{bulk}) were derived from kinetic equation (\ref{kin_eq_B_main})  with
 the departure term $-\gamma f$  and the magnetic field term $ \omega _c \partial f/ \partial \varphi $.  According to the above consideration, this implies that all electrons accounted  in this calculation are the travelling ones.
  After the scattering at edges, they  reach the opposite edges without inter-particle collisions
   (at $| |\varphi | - \pi/2|  \gg \gamma W  $) or undergo an inter-particle collision in the bulk
(at $| |\varphi | - \pi/2|  \lesssim   \gamma W   $).
 Herewith the main contributions to current~(\ref{j_0_gamma}) and Hall field~(\ref{E_H_main}) comes from the electrons
 with the angles $ \gamma W  \lesssim|\pi/2 - |\varphi| | \ll 1$.

 Let us consider the kinematics of such travelling electrons.
In the limit  $\omega_c \to 0$ and $E_0 \to 0$,  their trajectories $ \mathbf{r} (t) = [\, x(t) \, ,  \, y(t) \, ] $ are  almost straight lines. For the $x$ component of the velocity $ \mathbf{v} = \mathbf{\dot{r}}$ of an electron reflected  by the angle $\varphi$  from the left edge we have:
\begin{equation}
\label{v_cl}
 \begin{array}{c}
 \displaystyle
 v_x(t,\varphi )=  v_F \sin \varphi  + (eE_0/m) \, t \:,
   \end{array}
\end{equation}
where the time $t$ is counted from the moment of the reflection.  In Eqs.~(\ref{v_cl}) and until the end of this subsection  we again explicitly  write $v_F$ and $e$ for a better comprehension of the text.

According to the definition of the mean current density $j $ (the mean amount of charge passing through the sample section per unit time),
its value in the ballistic regime is calculated  by an analog of the Drude formula:
\begin{equation}
\label{j}
j=n_0 e^2 t_0E_0/m
\:,
\end{equation}
 where $t_0$ is   the mean time of a free motion of electrons. Such $t_0$ for the travelling electrons is
 calculated by the averaging  over all proper $\varphi$
 of
 the times $  t_{\pm}(y=\pm W/2, \varphi ) $ of collisionless motion
 of the electrons with initial velocity angles $\varphi$.
 Here the value $t_{\pm}(y ,\varphi)$   denotes the time of motion of an electron
 by the zero-field trajectory ($E_0, B =0$)
 with the initial angle $\varphi$   starting at
  $y_0=\mp W/2$, and ending in the point $y$, $-W/2 <y<W/2$:
 \begin{equation}
   \label{t}
 t_{\pm}(y,\varphi ) = \frac{ W/2 \pm  y  }{v_F \,| \cos \varphi |}\:.
\end{equation}
 To find $t_0$, one needs to integrate   $t_{\pm}(y=\pm  W/2,\varphi )$ by  $ \varphi $ up to the limiting values of
  $ \varphi _m^\pm  \approx \pm \pi/2 $  at which $ |\pi/2   -  |\varphi|  |$
   is equal to $ \delta_m^\pm (y= \pm W/2) = \gamma W $. Such limits correspond to the ballistic trajectories
    with the maximum length equal to the mean free path relative to the bulk scattering, $l = 1/\gamma$.
    After the integration, we obtain $j \approx j_\gamma$
   in the main order by $\gamma W $,    where $j_\gamma =(2/\pi) j_0  \ln [1/(\gamma W )]$ is defined in Eq.~(\ref{j_0_gamma}).

Next,  we calculate the Hall field $E_H $ in bulk region~(\ref{bulk}) within a similar approach.  We suppose $E_H $
 to be homogeneous in that region: $E_H (y)\approx E_H $.
We find $E_H$  from the $y$ component of Newton's equations:
\begin{equation}
  \label{N}
 m \dot{v}_y = e E_H - (eB/c) \, v_x \:.
\end{equation}
  After the averaging of this equation at any  $y$  over the travelling electrons
 with various  $\varphi$, only the contribution $\Delta v_x(t) = (eE_0/m) \, t$ in $v_x$ related
to the acceleration by the field $E_0$ [see Eq.~(\ref{v_cl})] remains non-zero.
  There is no acceleration along the $y$ direction of the ensemble of the travelling electrons in the bulk region:
\begin{equation}
  \label{a}
    \sum _\pm
   \int _{- y_{ne} } ^ {y_{ne}   }\frac{dy}{W}
   \; \mathfrak{R} \int _0^{2\pi} \frac{d\varphi }{2\pi}
   \: m \dot{v}_y [t_{\pm}(y,\varphi ),\varphi] =0 \:,
\end{equation}
where $ y_{ne} = W/2 - \omega_c / \gamma^2 $ is the boundary of the right near-edge region
and  the symbol $\mathfrak{R}$ denotes the exclusion the vicinities of the angles $\pm \pi /2$ of
 the size $\gamma W$ in the integral by $d\varphi $.
 This exclusion  corresponds to the neglect of the electrons scattered in the bulk on other electrons.
    From equality~(\ref{a}) and
 equation~(\ref{N}) averaged
  over all travelling electrons in the bulk region by the same way as in~(\ref{a}) we obtain:
    \begin{equation}
  \label{b}
E_H =  \frac{B}{c}\sum _\pm
  \int   _{- y_{ne} } ^ {y_{ne} }\frac{dy}{W}
 \; \mathfrak{R} \int  _0^{2\pi} \frac{d\varphi }{2\pi}
  \: v_x [t_{\pm}(y,\varphi ),\varphi]
\: .
 \end{equation}
  In the limit $E_0 ,B \to 0$ this equation, together with
 equation Eqs.~(\ref{j}) and (\ref{t}),  yields
  in the main order by~$\gamma W$: $E_H = (1/2)(B /c) \, j_\gamma /(en_0) $, which  is result~(\ref{E_H_main}).

In view of the above elementary consideration, it becomes apparent that the factor $1/2$ in Eqs.~(\ref{E_H_main}) and (\ref{R_H}) has a kinematic origin. Indeed, the above value
 of $E_H$ follows from  condition~(\ref{b}) of the compensation of the linearly increasing in time magnetic Lorenz force,
 $(eB/c)\Delta v_x(t)$,
 by the Hall field force $eE_H$. The resulting $E_H$  contains the mean time of motion $t_0$ of the traveling electrons between the edges  and the factor $1/2$ due to integrating of $\Delta v_x (t_\pm(y,\varphi))$ over  $y$ [see Eqs.~(\ref{v_cl}), (\ref{t}), and (\ref{b})]. Expression~(\ref{j}) for $j$ contains the same time $t_0$   with no numeric factors.  Thus the local ballistic Hall resistance  $\varrho_{xy}  = E_H (y) / j (y)$
 acquires the additional factor $1/2$ as compared with $\varrho_{xy}  $ in the Ohmic and the hydrodynamic regimes.
   We remind that
 in the last regimes for the flows
 of the same geometry ($j_y \equiv0$), the value $E_H$ is calculated from the balance of the  mean forces $eE_H $ and $(eB/c) \,j \, /(en_0)$ acting on fluid elements with given macroscopic values $ j = j_x (y)$.


\section{3. Ballistic-hydrodynamic and ballistic-Ohmic phase transitions}
\label{Sec:S3}

\subsection{3.1. Semiballistic solution near critical field}
\label{Sec:S3.1}
The model formulated in Sec.~1 allows one to consider the systems being the mixtures of the two cases:
 (i) there are no bulk defects and the electrons inside the sample are scattered only on each other, conserving momentum;
 and
 (ii) there are no inter-particle collisions, but inside the sample the scattering of electrons on a weak disorder, leading to a weak momentum relaxation, takes place. For simplicity, we consider only these unmixed types of samples  with only one of the scattering mechanisms: the electron-electron collisions ($\gamma = \gamma_{ee} $, $\gamma' =0$) or the scattering on disorder ($\gamma = \gamma' $, $\gamma_{ee} =0$). Herewith we always consider that the scattering rates are weak:
\begin{equation}
\gamma W   \, \ll   \:1\;,
\end{equation}
that is the main condition of applicability of our theory.

In this subsection,  we analyze
ballistic solution~(\ref{f_gen}) in the whole lower vicinity $ 0 < 2-\omega_c W \ll 1 $ of the critical  point,
including the region $2-\omega_c W \sim \gamma W $.
This distribution function
takes into account the departure term, $-\gamma \, f$, in the collision operator. Thus it
provides an exact description of the flow   at $ \omega_c W <2 $ in the disordered samples with no electron-electron scattering
 (that is, of the lower vicinity of the ballistic-Ohmic phase transition).
  In next subsections, we will also use function~(\ref{f_gen})
     in the region $ 0 < 2-\omega_c W \lesssim \gamma W $
       for pure samples with only the interparticle scattering
     as a trial solution
  to construct the mean-field model of the lower vicinity of the  ballistic-hydrodynamic phase transition.

The $\mathbf{E} \times \mathbf{B}$-drift flow $j_y^{(0)}$ in the $y$ direction, described by the space-homogeneous Drude part of the distribution function~(\ref{f_gen}), is compensated only by  the travelling electrons reaching the opposite edge,
 as the skipping electrons, provide
 no contribution in $j_y$ due to the symmetry of their trajectories (see Fig.~1 in the main text).

\begin{figure}[t!]
	\includegraphics[width=0.85\linewidth]{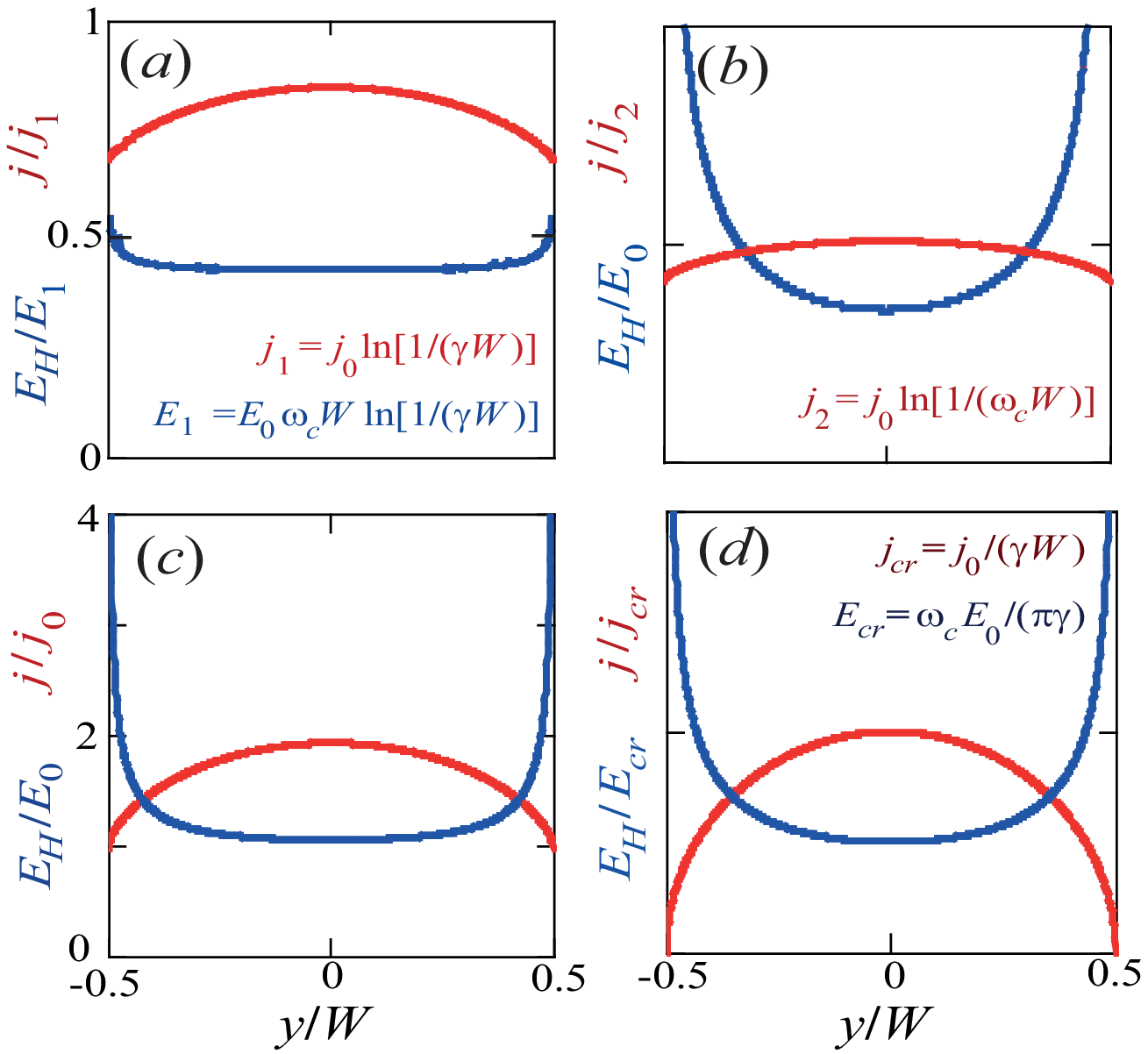}
	\caption{
 	Profiles of the current density (red curves) and the Hall electric field (blue curves)   at  different values of magnetic field:  (a)   the first ballistic subregime in the limit  $\omega_c \to 0 $ for  the ballistic
 	parameter $\gamma W = 0.08$ (only the bulk of the sample is shown); (b) the second ballistic subregime, corresponding to the small parameter $\omega_c W $, $\omega_c W  = 10^{-3} $ [herewith
 	$\omega_c W  \gg   (\gamma W)^2 $]; (c) the third ballistic subregime  at  the intermediate parameter  $\omega_c W $, $\omega_c W  = 1  $; and (d) the very point of the phase transition, when
 	$\omega_c W  = 2  $ [within approximation~(\ref{mean_field_subst}) of the proposed mean field model]. Curves are plotted by Eqs.~(\ref{def_j}) and (\ref{zero_harm_def}) with distribution
 	function~(\ref{f_pm}), (\ref{f_1}), (\ref{delta_f_1}) for~(a); with function~(\ref{f_bez_gamma}) for~(b) and (c); by formulas~(\ref{f_pm_crit}),  (\ref{mean_field_subst}), and  (\ref{av_crit})
 	for~(d).
 	}
 	\label{FigS4}
\end{figure}

The coefficients $I_{ll}$, $I_{lr}$  of the system of linear  equations (\ref{syst_c12_exact})
  tend to zero at $ 2-\omega_c W \ll 1 $. From Eqs.~(\ref{I_ll}) and   (\ref{I_lr}) for $I_{ll}$ and
  $I_{lr}$ in main order by the small parameters $2 - \omega_c W  \ll 1$ and $ \gamma / \omega_c \ll 1  $ we have:
\begin{equation}
   \label{I_ll_lr}
 I_{ll} = 2 - \omega_c W + 2\pi \gamma /\omega_c \: ,
 \qquad
 I_{lr} =  2 - \omega_c W
\:.
\end{equation}
Herewith, it follows from Eq.~(\ref{I_l}) that the right-hand side coefficients in Eq.~(\ref{syst_c12_exact})
 remains finite at $B \to B_c $:
$
I_{l} = -\pi \omega
 $.
Equations~(\ref{syst_c12_exact}) with the above values $I_{ll,\, lr, \, l}$
  yield for the parameters $c_{\pm}$ in distribution function (\ref{f_gen}):
\begin{equation}
 \label{c_pm_crit}
 c_{\pm} =  \pm  \,  \omega_c^2 \,  / \, (  \, 2    \gamma  +
 u  \, \omega_c   \, )
 \:,
\end{equation}
where the small  parameter $u =u(B)   \sim B_c-B $,
\begin{equation}
   u  =    (2/\pi)  \, (2 - \omega_c W )    \:,  \qquad 0<  u \ll 1  \, ,
\end{equation}
originates from $  I_{ll}  $, $I_{lr}  $ and
 characterizes the proximity to the transition point
  at the fields below the critical point  $ \omega_c^{cr} =2/W $.
It is proportional to the  small relative density,  $ \alpha_{tr} \sim u $,
 of the travelling electrons those compensate the $\mathbf{E} _0 \times \mathbf{B}$-drift contribution in $j_y$.
Coefficients   $c_{\pm} $~(\ref{c_pm_crit}) have a divergent behavior
 at  $\omega_c\to \omega_c^{cr}$,
 becoming  much larger than $ c_{\pm} | _ { \omega _c W \sim 1 } $,
whereas other terms in general solution $f_\pm$~(\ref{f_gen})
 have no divergence at $\omega_c\to \omega_c^{cr}$. Thus the distribution function (\ref{f_gen})
  in the man order by the small parameters   $2- \omega_cW $ and $\gamma W$  is given by:
\begin{equation}
 \label{f_pm_crit}
 f_{\pm} (y,\varphi) = \pm \frac{ E_0   }{ \displaystyle
  2 \gamma  +
u  \, \omega_c
   }
 \:.
\end{equation}
Such  $ f_{\pm}$ is a generalization of purely ballistic solution~(\ref{c_near_2}) and additionally
 accounts
 the electron scattering in the bulk.
 This scattering  provides much larger corrections in $f_\pm $ (of the relative order of $\gamma W/ u $)
in the upper part of the third ballistic regime,
  $\gamma W \ll u \ll 1 $, than in the intermediate fields, when $ u \sim 1 $.
   The effects from the scatterings in the bulk  and on the edges
     become comparable
   at $ u \sim \gamma W$ [see Eq.~(\ref{f_pm_crit})].

 Expressions~$ f_{\pm}$~(\ref{f_pm_crit}) describe a redistribution of the skipping and
  the travelling electrons between
  the left and the right regions (\ref{left}) and (\ref{right}).
   The redistribution  of the travelling electrons
   contribute  in $j_y$, compensating the  $\mathbf{E} \times \mathbf{B}$-drift.
  The redistribution  of the  skipping  electrons
  provides no contribution in $j_y$ due to the symmetry of their trajectories.
        As $ \alpha_{tr}  \sim u $ and $ c_\pm \sim 1 / u $,
   the     full density of the travelling electrons,  $ \sim \alpha_{tr} \, c_\pm$,
   corresponds    by the order of magnitude
    to the  $\mathbf{E} \times \mathbf{B}$-drift terms in  general solution~(\ref{f_gen}).
      The appearance of the skipping electrons
   with the relative density
   $\alpha_{sk}  \sim 1 $ and divergent full density, $ \alpha_{sk} c_\pm \sim 1/u$, originate from
 the equal probabilities of the scattering on the edges for all angles~$\varphi$.

Beside this, there is also a small group of electrons, with the proportion
  $ \alpha _ {\gamma}  \sim \gamma W$,
which would
 have returned to the same edge or reach the other one,
  but  due to the scattering in the bulk  have changed their trajectories.
  Such electrons  leads to an additional compensation of the $\mathbf{E} \times \mathbf{B}$-drift and,
 thus, weakens the imbalance between
  the left and
   the right  skipping electrons.
    So the divergence of $c_{\pm} $~(\ref{c_pm_crit})
 at $u \to 0$ is limited by the rate  $\gamma $.

The current density and the Hall field for function~(\ref{f_pm_crit}) in the main orders by $\gamma $ and $u$ takes the forms:
\begin{equation}
 \label{j_crit}
   j(y) =  \frac{4 nE_0}{ \pi m } \frac{1}{ 2\gamma +
   u  \, \omega_c
    }  \sqrt{1 - \omega_c^2 y^2 }
   \:,
   \end{equation}
\begin{equation}
 \label{fi_crit}
   E_H(y) =  \frac{2 E_0}{ \pi } \frac{1}{2 \gamma  +
   u  \, \omega_c
   } \frac{\omega_c }{ \sqrt{1 - \omega_c^2 y^2 }}
   \:.
\end{equation}
 These profiles  as well as $j(y)$ and $E_H(y)$ in the three ballistic subregimes are plotted in Fig.~\ref{FigS4}.
 For the averaged
 current density $ j = \int _{-W/2}^{W/2}dy\,j(y)/W $ and
 Hall field $E_H = \int _{-W/2}^{W/2}dy\,E_H(y)/W  $ we obtain:
\begin{equation}
\label{av_crit}
j  = \frac{ n _0 E_0 / m }{ 2 \gamma +
u  \, \omega_c
 }
\:,
\qquad
 E_H =  \frac{ E_0 \, \omega_c   }{ 2\gamma +
 u  \, \omega_c
   }
  \:.
\end{equation}

It is of importance to perform a more precise calculation of the current density $ j $   and the Hall electric field $E_H$ for the critical distribution function (\ref{f_pm_crit}).  In the next order by the small parameter $\sqrt{u }$ we obtain  the same equation~(\ref{av_crit}) for $j$, but the corrected result for $E_H$:
\begin{equation}
\label{av_crit_H2}
 E_H =  \frac{ E_0 \, \omega_c  }{ 2 \gamma +
 u  \, \omega_c
  } \, F(u) \:,
  \quad \;\;  F(u) =1- \sqrt{\frac{2}{\pi}\, u \,  }
 \:.
\end{equation}
The factor $F(u) $ describes the correction
 from the deviation of the shapes of the left and the right ballistic regions
  from their limiting form at $u=0$.

 For brevity, below we omit the factor  $n_0/m$ in all current densities.
  That is, we change the units of the current densities according to: $j \to j/[n_0/m]$.

The obtained values of $j$ and $E_H$  \eqref{av_crit}
 at the very critical point, $ u = 0$, turn out to be one half as compared
  with the Dude results for $j$ and $E_H$ of a bulk sample
  with the same width, the rate $\gamma'=\gamma$ of the scattering on disorder, and
     no interparticle collisions.
  This fact implies  that, for such disordered samples,
        the diffusive scattering on the  edges
      and  the scattering  on the bulk disorder at  $W =2R_c $
       provide the comparable
       contributions to the total momentum relaxation rate,  $2\gamma$.

\subsection{3.2. Mean field theory below critical field}
\label{Sec:S3.2}

In this section we develop a model of the  ``near-transitional''  flow
 in pure samples with interacting electrons in the nearest lower  vicinity of the critical  field,
 \begin{equation}
 \label{nearest_vis}
0< 2 - \omega_c W \lesssim \gamma  W
 \:.
 \end{equation}
At such $\omega_c$,
 the substitution of distribution function $f_{\pm}$~(\ref{f_pm_crit})
 into  the departure and the arrival terms,
 $\gamma f$ and $ \gamma \hat{P}[f] $,
 leads to the values  of the same order  of magnitude, $\sim E_0 $,
 as the other terms in kinetic equation~(\ref{kin_eq_with_gamma}).
  This indicates that both the terms  $\gamma f$  and $ \gamma \hat{P}[f] $ are equally important in the near-transition interval~(\ref{nearest_vis}).

  Instead of the exact solution of Eq.~(\ref{kin_eq_with_gamma}), we propose
   a mean-field  approach of the description of the electron dynamics in interval~(\ref{nearest_vis})
  based on an  approximate  accounting for
  the arrival term $\gamma \hat{P} [f]$.
  Namely, we treat   the part of this term:
\begin{equation}
\label{Proj}
\gamma  \, \hat{P}_{\sin} [f]  (y, \varphi  ) = \gamma \, j(y) \, \sin \varphi
\:,
\end{equation}
as the appearance of the addition ``internal'' field $\Delta E _0(y) \sim \gamma \, j(y) $
 in the truncated kinetic equation~(\ref{kin_eq_B_main}).
    Herewith we will
 omit the dependence  of $j(y) $ on the coordinate $y$, replacing
 in $ \Delta E _{0}    \sin \varphi $~(\ref{Proj})
  the current density $j(y)$ by its average value $ j = \int _{-W/2}^{W/2}dy\,j(y)/W $ .

The substitution of the function $f_{\pm}$~(\ref{f_pm_crit})
   into the arrival term $\gamma \hat{P} [f]$  leads also to the term $\gamma \, f^{m=0} (y)$ and
   the term $\gamma \, \cos \varphi \, j_y(y) $
    with a small non-physical current $j_y(y) \sim u $, appearing due to the non-exact form
    of solution~(\ref{f_pm_crit}).
      Such  terms with  $j_y(y)$ and  $f^{m=0} (y) $ also can be interpreted within the mean-field approach
      as the other internal fields in the
      truncated kinetic equation of the type of~(\ref{kin_eq_B_main}).
       However, our analysis shows that these terms induce  the contributions in $f_{\pm}$, $j$, and $E_H$
       which are relatively small by the parameters $\gamma W$ and $u$
       in interval~(\ref{nearest_vis}),  as compared with the effect from the $\sin$-term~(\ref{Proj}) with $j$
       corresponding to $f_{\pm}$~(\ref{f_pm_crit}).

It was shown in Refs.~\cite{we_6,we_6_2} that, for the first ballistic subregime, $\omega_c \ll \gamma^2/W $,
 the similar approach of the solution of Eq.~(\ref{kin_eq_with_gamma})
based on
the independent on $y$ approximation of  $\gamma \hat{P} [f]$
 is asymptotically exact by the parameter $\ln[1/(\gamma W)] \gg 1$ [see Eq.~(\ref{j_0_gamma})].
  Near the transition, the ballistic current density $j(y)$~(\ref{j_crit}) is strongly inhomogeneous,
  thus such approach    should lead to an inaccurate calculation of
  numerical coefficients in all values.
  However, in view of the success of similar mean-field methods for thermodynamic phase transitions,
  one can expect that this approach will provide a qualitatively correct results.
 Comparison in next section of our final results with the preceding numerical theory and experiments
  also testifies to the good applicability of our method.

The replacement of the term  $\gamma \hat{P} [f]$ by its value average  by $y$ can be interpreted as using of a special,  nonlocal by $y$, collision integral:
\begin{equation}
  \label{St_av}
  \mathrm{St}'[f]  = - \,  \gamma \,
   \big\{ \,  f (y, \varphi)- \int  _{-W/2} ^{W/2  }
  \frac {d  \tilde{y}  }{W} \:
   \hat{P} [f](  \tilde{y}   , \varphi) \Big\}  \, .
\end{equation}
Such operator conserves momentum and number of electrons only within the whole sample, but not at any  $y$.

In this way, the mean field equation for the averaged current  $j=j(u)$,
 which determines the state of the system,
 is constructed by the substitution:
\begin{equation}
\label{mean_field_subst}
 E_0  \: \rightarrow \: \tilde{E}_0=E_0 + \gamma j
\end{equation}
in the ballistic equation~(\ref{av_crit}) for $j$. After such  substitution, we obtain the self-consistent  equation for
 $j$ in the ballistic-hydrodynamic region~(\ref{nearest_vis}):
\begin{equation}
 \label{mean_f_eq_j_below}
  j  =   \frac{E_0 +\gamma j}{ 2  \gamma + u \, \omega_c
     }
  \: .
\end{equation}
Solution of this equation is:
\begin{equation}
 \label{mean_f_res_j_below}
  j  =   \frac{E_0}{ \gamma + u \, \omega_c }
  \: .
\end{equation}
In the very transition point, $u=0 $, this value is twice as large compared with the fully ballistic result (\ref{av_crit}), neglecting the arrival term $\gamma \hat{P} [f]$.

It was discussed in Sec.~3.1 that
semiballistic function~(\ref{f_pm_crit}) describe a shortage of the skipping electrons near one of the edge and the excess near the other one (as compared with  the equilibrium state).
  The mean-field current~(\ref{mean_f_res_j_below})
  additionally accounts for the conserving of momentum in interparticle collisions.
The magnitudes of~(\ref{f_pm_crit}) and~(\ref{mean_f_res_j_below})
  correspond to the compensation of the $ \mathbf{E} \times \mathbf{B} $-drift along the  $y$ direction [the first term in Eq.~(\ref{f_gen})]. This  compensation is realized due to: (i) motion of the few travelling  electrons along the trajectories connecting opposite edges and (ii)  the interparticle scattering of both the skipping and the travelling electrons in the bulk. These two processes correspond to the two terms,  $\sim u \, \omega_c $ and $\sim \gamma $, in the denominators of Eqs.~(\ref{f_pm_crit}), (\ref{av_crit}), and (\ref{mean_f_res_j_below}).

 In disordered samples,
    each scattering of an electron on a defect leads to the  loss of its  inequilibrium momentum and  to
 a shift of the center of its trajectory. In pure samples, each collision of two electrons
 induces only shifts of the centers of their trajectories. As a result,  the scattering on the bulk disorder is twice more effective for momentum relaxation  than the electron-electron scattering with the same rate $\gamma = \gamma '$
  within our  mean-field approach
  [compare Eqs.~(\ref{av_crit}) and (\ref{mean_f_res_j_below})].

The Hall electric field in the lowest vicinity~(\ref{nearest_vis})
 of the ballistic-hydrodynamic transition
   is calculated by the
   replacement $E_0 \rightarrow E_0 + \gamma j$
   in semiballistic  result~(\ref{av_crit_H2}) for the Hall field:
   \begin{equation}
 \label{mean_f_res_EH_below}
 E_H =  ( E_0 + \gamma j) \frac{ \omega_c  }{ 2 \gamma  +  u/W }\,  F(u)\:.
\end{equation}
     We emphasize that this equation, unlike Eqs.~(\ref{mean_field_subst}) and (\ref{mean_f_eq_j_below}),
      does not participate in the self-consistent procedure of determining the state of the system. Substituting
     the mean field current  $j$~(\ref{mean_f_res_j_below}) in Eq.~(\ref{mean_f_res_EH_below})    we obtain:
\begin{equation}
\label{mean_f_res_EH_below2}
 E_H =
  E_0  \, \frac{\omega_c}{ \gamma  +  u \, \omega_c  }
  \,  F(u) \:.
\end{equation}
 Due to the factor $F(u) $, this function, like as $E_H(u)$~(\ref{av_crit_H2}),
  has a strong square-root singularity at $\omega_c W \to 2 $.

The above results for the third ballistic subregime~$ \gamma W \ll 2- \omega _c W  \lesssim 1 $ and
 for the nearest vicinity of $B_c$~(\ref{nearest_vis}) are apparently valid also for not too long, $W\ll L \ll l$,
 and even for short samples, $W\sim L \ll l$. Indeed, the size of all the electron  trajectories in the $x$ direction
 at  $ 1 \lesssim \omega _c W  <2$
 is confined by the cyclotron diameter. Thus the distribution function in these regimes is
 formed in the regions of the stripe of the lengths $\Delta L \sim W \approx 2/\omega $.
 It follows from this circumstance that one can independently consider segments
 with the lengths
 $ \Delta L $ and
 account their contribution to the total resistance of a long sample
 by summation,
 like as for subsequently connected resistors.

\subsection{3.3. Mean field theory above critical field}
\label{Sec:S3.3}

When the  diameter of the cyclotron circle becomes smaller than the sample width, $W>2R_c$, there arises a group of the ``central electrons'' whose trajectories do not cross the sample edges [see Fig.~\ref{FigS1}(c) and Fig.~\ref{Fig1}(c) in the main text]. Such electrons spend a long time, $ \sim 1/\gamma \gg W$ on their  trajectories without collisions. We will show  in this Section that
 they form a pre-fluid collectivized fraction inside
 the dominant part of the semiballistic ``edge electrons'' those are scattered at the edges.
The central electrons are crucial for both the ballistic-hydrodynamic and the  ballistic-Ohmic phase transitions,  occuring in pure    and in disordered samples, respectively.

Below we formulate the two-component mean field model based on kinetic equation~(\ref{kin_eq_with_gamma})
  to describe the dynamics
 of the central and the edge electrons in the upper vicinities  of these
two  phase transitions:
 \begin{equation}
 \label{nearest_u_vis}
0<  \omega_c W -2 \ll 1
 \:.
 \end{equation}
 This model is a direct extension of the one-component mean field model formulated in Sec.~3.2.
 The relative density of the central electrons,
\begin{equation}
  \label{param_def}
 \alpha_c = (W-2R_c)/W  \sim   B-B_c \:, \quad
 \alpha_c \ll 1
 \:,
\end{equation}
  is the order parameter of these transitions at the fields above
 the critical point: $0<B-B_c\ll B_c$.
  The proportion of the edge electrons $ \alpha_e  = 1 - \alpha_c $
   is close to unity.

 The distribution functions  $f_e$ and $f_c$ of the edge electrons (``$e$'') and
  of the central electrons (``$c$'') are defined in the two distinct regions
   in the $(y,\varphi)$ plane  [see Fig.~\ref{FigS1}(c)].
   To describe the whole flow, one should   solve the exact kinetic equation
   (\ref{kin_eq_with_gamma}) in such regions with accounting of both the arrival and the departure terms.

In order to find the distribution $f_c$ of  the central electrons,
it is convenient to change the variables $y,\varphi$ on the new variables  $y_c,\varphi$,
 where $ y _c $ is the coordinate of the center of electron cyclotron orbits.
  The new variables of the central electron lie in the interval:
  \begin{equation}
   \label{ineq_c}
- W/2+R_c < y_c < W/2-R_c
\, ,\;\; \;\;\;
0< \varphi <2\pi
 \:.
\end{equation}
   For the coordinates of the centers of
    the  edge electrons trajectories we have:
 \begin{equation}
 |y_c| > W/2-R_c
 \:,
 \end{equation}
while their velocity  angle $\varphi$ lies in the diapasons depending on $y_c$.
  However, we do not need to describe in detail the distribution $f_e$ of the edge electrons,
   as their dynamics is similar
    to the one  of the skipping electrons
   in the nearest low vicinity of the transition~(\ref{nearest_vis})
    and therefore below we will apply for them the results of Sec.~3.2.

The solution~$f_{c,e}(y_c ,\varphi)$     of the exact kinetic equation~(\ref{kin_eq_with_gamma}) with
  the non-local arrival term   $ \gamma \hat{P}[f_e+f_c] $ (or $\gamma \hat{P}_0[f_e+f_c]$)
   is a continuous function in  diapason~(\ref{ineq_c}). Therefore the angular harmonics
    $ f_{c,m} \sim \int d \varphi \, f_c (y_c, \varphi)  \, e^{- i m \varphi } $  of $f_c$  rapidly decrease
     with the increase of $m$ (apparently, as a geometric progression).
     On the contrary,  the studied above ballistic solutions~$f_{\pm}$~(\ref{f_gen}), (\ref{f_pm_crit}) of Eq.~(\ref{kin_eq_B_main}),
 describing also the edge electrons, has discontinuities at the marginal trajectories shown in Figs.~S1.
   Thus functions $f_{\pm}$~(\ref{f_pm_crit})  has angular harmonic depending on $m$ slowly, as a power of $m$.
 In this way, the central electrons has the distribution function of a hydrodynamic type,
 being substantially different from ballistic function  $f_{\pm}$~(\ref{f_pm_crit}), therefore  they
   constitute a nucleus of the viscous  or the Ohmic flow.

Instead of an exact solution of kinetic equation~(\ref{kin_eq_with_gamma}),
 we propose, following to Sec.~3.2,
 a two-component mean field model which accounts for  the interparticle scattering
   with the neglect of  the inhomogeneity of the arrival term $ \gamma \hat{P} [f_e+f_c] $
  by $y$. This model is to be qualitatively applicable
   in the nearest upper vicinity of the phase transition~(\ref{nearest_u_vis}),
   where the relative part of the central electrons is small, $\alpha_c \ll 1 $.
The exact form of such model is different for the pure and the disordered samples.

First, we study the ballistic-hydrodynamic transition
 in pure samples when only the  electron-electron scattering takes place.

When the fraction of the central  electrons is small,  $\alpha_c \ll 1$, the centers $y_c  $
 of their cyclotron orbits lie approximately in the center of the sample,  $y=0$ [see Eq.~(\ref{ineq_c})
 and  Fig.~\ref{Fig1}(c) in the main text].   Herewith
 central electrons most often scatter on the edge electrons, which have
 the distribution $f_e (y_c, \varphi)$  varying by $y_c$ on the scale of the order $W$. Thus the properties of all central electrons in the main orders
  by $\gamma $ and $\alpha_c $  are almost identical  and are described by the function $f_c $  weakly depending  on  $y$:
 \begin{equation}
    \label{f_c}
 f_c (y_c, \varphi)  \approx  f_c (0, \varphi)
 \:.
  \end{equation}
Within our mean-field approach, we use the only one  parameter to describe the state of the central electrons.
 It is their contribution, $j_c \sim f_c $, to the averaged current density  $j$. Correspondingly,
 for the description of the state of the edge electrons we also use the similar mean field parameter,
  $j_e$, being the second contribution to the averaged current $j$.

To find $j_e$,  we note that at $\alpha_c \ll 1 $  kinetic equation~(\ref{kin_eq_with_gamma})
 with the arrival term simplified  according to Eqs.~(\ref{St_av}) and   (\ref{mean_field_subst})
 are applicable also
   for the edge electrons in  their regions on the $(y,\varphi)$-plane [see Fig.~\ref{FigS1}(c)].
 Thus  in the main order by $\gamma W \ll 1 $ the distribution $f_e$ of these electrons at
 $\alpha_c \ll 1$
 is  the function  $f_{\pm}$~(\ref{f_pm_crit})
 with  $u =0 $,
 the proper shifts of the variable $y$,  and
 the substitutions:
 \begin{equation}
\label{E0_above}
 E_0 \:  \rightarrow \: \tilde{E}_0 = E_0 + \gamma (j_e +j_c) \:,
\end{equation}
and $ W \to \tilde{W}  = 2R_c   $. The last change  accounts for
  the decrease with $B$ of the total width $ \tilde{W} $
of the regions containing the edge electrons.
 The integration of such function   $f_e$ with the factor $\sin \varphi $ over the left and the right semiballistic edge regions
   of the width $\tilde{W}$ (see Fig.~S1)
  yields
  the   formula for the edge electron contribution $j_e$
  to to full current $j$, which contains  $\tilde{E} $~(\ref{E0_above}) and
  is
  similar to Eq.~(\ref{mean_f_eq_j_below}) obtained form
  the integration of  $f_{\pm}$~(\ref{f_pm_crit})  with $\tilde{E} $~(\ref{mean_field_subst})
    over the whole sample.
     In the main order by $\alpha_c$ and $\gamma W $    we obtain:
\begin{equation}
 \label{j_e_mean_f}
 j_e =\alpha_e  \frac{ E_0  + \gamma (j_e +j_c)}{2 \gamma}
 \:,
\end{equation}
where the factor $\alpha_e$ takes into account the relative edge electron density
  [compare Eqs.~(\ref{j_e_mean_f}) and (\ref{mean_f_eq_j_below})].

 In order to calculate $j_c$,  we multiply the kinetic equation   (\ref{kin_eq_with_gamma}) expressed in the variables $\varphi , y_c$ on the factor $\sin \varphi $ and integrate it  by $\varphi $ and $y_c$ over  $-\pi/2<\varphi <3\pi/2$ and $-W/2+R_c<y_c<W/2-R_c$. In this diapason  of $y_c$,  the distribution function $f_c$  describes the central electrons and is approximately independent on $y_c$, according to
  Eq.~(\ref{f_c}).   As the result, we arrive to   the  formula:
\begin{equation} \label{j_c_mean_f}
 j_c =\alpha_c  \frac{ E_0  + \gamma (j_e +j_c)}{ \gamma}
 \:.
\end{equation}
This is actually the Drude formula for the contribution to the total current from central electrons,
 which are scattered on the edge electrons with the rate $\gamma$  and accelerated be the effective field~(\ref{E0_above}).

Solving together the resulting  mean-field system of equations (\ref{j_e_mean_f}) and (\ref{j_c_mean_f}),
 keeping in mind that $\alpha_e+\alpha_c = 1, $, we obtain in the zero and first orders by $\alpha_c$ :
\begin{equation}
\label{mean_f_res_above_je_jc}
j_e =\frac{ E_0  }{ \gamma}  , \; \; \;\;\; j_c = 2 \alpha_c   \frac{ E_0  }{ \gamma}\:.
\end{equation}
For the total current $j=j_e+j_c$ we obtain from Eq.~(\ref{mean_f_res_above_je_jc}):
\begin{equation}
   \label{mean_f_above_res_j}
 j =  (1+ 2 \alpha_c)   \frac{ E_0  }{ \gamma}
 \:.
\end{equation}
According to Eq.~(\ref{param_def}) for $\alpha_c $,
 this function grows linearly with the difference   $B-B_c$.

The physical picture implied under the second equation   in formulas~(\ref{mean_f_res_above_je_jc}) is as  follows. The dynamics of the central electrons in the reference  frame moving with the effective drift velocity of the edge electrons  [$v_{d,e} = 1/\gamma$, see Eq.~(\ref{mean_f_res_above_je_jc})] is similar to the scattering on static defects.
 Namely, the  edge electrons as a whole are looked like static defects for the central electrons in this moving frame. The effective momentum relaxation time of the edge electrons  due scattering on the edges and on other
  electrons is $1/\gamma $  [see the first equation in formula~(\ref{mean_f_res_above_je_jc})]. The scattering time of the
   central electrons on the edge electrons
   is also  $1/\gamma$. As the drift velocity of the central electrons $v_{d,c}$ is the sum of
   and their drift velocity in the moving frame
    $v_{d,c}' =1/\gamma $
    and the velocity of the frame $v_{d,e} $,
    a doubling of the scattering time $1/\gamma$
    in the equation for $j_c$ in Eqs.~(\ref{mean_f_res_above_je_jc}) arises.

 The Hall field above the transition  also contains the contributions  from
 the edge electrons and from the central  ones.
 As the ballistic solution $f_{\pm}$~(\ref{f_pm_crit}) at $u=0$ with  the effective field $\tilde{E}_0$ and the change $ W \to 2/\omega_c $
   provides the distribution $f_e  $ of the edge electrons in interval~(\ref{nearest_u_vis}),
 the edge electrons contribution to the Hall field
 is given by the  semiballistic value $E_H$~(\ref{av_crit_H2})
 with $u=0$ and the changes  $E_0 \to \tilde{E}_0$~(\ref{E0_above}) and $ W= 2/\omega^{cr}_c \to 2/\omega_c $.
  Herewith we should
  use  Eq.~(\ref{mean_f_above_res_j}) for the current $j$ in $\tilde{E}_0$~(\ref{E0_above})
  and add the factor $\alpha_e$ accounting for the relative density
   of the edge electrons. We obtain:
\begin{equation}
\label{E_H_e}
 E_{H,e} = \alpha_e \, (1+\alpha_c )\,
\frac{\omega_c}{\gamma} \, E_0
 \:,
\end{equation}
where  we should put: $\alpha_e \, (1+\alpha_c )  =  1- \alpha_c ^2 \approx 1 $.

The contribution to the Hall field from the central electrons is calculated using the $y$-independent distribution function $f_c $~(\ref{f_c}). Multiplying kinetic equation (\ref{kin_eq_with_gamma}) on $\cos \varphi $ and integrating over $-\pi/2<\varphi<3\pi/2$ and $-W/2 +R_c <y_c< W/2-R_c$, we obtain the contribution to the Hall field, similar to the Drude model result:
\begin{equation}
 \label{E_H_c}
  E_{H ,c} =  \omega_c \, j_c  \:.
\end{equation}
Substituting  Eq.~(\ref{mean_f_res_above_je_jc}) for $j_c$ in this formula,
 we get for the total averaged Hall field $E_H =E_{H,e}+ E_{H,c} $:
\begin{equation}
\label{mean_f_above_res_EH}
E_H =  (1+ 2 \alpha_c)   \frac{ \omega_c   }{ \gamma} \, E_0
\:.
\end{equation}

\begin{figure}[t!]
	\includegraphics[width=1.0\linewidth]{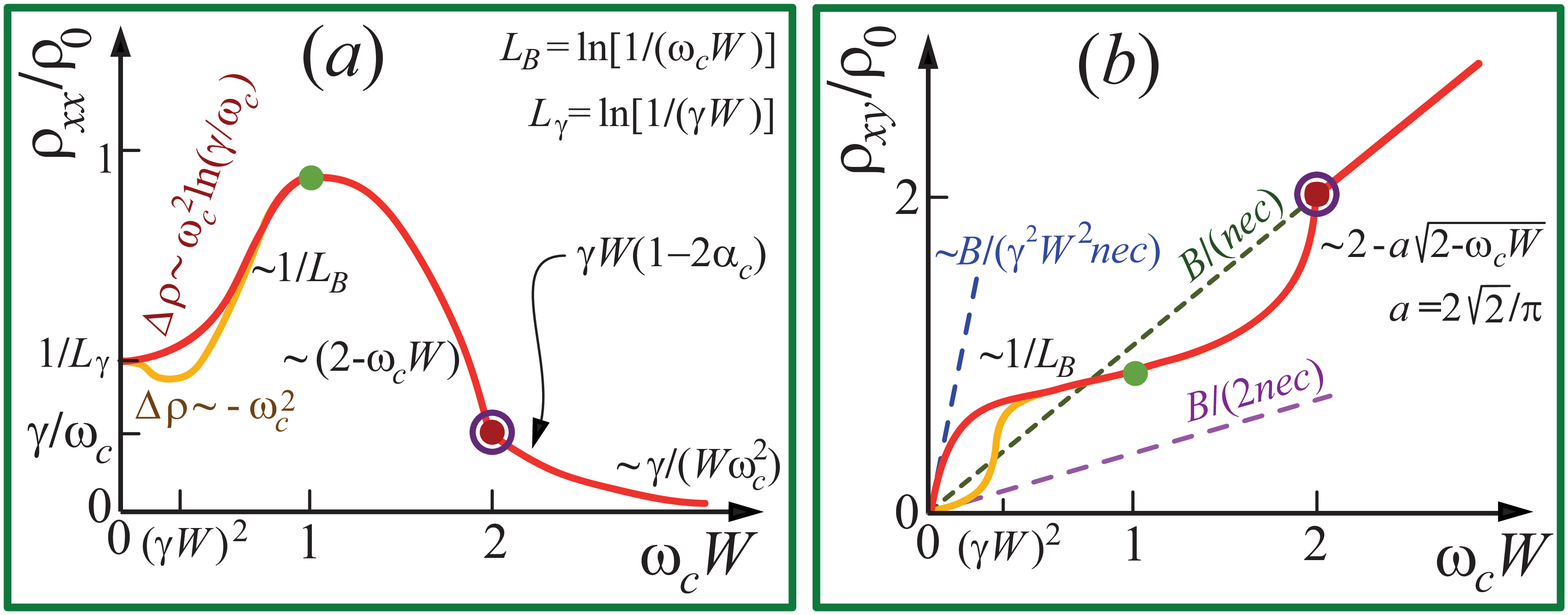}
	\caption{
	Longitudinal $\varrho_{xx}$ (a) and Hall $\varrho_{xy}$ (b) resistances  (schematically) as functions of magnetic field
for a defectless sample
normalized
 	on  the nominal ballistic resistance $\varrho_0 = m /(ne^2 W) $. In the first ballistic region ($ R_c \gg l^2 /W $) the resistances of   the whole sample with straight edges, including the near-edge
 	regions, (red curves) as well as of the bulk part of the sample without accounting the near-edge regions (orange curves) are drawn.
 	}
 	\label{FigS5}
\end{figure}

For the longitudinal $\varrho_{xx} = E_0/j$ and the Hall $\varrho_{xy} = E_H/j$  resistances  we obtain from Eqs.~(\ref{mean_f_res_j_below}), (\ref{mean_f_res_EH_below2}), (\ref{mean_f_above_res_j}), and (\ref{mean_f_above_res_EH}) in the both vicinities  of  the ballistic-hydrodynamic  phase transition, $ | \omega_c W - 2 | \ll 1 $:
\begin{equation}
\label{ro_xx_abov}
\varrho_{xx} (B) =
 \frac{m \gamma }{n_0e^2}
\left\{ \begin{array}{r}
1 + 2 u\, / \, (\gamma W)
 \,  ,\quad  B <B_c
 \\
1-2\alpha_c
 \, ,\quad  B>B_c
 \end{array}\right.
\end{equation}
and
\begin{equation}
\label{ro_xy_abov}
\varrho_{xy}(B)
  =
\frac{B}{n_0ec}
\left\{ \begin{array}{r}
1- \sqrt{ u / \pi }
\,  ,  \quad B <B_c
\\
1
\, , \quad  B>B_c
 \end{array}\right.
 \:.
\end{equation}
 In these  resulting formulas we again use the usual units for clarity.
We remind that  $u(B) =(2/\pi) (2 - \omega_c W  )  \sim B_c -B $ and $\alpha_c(B) = 1- \omega_c W/2 \sim B -B_c$.
 It is noteworthy that the Hall resistance  $\varrho_{xy} $  above the transition, $B>B_c$,
  coincides with its value in the Ohmic regime at zero temperature,  $ \varrho _{xy} ^{(0)} = B/(n_0ec)  $, at least,  in the main order by $\gamma W$ in which all our calculations are performed.

In Fig.~\ref{FigS5} we schematically draw the longitudinal and the Hall resistances  $\varrho_{xx} (B)$ and  $\varrho_{xy}(B)$ in the whole considered
 diapason of $B$: the three
ballistic subregimes
as well as
the upper and lower vicinities  of the ballistic-hydrodynamic transition,

Second,
let us consider the ballistic-Ohmic phase transition in a sample in which the scattering only  on disorder in
the bulk is substantial.
For clarity, below we will  use the explicit notation $\gamma'$ for the rate of the electron scattering on disorder.

In such system, the  contribution to the current from the edge and from the central  electrons are independent. For the contributions to $j$ and $E_H$ from edge electrons   one should use the pure ballistic formulas (\ref{av_crit})  at $u = 0 $, multiplied on the fraction of the edge electrons $\alpha_e$:
 \begin{equation}
 j_e= \alpha_e \, \frac{E_0  }{2\gamma'}
 \:, \quad\;\;
 E_{H,e} = \alpha_e \, \frac{\omega_c  }{2\gamma'}  \, E_0
 \:.
 \end{equation}
For the contribution to the current and the Hall field from the central electrons  we should use the Drude formulas with
 the ``bare'' field $E_0$ and
the factor $\alpha_c$   accounting their relative density:
 \begin{equation}
\label{Ohm_res_jc_Delt_EH_above}
j_c =   \alpha_c  \,  \frac{ E_0  }{ \gamma'} \:,  \:\;\;\;\;
 E_{H,c} =   \alpha_c  \, \frac{ \omega_c E_0  }{ \gamma'}
 \:.
\end{equation}
 Keeping in mind that $\alpha_c+\alpha_e =1$, for the total averaged current density
  and the averaged  Hall field we obtain:
\begin{equation}
\label{Ohm_res_j_EH_above}
 j =   \frac{1+ \alpha_c }{2}   \, \frac{ E_0  }{ \gamma'} \:,  \:\;\;\;\;\;
 E_{H} =   \frac{1+ \alpha_c }{2}  \, \frac{ \omega_c E_0  }{ \gamma'}
 \:.
\end{equation}
Unlike the pure samples where only the electron-electron scattering takes place in the bulk, results (\ref{Ohm_res_j_EH_above}) remains valid in the disordered samples  at any relation between the fractions $\alpha_c $ and $\alpha_e $ of electrons of the both  groups.
 In particular, in the limit of very magnetic large fields,
  when $W \gg R_c$ and $ \alpha_c \approx 1 $, equations~(\ref{Ohm_res_j_EH_above})
  turn into the usual Drude formulas for long Ohmic samples.

\begin{figure}[t!]
	\includegraphics[width=1.0\linewidth]{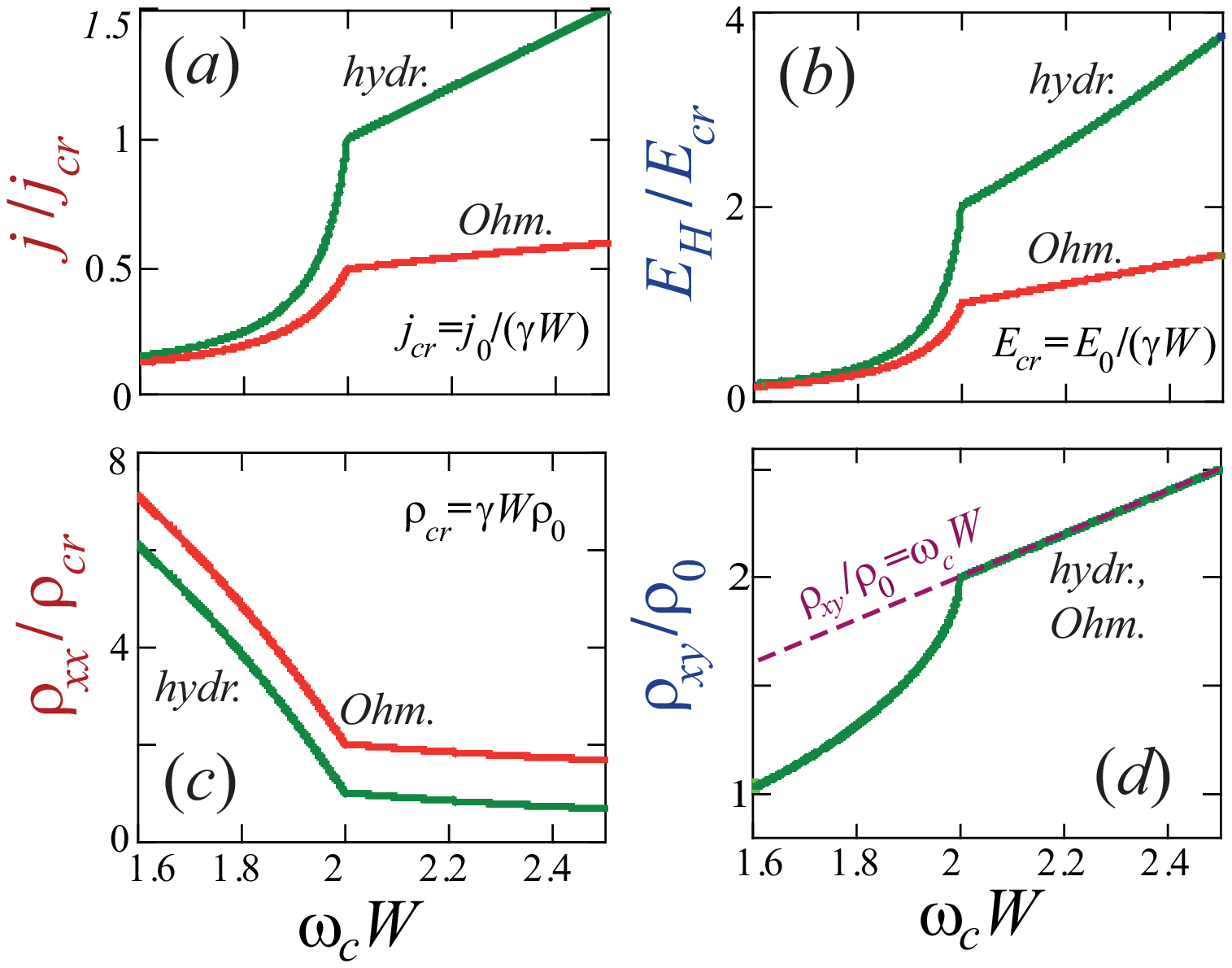}
	\caption{
	Current (a), Hall electric field (b), longitudinal (c) and Hall (d) resistances as functions of  magnetic field in a wide region around the critical field,  $\omega_c^{cr} =2/W $, for the ballistic-hydrodynamic phase
	transition [green curves] and for the ballistic-Ohmic phase transition [red curves].  In
	panel (d) for the Hall resistances,  the green and the red curves referring to these to transitions coincide.
All curves are plotted for the same values of the interparticle and the disorder scattering rates: $\gamma = \gamma' $.
	}
	\label{FigS6}
\end{figure}

Equations (\ref{av_crit}), (\ref{av_crit_H2}), and  (\ref{Ohm_res_j_EH_above})
 lead to the following result for the longitudinal resistance in the both vicinities of the ballistic-Ohmic transition, $ | \omega_c W - 2 | \ll 1 $:
\begin{equation}
\label{ro_xx_abov_O}
\varrho_{xx} (B) =
 \frac{2m  \gamma ' }{n_0e^2}
\left\{ \begin{array}{r}
  1+ u \, / \,(  \gamma ' W)
 \,  ,\quad  B < B_c
   \\
 1 - \alpha_c
 \, ,\quad  B > B _c
 \end{array}\right.
 \:,
\end{equation}
while  to the Hall resistance identical with result~(\ref{ro_xy_abov}) for the  ballistic-hydrodynamic transition.

\begin{figure}[t!]
	\includegraphics[width=0.99\linewidth]{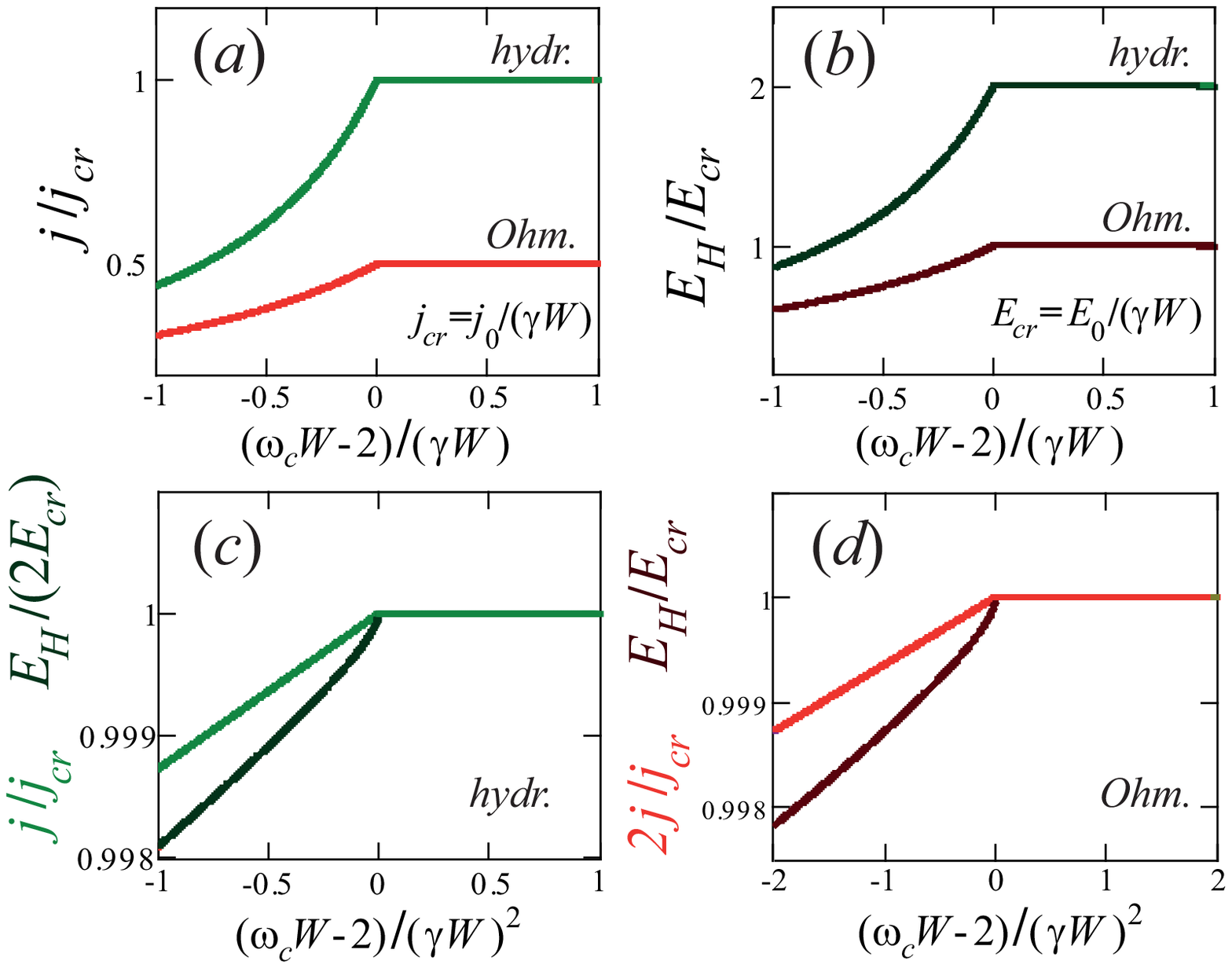}
	\caption{
	Current and  Hall electric field as functions of  magnetic field in narrow (a,b) and  very narrow (c,d) vicinities of the critical point $\omega_c^{cr} =2/W$  for the ballistic-hydrodynamic and for the ballistic-Ohmic
	phase transition. All curves are plotted for the same values of the interparticle and the disorder scattering rates: $\gamma = \gamma' $.
	}
	\label{FigS7}
\end{figure}

In Fig.~\ref{FigS6} and Fig.~\ref{FigS7}  we draw the values  $j$ and $E_H $ in the fields above and below the critical point $ \omega_c ^{cr} =2/W $ [Eqs.~(\ref{mean_f_res_j_below}), (\ref{mean_f_res_EH_below2}), (\ref{mean_f_above_res_j}),  (\ref{mean_f_above_res_EH}), and Eqs.~(\ref{Ohm_res_j_EH_above})], together  with the corresponding resistances $\varrho_{xx} $ and $\varrho_{xy}$ [Eqs.~(\ref{ro_xx_abov}), (\ref{ro_xy_abov}), and (\ref{ro_xx_abov_O})], for both the ballistic-hydrodynamic and the ballistic-Ohmic transitions in wide  and narrow vicinities the critical magnetic field  $\omega_c ^{cr} =2/W$.
 We see from Figs.~\ref{FigS5}-\ref{FigS7} that, in  both the ballistic-hydrodynamic and the ballistic-Ohmic phase  transitions,
 the current  $j(\omega_c)$ and the longitudinal resistance $\varrho _{xx}(\omega_c)$ have a jump of its derivative
 (a kink) at the  critical field $\omega_c^{cr} =2 /W$, while  the Hall field $E_H(\omega_c)$  and resistance $\varrho _{xy}(\omega_c)$ have even a square-root singularity at $\omega_c \to \omega_c^{cr} -0$ (see Figs.~\ref{FigS6} and \ref{FigS7}).

\begin{figure}[t!]
	\includegraphics[width=1.0\linewidth]{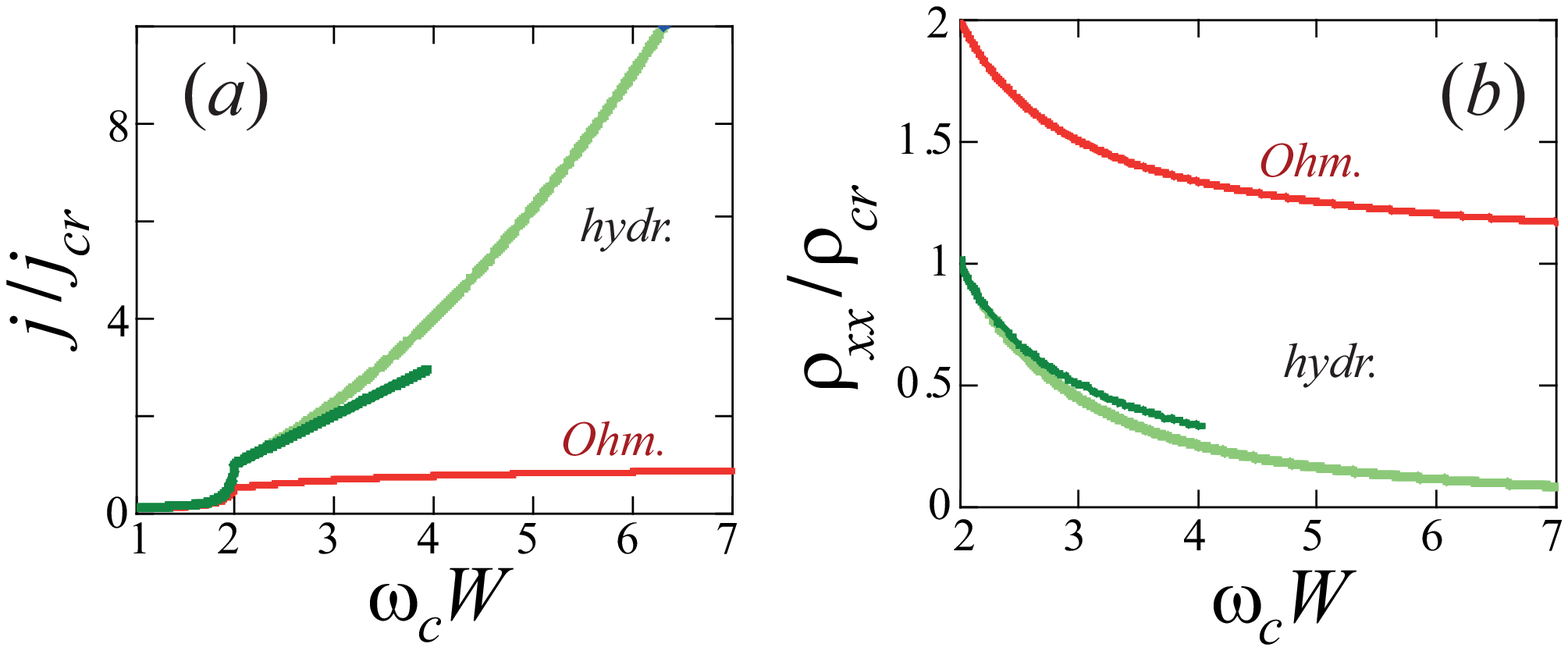}
	\caption{
	Current (a) and longitudinal resistance (b) as functions of magnetic field near  and  far beyond   the critical field
	$\omega_c^{cr} = 2/W $ for  the ballistic-hydrodynamic  and the ballistic-Ohmic  phase transitions at the coinciding interparticle and disorder scattering  rates: $\gamma_{ee} = \gamma' = \gamma$. Red curves for the  Ohmic samples are plotted by formulas~(\ref{Ohm_res_j_EH_above}).
 Dark green curves for $j$ and $\varrho_{xx} $  in the vicinity  of the ballistic-hydrodynamic   phase transition are drawn  by formulas~(\ref{mean_f_above_res_j}) and (\ref{ro_xx_abov}),  which do not take into account the
	inhomogeneous distribution of central electrons.   Light-green curves schematically  show the results for the inhomogeneous  hydrodynamic flow.
The asymptote of light-green and red curves at $\omega_c \gg 1/W$ correspond
 to the bulk hydrodynamic~(\ref{P})  and  Ohmic~(\ref{D})  resistances.
 	}
 	\label{FigS9}
\end{figure}

Third, we briefly discuss the behavior of the hydrodynamic and Ohmic flows far beyond the transition, when  $W-2R_c \sim W$  and $W-2R_c \approx W \gg R_c$. In this cases, the fraction of the central electrons is comparable or greater than the one of the edge electrons:  $\alpha_c \sim \alpha_e \sim 1$ or $\alpha_c \approx 1,\; \alpha_e \ll 1$. The collisions between the central electrons becomes important. The central electrons located farther from the edges are less likely  to collide with edge electrons than central electrons located closer to the edge. Such character of
 the scattering leads  to the formation of an inhomogeneous, parabolic by $y$, profile $j(y) \sim [(W/2) ^2 - y^2 ]$ of the Poiseuille flow.

The region $ W/R_c -2 \gtrsim  1$  of the mixed hydrodynamic-ballistic  and the Ohmic-ballistic flows were studied by the numeric solution of the kinetic equation in Refs.~\cite{Scaffidi,pohozaja_statja}. For the ballistic-Ohmic transition, formulas~(\ref{Ohm_res_j_EH_above}) for $j$ and $E_H$, as it was mentioned above, are valid at any relation between  $R_c $ and $ W$.

At large magnetic fields, $\omega_c \gg 1/W$, in pure  samples  the hydrodynamic contribution
from the central electrons to the current dominates. The averaged resistance  is given by the formula:
$ \varrho_P = 12  \eta_{xx} / W^2 $~\cite{je_visc},  where $\eta _{xx}$ is the diagonal viscosity coefficient,
 being equal in our notations to $\eta _{xx}= \gamma/(16\omega_c^2)$ at  $\gamma \ll \omega_c$.
   Thus,  we obtain for the averaged sample resistance at $W\gg 1/\omega_c$:
 \begin{equation}
 \label{P}
 \varrho_P (\omega_c) = \frac{  3\gamma }{4 \omega_c^2 W^2 } \:.
 \end{equation}
In this formula we neglect
 the contribution from the edge electrons in the narrow near-edge regions $W/2 - |y| \gtrsim R_c $~[see Fig.~2(f) in the main text].

In the limit  $\omega_c \gg 1/W$  for disordered samples
 we obtain from  Ref.~(\ref{Ohm_res_j_EH_above}) in the limit $W\gg R_c$   the usual Drude result for magnetoresistance of long samples,
 which is independent on $\omega_c$:
 \begin{equation}
 \label{D}
 \varrho_D (\omega_c) = \gamma' \,.
 \end{equation}

It is noteworthy that both the hydrodynamic and the Ohmic resistances   $ \varrho_P $ and  $ \varrho_D $ are proportional to the scattering rates $\gamma$ and $\gamma'$, but the second one decreases with the magnetic field  as $\sim1/(\omega_c W)^2$. In Fig.~\ref{FigS9} we schematically plot the dependencies $j(\omega_c)$ and $\varrho_{xx}(\omega_c)$ near and far from the transition ballistic-hydrodynamic and the ballistic-Ohmic transitions. At $0 < \omega_cW - 2 \ll 1  $ these curves are  described by   Eq.~(\ref{mean_f_above_res_j}), (\ref{ro_xx_abov}), (\ref{Ohm_res_j_EH_above}), and (\ref{ro_xx_abov_O}). At $\omega_c \gg 1/W$ they follow the asymptotes~$ \varrho_P $~(\ref{P}) and  $ \varrho_D $~(\ref{D}).

\begin{figure}[t!]
	\includegraphics[width=1.0\linewidth]{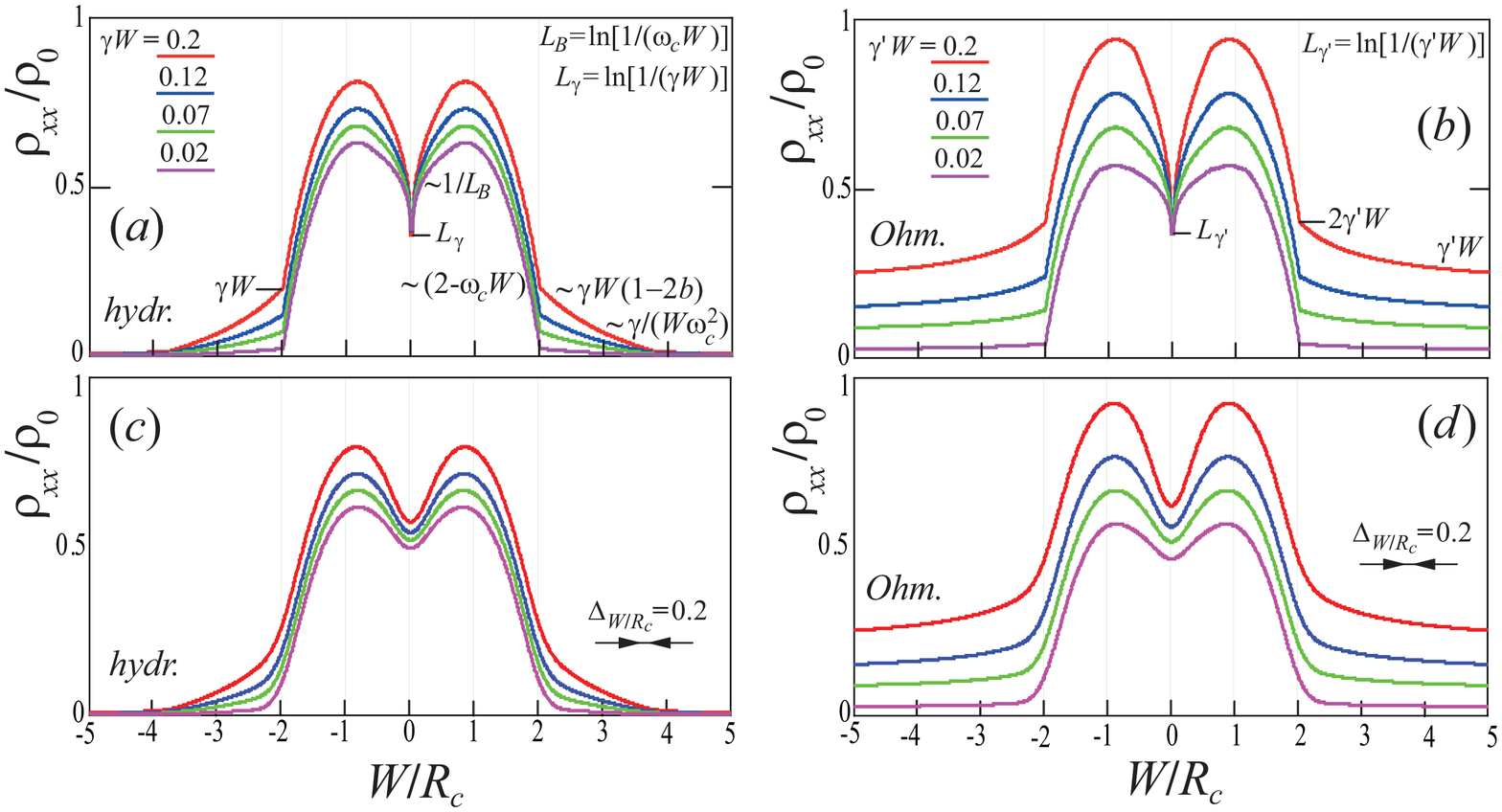}
	\caption{
		Longitudinal $\varrho_{xx}$ resistance  as functions of the parameter $W/R_c \propto B$,
 normalized  on  the nominal ballistic resistance  $\varrho_0 = m /(ne^2 W) $,
		for pure samples at different intrparticle scattering rates $\gamma $ (a,c)
and
for the disordered samples  with no intrparticle scattering
		at different  disorder scattering rates $\gamma'$ (b,d).
Curves in panels (c,d) are plotted  by convolution of curves in panels (a,b),
		with a Gaussian with a width $\Delta_{  W/R_c}  =0.2 $,
 that   qualitatively accounts sample imperfections.
		}
	\label{FigS8}
\end{figure}

In Fig.~\ref{FigS8}  we plot the longitudinal resistance $\varrho_{xx} (B)$ in the whole interval of magnetic fields for a sample with no disorder at different interparticle scattering rates $\gamma $ [panels (a,c)] and for a sample with no interparticle scattering at the same values of  the scattering rate on bulk disorder $\gamma '$ [panels (b,d)].
 The results presented in Fig.~\ref{FigS8}  are a more precise and detailed version for of the results in Fig.~\ref{FigS5}.
Curves in panels (a,b) are plotted by the interpolation formulas based of the presented above asymptotes
for $\varrho_{xx}$  in the three ballistic subregimes (Sec.~2),
 in the upper and the lower vicinities  of the transition point $B=B_c$ (Sec.~3.3),
  and in the bulk limit (Sec.~3.3).
  In panels (c,d) we plot the  curves from panel (a,b),
  smoothed by convolution with a Gaussian weight function $G_\Delta  (B)$
  with a width $\Delta_{  W/R_c}  =0.2 $, that   simulates the contribution of several sections of a
  long sample with varying widths and other imperfections.

The ballistic-hydrodynamic and the ballistic-Ohmic
 phase transitions can be distinguished in the experimental data, primarily, by the relationship between the value of the longitudinal resistance at the transition point $\varrho_{xx}|_{\omega_c W = 2 }$ and in the limit of high magnetic fields $\varrho_{xx}|_{\omega_c W \gg 1  }$. For the ballistic-hydrodynamic transition, the latter value decreases up to zero as $1/B^2$ with increasing magnetic field [see Fig.~\ref{FigS8}(a)] and quickly becomes much less than  $\varrho_{xx}|_{\omega_c W = 2 }$, while for the ballistic-Ohmic transition the value $\varrho_{xx}|_{\omega_c W = 2 }$ is twice the value $\varrho_{xx}|_{\omega_c W \gg 1  } $ [see Fig.~\ref{FigS8}(b)].

Of particular interest is the behavior of the Hall resistance $\varrho_{xy} (B) =E_H/j$.
 First, we saw that the dependencies $\varrho_{xy}(B)$  for both the  ballistic-hydrodynamic and
 ballistic-Ohmic transitions coincide one with another [see Fig.~\ref{FigS6}(d)].   Second, above the transition point, $\omega_c > \omega_c^{cr}$,  the function $\varrho_{xy} (B)$ is identically equal  to its ``standard'' value $\varrho_{xy}^{(0)} =B/n_0ec$,   corresponding to the   Ohmic regime at low temperatures. This result is obtained in the main order by the small parameters  $\gamma W$,  $\gamma ' W $ and the calculation of the corrections  to this main part of  $\varrho_{xy}$ by the powers of $\gamma W$,  $\gamma ' W$
 will apparently  lead to corrections to the standard value  $\varrho_{xy} = \varrho_{xy}^{(0)}$
 at high field $\omega_c W \gg  1 $. Such corrections was obtained Ref.~\cite{Scaffidi}
  in numerical solution of the kinetic equation and were experimentally studied in Refs.~\cite{Gusev_2_Hall,graphene_3}.

\section{4. Comparison with preceding  theoretical and experimental works}
\label{Sec:S4}
 \subsection{4.1. Related theoretical works}
 \label{Sec:S4.1}

\begin{figure}[t!]
	\includegraphics[width=1.0\linewidth]{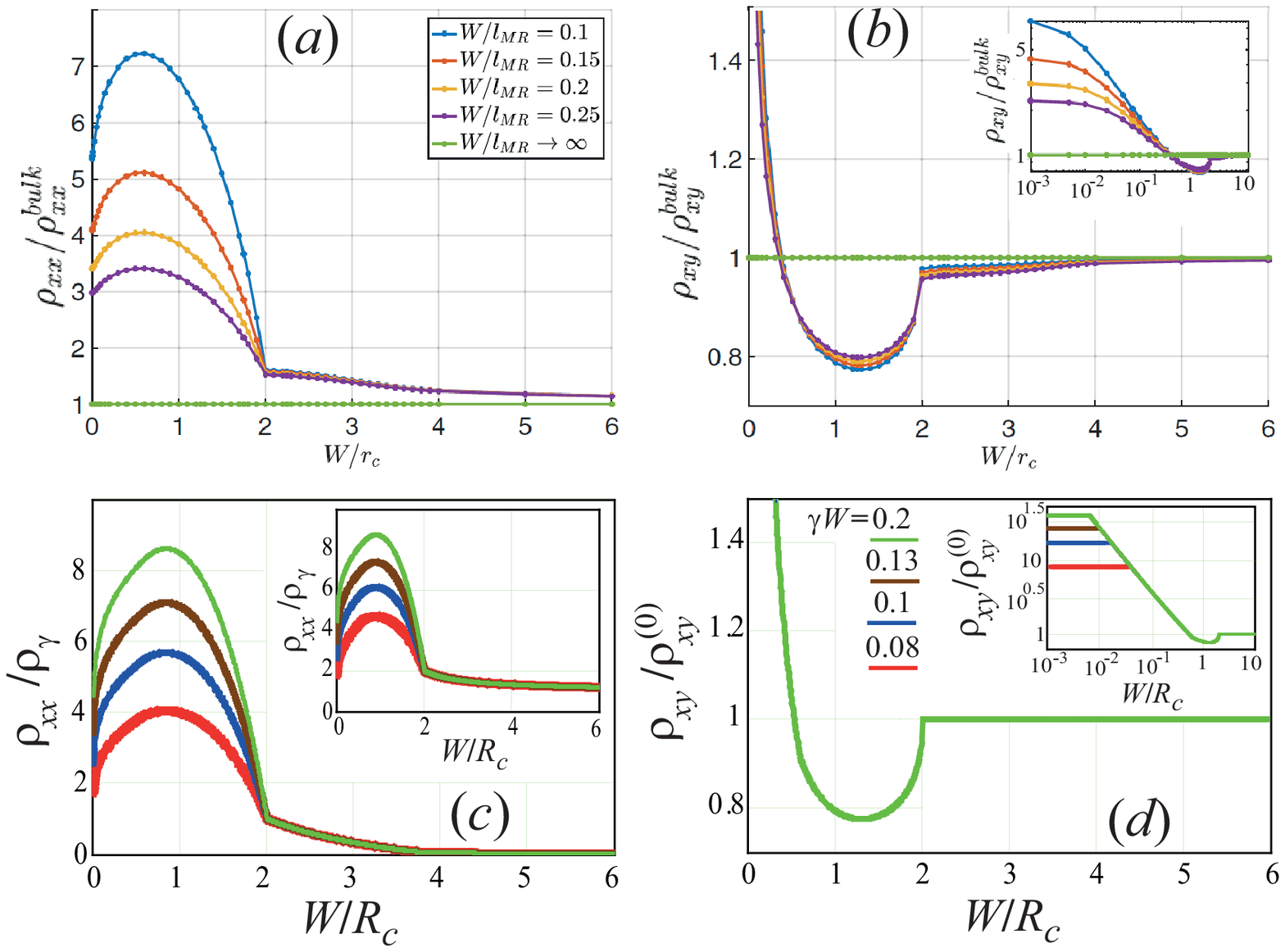}
	\caption{
  	Comparison of  the results of the numeric solution of kinetic equation~(\ref{kin_eq}) in Ref.~\cite{Scaffidi} with the results of our analytical theory. Panels (a,b), taken from Ref.~\cite{Scaffidi}, show the longitudinal
  	$\varrho_{xx}$ and the Hall $\varrho_{xy}$ resistances as function of magnetic field for the narrow  samples, $W<l_{MR}$,
  the where scattering of 2D electrons on disorder dominates ($l_{MR}$ is
  	corresponding scattering length). Panels (c,d) present our results for $\varrho_{xx}$ and  $\varrho_{xy}$ plotted in the same units as the ones in panels (a,b). Main graphics in panels (c,d) correspond to
  	pure samples with only the interparticle scattering with several rates $\gamma$ shown in panel~(d). Inset in panel (c) present
  $\varrho_{xx}$ for disordered samples with no interparticle scattering  and the
  	disorder scattering rates  $\gamma'$ equal to the values $\gamma$ used for  the  curves in main panel~(c).
  Results for $\varrho_{xy}$, shown in panel (d),
   are identical in the considered main approximation by $\gamma W$ for both
   pure samples and disordered samples
   at $\gamma = \gamma  ' $. Inset in panels (b) and (d) show the function $\varrho_{xy} (B)$ in the logarithmic scales.
  	}
  	\label{FigS10}
\end{figure}

In Ref.~\cite{Scaffidi} a numerical solution of the kinetic equation (\ref{kin_eq})
 was carried out for the same system as we studied in this work:
 2D electrons in
 a long sample with rough edges and various rates of interparticle and disorder scattering.
 The kinks in the dependencies of the resistances $\varrho_{xx}$ and $\varrho_{xy}$
  on magnetic field $B$ at the transition point $ B = B_c $ were obtained
 in~\cite{Scaffidi}
for  the case when the scattering length is longer than the sample width.
In Figs.~\ref{FigS10}(a,b) we cite the results of Ref.~\cite{Scaffidi}
for the sample with a very weak electron-electron scattering
 and
 the more substantial (but also weak) scattering of electrons
 on bulk disorder, corresponding  the mean free paths $l=l_{MR}$  much longer than the sample width.

Although the authors of Ref.~\cite{Scaffidi} discussed the emergence
of the  central electrons above $B_c$
those do not scatter on the edges,
 they  did not considered the peculiar electron dynamics in the regimes just below the transition [in the third ballistic region,
$ \gamma W  B_c \ll B_c  - B  \ll B_c  $,
and in the semiballistic pre-transition  region $  0< B_c  - B  \lesssim \gamma W B_c $] as well as    the interaction between the edge and the central electrons just above the transition,   $0< B-B_c \ll B_c$.  The shapes
of the dependencies $\varrho_{xx}(B)$ and $\varrho_{xy}(B)$ obtained within our model for the ballistic-hydrodynamic and the ballistic-Ohmic transitions by analytical calculations are similar to the one obtained in Ref.~\cite{Scaffidi}
for
 the ballistic-Ohmic transition
in
the narrow samples, $W \ll 1/\gamma$ (see Fig.~\ref{FigS10}).
Curves in  Figs.~\ref{FigS10}(c,d)
 are plotted by the interpolation formulas based of the  obtained above  and cited above asymptotes
 for $\varrho_{xx}$ and  $\varrho_{xy}$ in the three ballistic subregimes, in both the upper and the lower vicinities
  of the transition point $B=B_c$, and in the bulk limit $W \gg R_c$.

 Note that the results for
  $\varrho_{xx}$ and $\varrho_{xy}$
  in pure samples
  with long interparticle scattering lengths and no scattering on disorder,
  were not presented in  Ref.~\cite{Scaffidi},   although, apparently,
  they could  be
  calculated in the same method   as was used in~\cite{Scaffidi} for the obtaining
  of the resistances $\varrho_{xx}$ and $\varrho_{xy}$  in a disordered sample
  [see Figs.~\ref{FigS10}(a,b)]. In the current work we succeeded
in obtaining the dependencies  $ \varrho_{xx}(B)$ and $\varrho_{xy}(B)$  in both the pure samples
  ($\gamma \gg \gamma ' $, $\gamma  W \ll 1 $)
as well as of the disordered samples ($\gamma \ll \gamma ' $, $\gamma ' W \ll 1 $)
 in an exact form in the limiting magnetic fields
   $B \ll B_c$ and $|B-B_c| \ll B_c$.

In Ref.~\cite{pohozaja_statja} the ballistic and the hydrodynamic transport of 2D electrons in a long stripe was theoretically studied. It that work, similar general  formula for the distribution function in the ballistic regime [Eq.~(\ref{f_gen})] was obtained. Also the simplified form of the distribution function in the second ballistic subregime [Eq.~(\ref{f_bez_gamma})] was derived and the resulting anomalously large Hall field, $E_H \sim E_0$, was deduced. Additionally a numeric description of the hydrodynamic-ballistic transport regime above the transition field at $\omega_cW -2 \sim 1 $, in particular, the detailed studies of the Hall field profiles, were performed.

In  Ref.~\cite{pohozaja_statja}  the evolution  of the dynamics of individual 2D electrons in a long sample with the increase of a  magnetic field (in particular, the emergence of the edge and the central electrons at $\omega_cW >2$)
was generally discussed.
  However the curvature of the Hall field profile $E_H(y)$, characterizing the evolution  of the type of a flow,
 was attracted the main attention,
 whereas the presence of a genuine phase transition at $\omega_cW = 2$ between the ballistic almost independent electrons
 and a collectivized electron fluid phase
 as well as the nontrivial semi-ballistic dynamics
 of electrons in the first and third ballistic sub-regimes
  were not studied.
  In our work,
   we provide a proper  description of the flow in the first
   and the third ballistic subregimes;
 explain the origin of the bulk contribution to the Hall resistance $\varrho _{xy}= \varrho _{xy}^0/2$ in the limit $B \to 0$;
 as well as reveal and study
  the ballistic-hydrodynamic and  the ballistic-Ohmic phase transitions at $\omega_cW = 2$.

To conclude this section, we note that the mean field model developed in Sections~3.2  and~3.3, apparently,
has the usual accuracy
of the methods of this type, namely,
 describes all quantities in the vicinities of the phase transition up to numerical factors of the order of unity.
 Such description of the phase transition within the basic equations of Sec.~1, apparently,
 can be refined only by numerical calculations similar to the ones performed in Refs.~\cite{Scaffidi,pohozaja_statja}.
 However, the collision operator~(\ref{St_N__St_U}), used in this work and in~\cite{Scaffidi,pohozaja_statja}
 is already a significant simplification of the exact operator of electron-electron collisions.
  For example, even for
  the 2D electron systems with
  the simplest quadratic electron spectrum,
  like GaAs quantum wells, the relaxation rates $\gamma_m $
  of the distributions proportional to various harmonics $e^{im\varphi}$
  have different orders of magnitudes~\cite{Alekseev_Dmitriev}, in contrast to operator~(\ref{St_N__St_U}) leading to the same relaxation rates,  $\gamma$, of the second and the all  higher harmonics.

  We also remind that the proposed mean field model may be considered as
  the application of the additionally simplified form~$\mathrm{St}'$~(\ref{St_av})
  of the  electron-electron    collision operator $\mathrm{St}$~(\ref{St_N__St_U}).
  Collision operator~(\ref{St_av})
  does not take into account the spatial peculiarities of the flow,
   but seems to be sufficient for a qualitative description of
   the average characteristics of the flow and
   their dependencies on magnetic field.

\subsection{4.2. Related experiments}
\label{Sec:S4.2}

In this subsection we compare
  the results of experiments~\cite{rrecentnest,rrecentnest2,Gusev_2,exps_neg_3,exps_neg_1,exps_neg_2}
  with
  our results on the ballistic-hydrodynamic phase transition   and on the magnetoresistance in different regimes.

In Ref.~\cite{rrecentnest} a flow of 2D electrons in a graphene stripe was experimentally studied. Magnetoresistance and distributions
 of the Hall electric field over the cross section of the stripe were measured at various magnetic fields,  electron densities, and temperatures
  corresponding to the ballistic and the hydrodynamic regimes.
 The obtained experimental  dependencies $\varrho_{xx}(B)$  are very similar to the theoretical ones presented
  in Figs.~\ref{FigS5}(a) and \ref{FigS8}.
 In Fig.~\ref{Fig3} in the main text
 we cite
  the magnetoresistance curves measured  in Ref.~\cite{rrecentnest} and
  present our theoretical results for reasonable values of the parameter $\gamma W$.
   Both the  experimental and the
  theoretical dependencies $\varrho_{xx}(B)$ exhibit a positive magnetoresistance at  $ B \ll B_c$,  a nonmonotonous behavior
   at $B \lesssim B_c$ with a maximum at $B \sim B_c/2$,
   a smeared kink at the critical field $B=B_c$,
   the dependence similar to $\varrho_{xx}(B) - \varrho_{xx}(B_c) \sim B_c-B$ at $ 0 <  B_c-B \ll B_c $,
  and a monotonous decrease at $B>B_c$.
  This coincidence signifies that the   ballistic subregimes and
  the ballistic-hydrodynamic phase transition, theoretically revealed in our work,
    were apparently realized and observed  in experiment~\cite{rrecentnest}.

In Ref.~\cite{rrecentnest2} an observation of the profile of the current density $j(y)$  for a flow of 2D electrons in a graphene stripe  was performed   by measurements of the local magnetic field induced by the distribution of $j(y)$. Additionally, magnetoresistance  of a graphene stripe, being very similar  to the magnetoresistance reported Ref.~\cite{rrecentnest},   was observed. In Fig.~\ref{Fig3}(b) in the main text we  cite the magnetoresistance curves presented in Supplemental information of Ref.~\cite{rrecentnest2}. These curves  exhibit
 a well-pronounced kink at $B=- B_c$, a maximum at $B \sim \pm B_c $, and the behavior
 $ \varrho_{xx}(|B|) - \varrho_{xx}(B_c) \sim B_c-|B|$ at $ 0 <  B_c-|B| \ll B_c $.
 The kink  at the critical field $B=B_c$, apparently indicating the  phase transition,
  is observed in \cite{rrecentnest2}  more clearly, than in \cite{rrecentnest}.
Note that the dependencies $ \varrho_{xx}(B)  $ from  Ref.~\cite{rrecentnest2}
 are substantially asymmetric by $B \to -B$ [see Fig.~\ref{Fig3}(b) in the main text]. The last fact can be related with an admixture of the Hall resistance in the observed resistance. This is especially clearly
  evidenced by the asymmetry of the kink at $B= \pm B_c$.

In Ref.~\cite{rrecentnest} the Hall field profiles $E_H(y)$
 were measured
 in a most part, but not in the entire section of each studied stripe.
  In the ballistic regime these profiles turned out to be nearly flat,  while in the hydrodynamic regimes they were strongly curved, similarly to the parabolic hydrodynamic profiles corresponding to the Poiseuille flow.
 Based on our theoretical results, we
  predict that the sharp features in $E_H(y)$ near  the sample edges, $y=\pm W/2$, can be observed at $B<B_c$
    in future studies similar to experiment~\cite{rrecentnest},   provided  the profiles  $E_H(y)$ will be measured
    up to
      the very sample edges.
 The divergent features in $E_H(y)$  were obtained     in our theory for all three ballistic subregimes,
   however  they have largest width and amplitude in the lower  vicinity of $B_c$
  [see Fig.~\ref{FigS4} and Fig.~\ref{Fig2} in the main text]

\begin{figure}[t!]
	\includegraphics[width=1.0\linewidth]{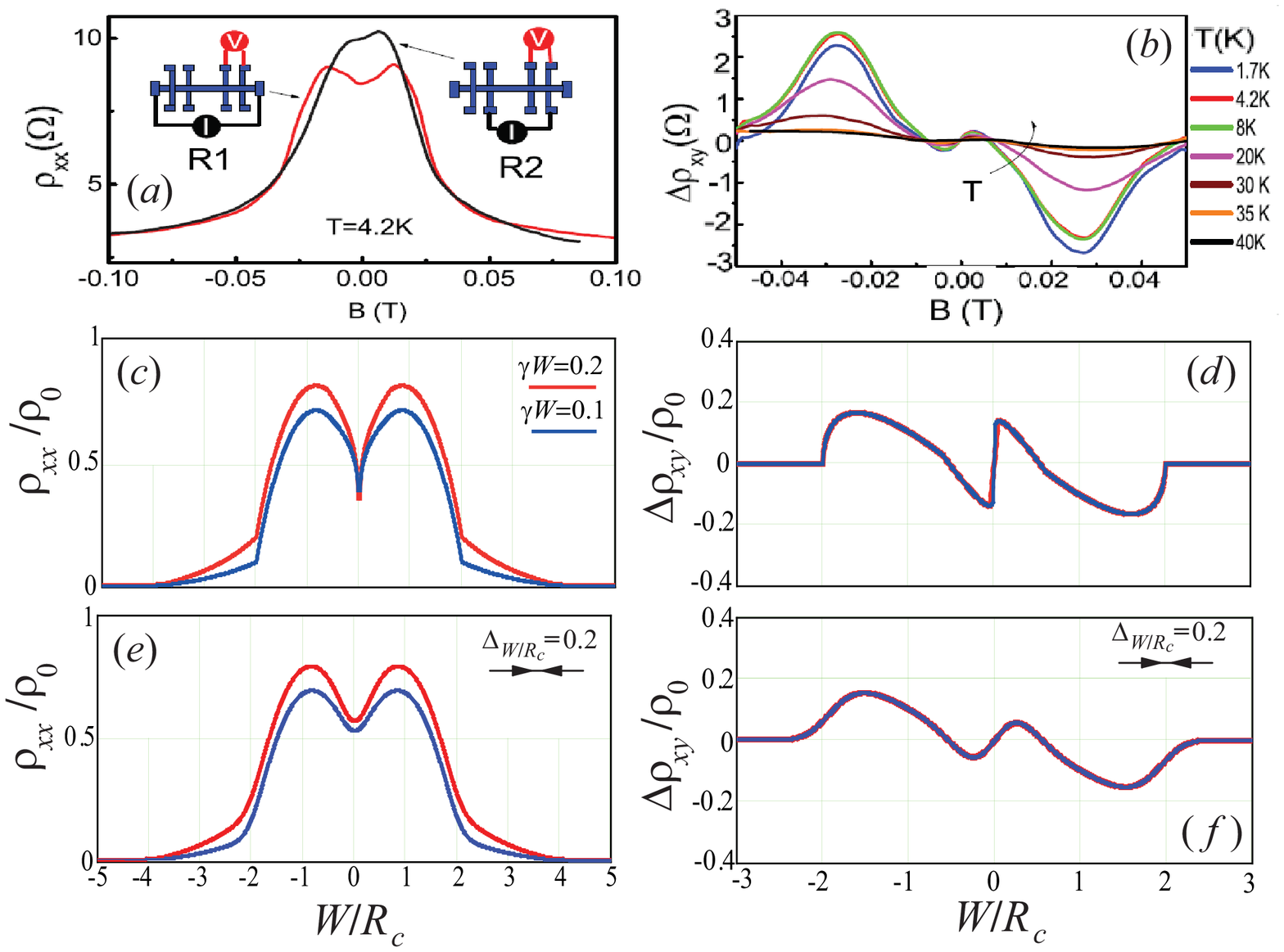}
	\caption{
	 Longitudinal $\varrho_{xx}$ and Hall resistances $\varrho_{xy}$ of long GaAs quantum wells samples
 [in panels (b,d,f) the value $\Delta \varrho_{xy} =\varrho_{xy} - \varrho_{xy} ^{(0)} $ is drawn].
  Panels (a,b) present the experimental results of  Ref.~\cite{Gusev_2_Hall}.
 Panels (c,d) show our results for $\varrho_{xx}$
	 and  $\varrho_{xy}$ for pure long samples with no disorder and the two values
of the interparticle scattering rates $\gamma$. In panels (e,f) we plot the  curves from panel (a,b), smoothed by convolution with a Gaussian weight function $G_\Delta(B)$
	 with a width $\Delta_{  W/R_c}  $, that   simulates the contribution of several sections of a long sample with varying widths
 as well as
other sample imperfections. The magnetic fields $0.05$~T at the right and the left edges of panel~(b), where the
	 resistance $\varrho_{xy}$  becomes close to the standard value $\varrho_{xy}^{(0)}$,
 well correspond to the equality $2R_c \approx  W  $, where $W=5~\mu$ is the width of the sample studied
    in Ref.~\cite{Gusev_2_Hall}.
 	}
 	\label{FigS11}
\end{figure}

In Ref.~\cite{Gusev_2_Hall}, a flow  of 2D electrons
in long samples of high-quality GaAs quantum
  wells was experimentally studied.
  The dependencies of the resistances $\varrho_{xx}$ and $\varrho_{xy}$
 on magnetic field were measured. We cite these results of  Ref.~\cite{Gusev_2_Hall} in Figs.~\ref{FigS11}(a,b).
  Apparently, in the 2D electron system   studied  in  Ref.~\cite{Gusev_2_Hall}
  the ballistic transport regime   at $W < 2R_c$
  and
  the hydrodynamic regime at $W > 2R_c$
  were realized.
  The experimental dependencies  $\varrho_{xx}(B)$ is rather similar to the magnetoresistance observed in Refs.~\cite{rrecentnest,rrecentnest2} in graphene stripes
  and to our theoretical result for $\varrho_{xx}(B)$ presented in Fig.~\ref{FigS5}(a) and \ref{FigS8}.
  The experimental result of Ref.~\cite{Gusev_2_Hall} for  $\varrho_{xy}(B)$    is also rather similar  the the theoretical  one obtained in the current work [see Fig.~\ref{FigS5}(b)]:
  the curve $\varrho_{xy}(B)$  exhibits positive and negative deviations from the standard Hall resistance  $\varrho_{xy}^{(0)} =B/(n_0ec)$ below the transition point, $W<2R_c$, and an almost exactly  coincidence with  $\varrho_{xy}^{(0)} $ above the transition point, $W>2R_c$.
  In Figs.~\ref{FigS11}(c,d) we plot the dependencies $\varrho_{xx}(B)$
  and $ \Delta \varrho_{xy}(B) = \varrho_{xy}(B) - \varrho_{xy} ^{(0)}(B)$
  obtained by the interpolation of the obtained and cited above asymptotes of these functions in the limiting diapasons of $B$,
 as it was done in Figs.~\ref{FigS8} and  \ref{FigS10}.

\begin{figure}[t!]
	\includegraphics[width=1.0\linewidth]{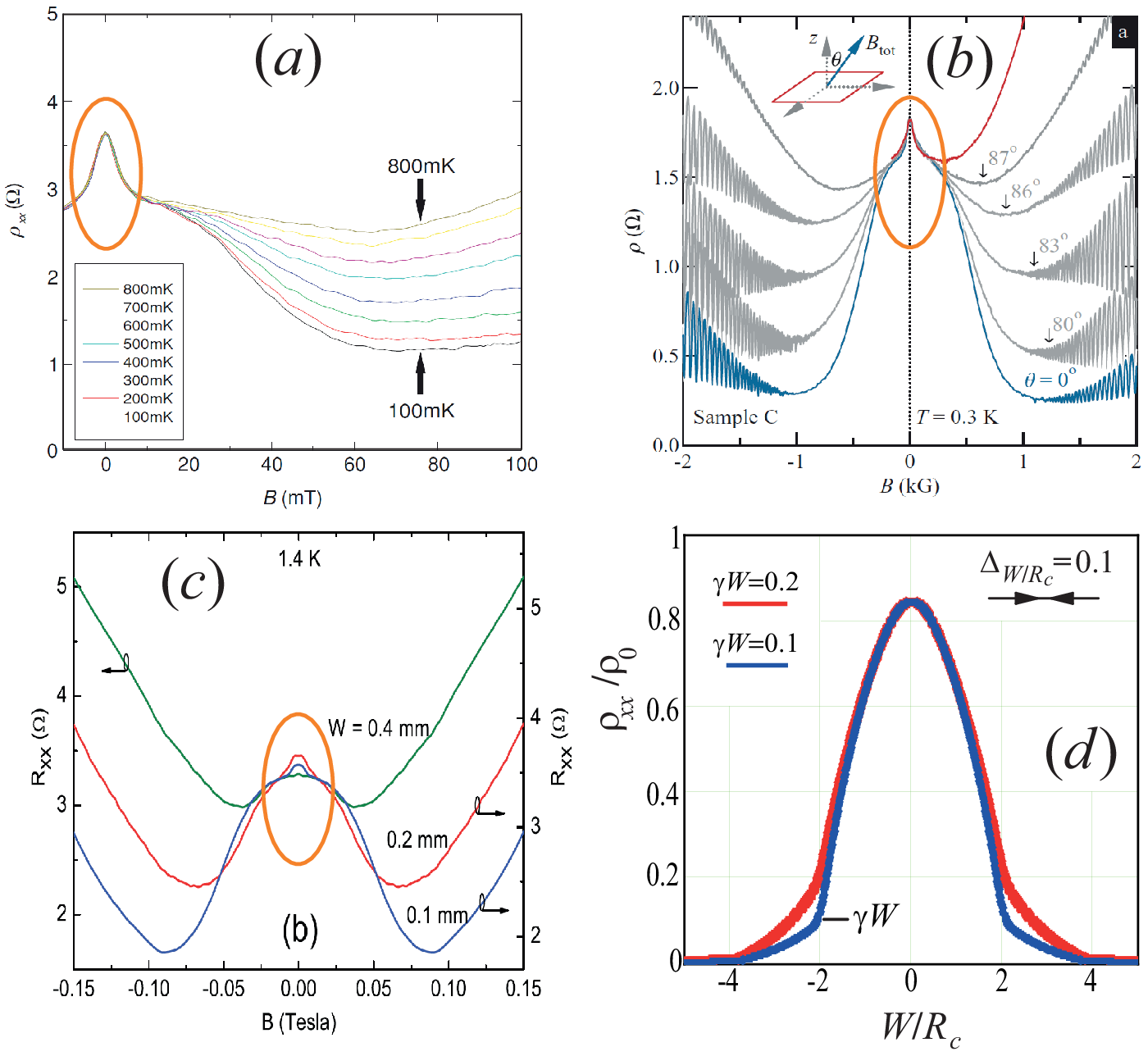}
	\caption{
	(a-c) Longitudinal ``bell-shape''  magnetoresistance of high-mobility GaAs quantum wells
 at low temperatures. Panels (a,b,c) are taken from experimental articles~\cite{exps_neg_3},\cite{exps_neg_1},\cite{exps_neg_2},
	respectively.
	In the samples studied in Ref.~\cite{exps_neg_3} (and possibly, in others) macroscopic defects present,
which  divide the 2D electron flow in the sample  on subregions of smaller sizes.
 Apparently, in all experiments~\cite{exps_neg_3}-\cite{exps_neg_2} the effective length and width
 of the sample  subregions were comparable one with other,
being
 much smaller than the interparticle scattering length and
  the length of the scattering on other disorder.
 In panel~(d) we plot the ballistic-hydrodynamic
  magnetoresistance given by Eqs.~(\ref{MR_B_straight_gamma_shrt}) and (\ref{ro_xx_abov}),
  obtained within our model
 for the case of short samples with $W\sim L \ll l$.
  The curves are smoothed by convolution with a Gaussian weight function $G_\Delta(B)$ with
 a width $\Delta_{  W/R_c}   $  in order to take into account the contributions
 from several  sample  subregions.
  	}
  	\label{FigS12}
\end{figure}

It is noteworthy that the magnitude and the sharpness of the features of $\varrho_{xx}(B)$
 and $\varrho_{xy} (B)$   observed in Ref.~\cite{Gusev_2_Hall} at $\omega_c W \lesssim 2 $
 are smaller than the ones observed for $\varrho_{xx}$  in Refs.~\cite{rrecentnest,rrecentnest2}. This can  be associated with a larger
  magnitude
 of imperfections of  the samples of  Ref.~\cite{Gusev_2_Hall} as compared with the graphene stripes examined in Refs.~\cite{rrecentnest,rrecentnest2}.
 In  Figs.~\ref{FigS11}(e,f) we plot
 the  curves  $\varrho_{xx}(B)$ and $\varrho_{xy}(B)$ from panel (c,d),
 smoothed by convolution with a Gaussian weight function with a width $\Delta_{  W/R_c}  =0.2 $,
 that   simulates the contribution of several sections of a long sample
  with varying widths, the samples corners,  and other imperfections.
   It is seen that the curves in panels~(e,f) much better fit
   the experimental curves in panels (a,b), than the ones in panels~(c,d).

We conclude that  the predicted magnetic field dependencies
 of the longitudinal and of the Hall resistances $\varrho_{xx,xy} (B)$ in the ballistic subregimes
 as well as
around and far above the ballistic-hydrodynamic phase transition
 are in a good agreement with results of experiments~\cite{rrecentnest,rrecentnest2,Gusev_2_Hall}.

 Additionally, we analyze experiments~\cite{exps_neg_3,exps_neg_1,exps_neg_2}
  on magnetotransport of 2D electrons in high-mobility GaAs quantum wells.
 In Figs.~\ref{FigS12}(a-c) we cite the magnetoresistance measured
  in those works. Apparently, the observed  negative magnetoresistance
  in the very low magnetic fields has a purely ballistic origin.
  Indeed, it is independent on temperature [see Fig.~\ref{FigS12}(a)];
 is robust to the effect of in-plane magnetic field [see Fig.~\ref{FigS12}(b)];
 and
 occurs  in sufficiently narrow samples [see Fig.~\ref{FigS12}(c)].

We argue that such negative magnetoresistance is qualitatively explained
by our result~(\ref{MR_B_straight_gamma_shrt}) obtained  for the first ballistic subregime
in the case of the short samples, $W\sim L \ll l$. We remind that
formula~(\ref{MR_B_straight_gamma_shrt})
 originates from
  the bending of the ballistic trajectories
 of the travelling electrons, leading  on the increase their average  length~\cite{we_6}.
   Similar negative ballistic
   magnetoresistance~(\ref{MR_B_straight_gamma}) is also possible in long samples at $ \omega_c \ll \gamma ^2 W $,
   when the effect of the near-edge layers with the skipping  is suppressed [the small orange peak near  $B = 0$ in Fig.~\ref{FigS5}(a)].
   However, in the narrow samples such magnetoresistance is much more pronounced and therefore
   can be more easily observed.

Due to the presence of the macroscopic oval defects in the samples examined in Ref.~\cite{exps_neg_3}  (see details in Ref.~\cite{example})
    and, possibly, in Refs.~\cite{exps_neg_1,exps_neg_2},
 the characteristic width and length of the flows
 can differ from $W$ and $L$ of the those samples.
   Such effective width  $W_{eff}   $ is to be
 of the order of the mean distance between the oval defects and
  becomes smaller than $ W  $
(see discussions in Refs.~\cite{je_visc},\cite{we_6}).
  In this case, the purely  ballistic regime  is realized when
  the cyclotron diameter $2R_c$
  and
  the bulk scattering  length $l$, related to
 the scattering on other disorder
 and/or interparticle collisions,
  are larger than the effective width $W_{eff} $ and length $L_{eff} $.
 At $W_{eff} \sim L_{eff} $,
  the ballistic negative magnetoresistance~(\ref{MR_B_straight_gamma_shrt})
   should appear  at $\omega_c \ll 1/W_{eff}$,
 unlike  the case of long stripes, in which
 such magnetoresistance is realized in the subregimess
  $\omega_c \ll \gamma^2 W $ or $\omega_c \ll W / L   ^2 $ (see Sec.~2.3).

Our estimates shows that  the temperature-independent  negative  magnetoresistance in Figs.~\ref{FigS12}(a-c)
 occurs in the magnetic fields $B \ll  B_{c}^{eff}$, where $ B_{c}^{eff}$ corresponds to the cyclotron radius $R_c$
  of the order of   the mean distances between the oval defects and,  thus, of the effective width  $W_{eff}  $
  (see Ref.~\cite{we_6}  for details). In the larger magnetic fields,  $  B \sim  B_{c}^{eff} $,
 a smeared ballistic-hydrodynamic transition is to be realized  by the mechanism similar to the one studied in Sec.~3
  for the long samples (see  also the discussion   in the end of Sec.~3.2).
 At $ B \gg  B_{c}^{eff} $ the hydrodynamic  temperature-dependent magnetoresistance   arises due to the formation
  of a viscous flow.

Finally, we note that the sharp ballistic-hydrodynamic transition in 2D electron  flows
can be
possibly realized not only in stripes, but also in
 the bulk samples similar to the discussed above GaAs quantum wells with oval
 defects, but containing thermodynamically large numbers
 of such  defects.
  Magnetotransport in such systems was theoretically considered in many works
 without taking into account the hydrodynamic effects  (see, for example,  Refs.~\cite{Entin,Bobylev1,Dmitriev,Beltukov}).
 Accounting for the interparticle collisions, apparently,
  should lead to the realization of the hydrodynamic regime
  at the  magnetic fields $B$ larger than some exact critical field $B_c^{bulk}$,
  corresponding to the cyclotron radius of the order of the mean distance between the defects
  and, additionally,
  to the formation of a connected region (``cluster'') with the 2D fluid nucleus.
  The description of such ballistic-hydrodynamic phase
  transition is to
  require more complex methods than the ones used in this work,
   first of all,  the percolation theory.

\end{document}